\documentclass[12pt,preprint]{aastex}

\usepackage{amssymb}
\usepackage{graphicx}
\usepackage{lscape,longtable,multirow}

\newcommand{\bex}{Be/X-ray}

\newcommand{\smc}{SMC}

\newcommand{\mw}{Milky Way}
\newcommand{\xray}{X-ray}

\newcommand{\xrays}{X-rays}

\newcommand{\lumx}{{\em L}$_{\rm x}$}

\newcommand{\xte}{{\em XTE}}
\newcommand{\rxte}{{\em RXTE}}
\newcommand{\pca}{PCA}
\newcommand{\pcu}{PCU}

\newcommand{\asm}{ASM}
\newcommand{\batse}{{\em BATSE}}
\newcommand{\rosat}{{\em ROSAT}}

\newcommand{\asca}{{\em ASCA}}

\newcommand{\beppo}{{\em BeppoSAX}}
\newcommand{\xmm}{{\em XMM}}

\newcommand{\chandra}{{\em Chandra}}

\newcommand{\einstein}{{\em EINSTEIN}}

\newcommand{\sas}{{\em SAS-3}}
\newcommand{\heao}{{\em HEAO 1}}
\newcommand{\macho}{MACHO}
\newcommand{\ogle}{OGLE}
\newcommand{\counts}{{\rm\thinspace counts}}

\newcommand{\erg}{{\rm\thinspace erg}}

\newcommand{\G}{{\rm\thinspace G}}

\newcommand{\keV}{\hbox{{\rm\thinspace keV}}}
\newcommand{\kev}{\hbox{{\rm\thinspace keV}}}

\newcommand{\m}{{\rm\thinspace m}}

\newcommand{\s}{{\rm\thinspace s}}
\newcommand{\pers}{{\rm\thinspace s}$^{-1}$}
\newcommand{\ks}{{\rm\thinspace ks}}

\newcommand{\yr}{{\rm\thinspace yr}}

\newcommand{\dy}{{\rm\thinspace d}}
\newcommand{\hr}{{\rm\thinspace h}}

\newcommand{\ergps}{\mbox{$\erg\s^{-1}$}}

\newcommand{\Msun}{\thinspace M_\odot}                                      

\newcommand{\asec}{\hbox{\rm\thinspace arcsec}}

\newcommand{\ts}{\thinspace}
\newcommand{\supers}[1]{$^{\rm{#1}}$}                                       
\newcommand{\supersm}[1]{\ensuremath{^{#1}}}                                
\newcommand{\subs}[1]{$_{\rm{#1}}$}                                         

\newcommand{\nd}[0]{\supers{nd}}
\newcommand{\rd}[0]{\supers{rd}}

\newcommand{\tref}[1]{Table~\ref{#1}}                                       
\newcommand{\aref}[1]{Appendix~\ref{#1}}                                    
\newcommand{\fref}[1]{Fig.~\ref{#1}}                                        
\newcommand{\eqref}[1]{(\ref{#1})}                                          
\newcommand{\eref}[1]{Eq.~\eqref{#1}}                                       
\newcommand{\sref}[1]{\S\ts\ref{#1}}                                        
\newcommand{\sd}{\ts --\ts}                                                 
\newcommand{\psps}{P$^2$S$^2$}
\newcommand{\x}{\ts$\times$\ts}

\newcommand{\degree}{\ensuremath{^\circ}}
\newcommand{\microsec}{\rm\ts \ensuremath{\mu}s}
\newcommand{\aprox}{\ensuremath{\sim}\ts}
\newcommand{\cpcus}{\counts \ts \pcu$^{-1} \s^{-1}$}

\newcommand{\kg}{\ts {\rm kg}}

\newcommand{\halpha}{H$\alpha$}
\newcommand{\HI}{H{\rm\thinspace {\footnotesize I}}}
\newcommand{\porb}{\textit{P}\subs{orb}}                                    
\newcommand{\ps}{\textit{P}\subs{s}}                                        
\newcommand{\pdot}{$\dot{P}$}
\newcommand{\tp}{\textit{T}\subs{P}}

\newcommand{\msunt}{\ts \textit{M}\subs{\ensuremath\odot}}                  

\newcommand{\gauss}{\ts G}
\newcommand{\his}{\noindent\textbf{History:} }
\newcommand{\sur}{\noindent\textbf{Survey Results:} }
\newcommand{\fpul}{\textit{F}\ensuremath{_{{\rm x}_{\rm{pul}}}}}            
\newcommand{\pmt}{\ts$\pm$\ts}                                              
\newcommand{\expt}[1]{10\supersm{#1}}                                       
\newcommand{\pdoteq}[2]{$\dot{P} = {#1}\times 10^{#2}$\s\pers}
\newcommand{\lumxeq}[2]{{\em L}\subs{x} = {#1}\x \expt{#2}\ergps}
\newcommand{\lumxge}[2]{{\em L}\subs{x} \ensuremath{\geq} {#1}\x \expt{#2}\ergps}

\newcommand{\ble}[2]{{\em B} \ensuremath{\leq} {#1}\x \expt{#2}\gauss}

\newcommand{\minus}{$-$}

\newcommand{\pcom}[1]{{#1}, private communication}

\shorttitle{A Long Look at the Be/X-Ray Binaries of the Small Magellanic Cloud}
\shortauthors{Galache et al.}
\bigskip

\begin{document}

\title{A Long Look at the Be/X-Ray Binaries of the Small Magellanic Cloud}

\author{J.L. Galache}
\affil{Harvard-Smithsonian Center for Astrophysics, 60 Garden Street, Cambridge, MA 02138}
\author{R.H.D. Corbet\altaffilmark{1}}
\affil{NASA Goddard Space Flight Center, Greenbelt, MD 20771}
\author{M.J. Coe}
\affil{School of Physics and Astronomy, University of Southampton, Southampton SO17 1BJ, UK}
\author{S. Laycock}
\affil{Harvard-Smithsonian Center for Astrophysics, 60 Garden Street, Cambridge, MA 02138}
\author{M.P.E. Schurch}
\affil{School of Physics and Astronomy, University of Southampton, Southampton SO17 1BJ, UK}
\author{C. Markwardt\altaffilmark{2}}
\affil{NASA Goddard Space Flight Center, Greenbelt, MD 20771}
\author{F.E. Marshall}
\affil{NASA Goddard Space Flight Center, Greenbelt, MD 20771}
\and
\author{J. Lochner\altaffilmark{3}}
\affil{NASA Goddard Space Flight Center, Greenbelt, MD 20771}

\altaffiltext{1}{University of Maryland, Baltimore County, MD 21250}
\altaffiltext{2}{Department of Astronomy, University of Maryland, College Park, MD 20742}
\altaffiltext{3}{Universities Space Research Association}

\begin{abstract}
We have monitored 41 \bex\ binary systems in the Small Magellanic Cloud over \aprox9 years using \pca -\rxte\ data from a weekly survey program. The resulting light curves were analysed in search of orbital modulations with the result that 10 known orbital ephemerides were confirmed and refined, while 10 new ones where determined. A large number of \xray\ orbital profiles are presented for the first time, showing similar characteristics over a wide range of orbital periods. Lastly, three pulsars: SXP46.4, SXP89.0 and SXP165 were found to be misidentifications of SXP46.6, SXP91.1 and SXP169, respectively.
\end{abstract}


\section{Introduction}

The Small Magellanic Cloud (\smc) has become a fertile orchard of High-Mass X-ray Binaries (HMXBs), with 49 confirmed systems \citep{coe2005,mcgowan2007}. Given that, from extrapolation of the Milky Way's population (and even correcting for the higher Be/B ratio in the \smc\ \citep{maeder99}), one would expect to find only 3--4 systems, it is clear that the \smc\ is a special place where recent star formation has provided an abundance of HMXBs; indeed, the SFR/M (Star Formation Rate/Mass) of the \smc\ is 150 times that of the \mw\ \citep{grimm2003}. Of particular significance is the fact that only one of these binary systems is not a \bex\ binary (SMC~X-1 is the sole supergiant system discovered so far). This large number of \bex\ binary systems, conveniently located within the 3\degree$\times$\ts3\degree area of the \smc, provides an unrivaled opportunity to study this population as a whole, as well as individually. With a 2\degree\ FWZI field of view, high timing resolution (1\microsec), and sensitive enough to detect the \expt{36}\sd\expt{38}\ergps\ luminosities typical of these systems when in outburst, the \pca\ instrument \citep{jahoda96,jahoda2006} on board \rxte\ is well suited for a long-term monitoring survey of the \smc.

The different types of \xray\ activity displayed by \bex\ transient systems were classified by \citet{stella86} into the following categories:

\begin{list}{\textbf{--} }{}
  \item Persistent low-luminosity \xray\ emission (\lumx\ $\lesssim$ 10\supersm{36}\ergps) or none detectable (in which case the system is said to be in quiescence).
  \item Type~I outbursts: Outbursts of moderate intensity (\lumx\ $\simeq$ 10\supersm{36}\sd10\supersm{37}\ergps) and short duration (a few days) generally recurring with the orbital period of the system and taking place at, or close to the time of periastron passage.
  \item Type~II outbursts: Giant \xray\ outbursts (\lumx\ $\gtrsim$ 10\supersm{37}\ergps) lasting for weeks or months that generally show no correlation with orbital phase.
\end{list}

The data presented in this paper spans November 1997 -- November 2006, and builds upon the work of \citet{laycock2005}, who analysed the first 4 years of data. Following is a brief description of the survey so far, which is still ongoing as of June 2007.

\subsection{The survey}

The initial observations of the \smc\ with \rxte\ began in 1997. The first observation took place in November of that year when an outburst detected by the \asm\ was missidentified as \smc~X-3. It was \rxte 's second year of operation, and only \smc~X-1, 2 and 3, SXP2.76 and SXP8.88 were known. From these initial observations it soon became apparent that there were more than 5 X-ray pulsars in the \smc. The observations carried out within the next year brought about the discovery of 5 new systems: SXPs 46.6, 59.0, 74.7, 91.1 and 169. Other pulsars were also discovered or detected with \einstein, \asca, \beppo\ and \rosat.

\begin{figure*}
  \includegraphics[]{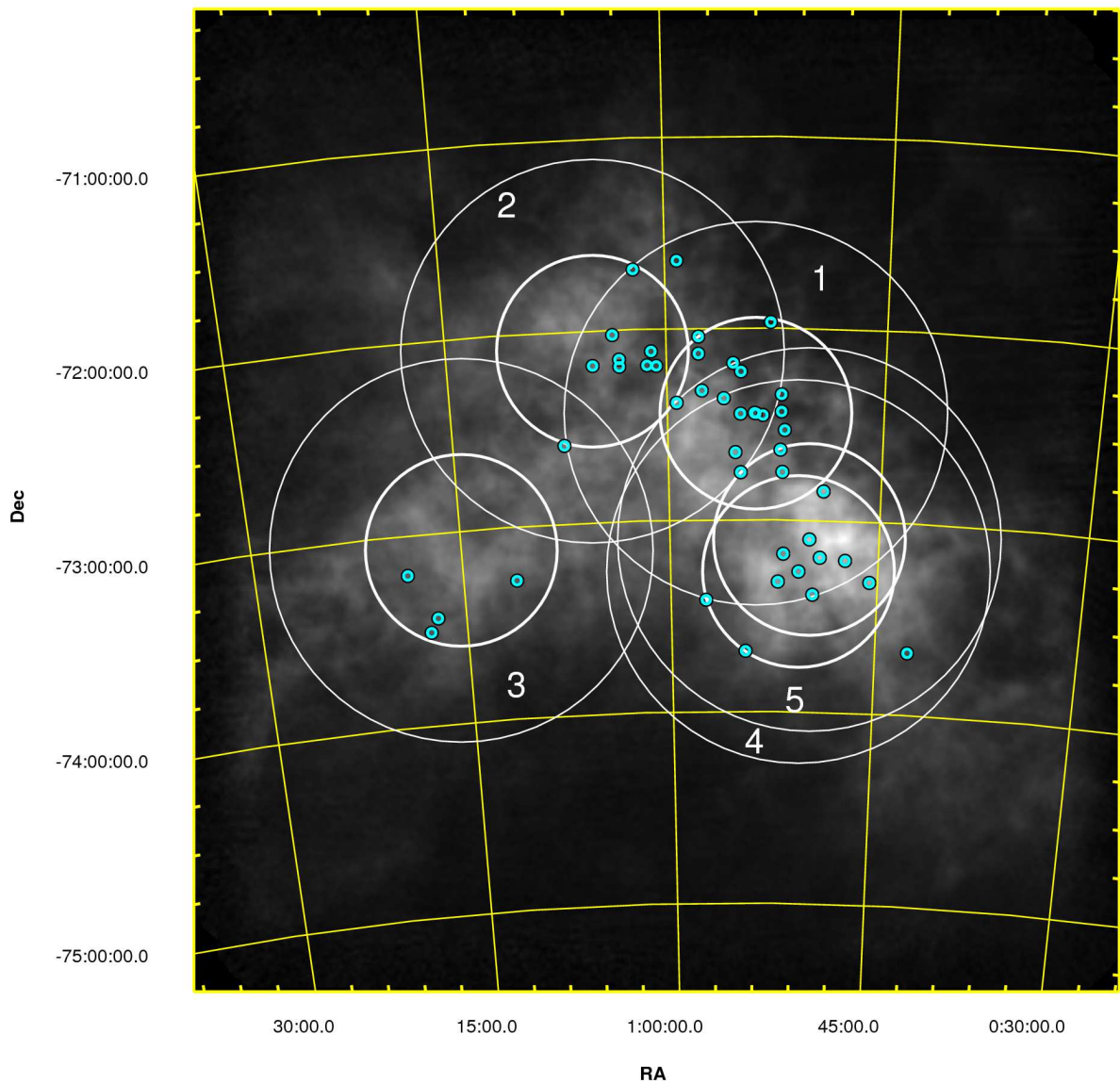}
  \caption{Map of the \smc\ \protect\HI\ distribution with the 5 main pointing positions of \rxte\ during the survey shown as numbered white circles. For each pointing, the inner circle has a diameter of 1\degree, the outer of 2\degree. For clarity, Positions A, B, C, D and X have not been plotted, as they are very close to Positions 1, 2, 3, 4 and 2, respectively. Pulsars with known positions are marked by small circles. \protect\HI\ data from \citet{stanimirovic99}.}
  \label{fig_smc_point_pos}
\end{figure*}

1999 marked the beginning of a coordinated survey of the \smc\ using the \pca. The \pca 's 2\degree\ field of view provides coverage of a wide area of the \smc, which allows many pulsars to be monitored with just one pointing. A number of different pointing positions have been used throughout the years and are given in \tref{table_pointings} (see also \fref{fig_smc_point_pos}); some of the less observed ones never received a name and are omitted from the table. The most frequently observed is Position~1 (later renamed to A), which has been the main pointing position since AO4, except during AO6 and AO7, when Position~5 was the main target. In total, the collected data spans \aprox9 years.

\input{tab1.tex}

The survey has gone through various phases characterised by different observing positions and/or observing modalities. Phases 1\sd4 have already been described in \citet{laycock2005} but they are outlined here again, together with the two latest phases not included in previous studies.

\begin{description}
  \item [Phase 1] (AO2\sd AO3): These observations used Positions 1a, 1b and 1c and are described in \citet{lochner99a,lochner99b}. Only 30 observations were carried out, their main purpose being to monitor the 5 newly discovered pulsars in those regions.
  \item [Phase 2] (AO4): Positions 1\sd4 were defined; Position~1 overlaps most of Positions 1a\sd c and contained the new pulsars, Positions 2 and 4 cover the rest of the bar of the \smc\ while Position~3 covers the area of the wing containing \smc~X-1. A continued survey began at this point in time, with \aprox3\ks\ observations being made once a week, mostly of Position~1, with some occasional looks at the other positions.
  \item [Phase 3] (AO5\sd AO6): Only Position~5 was monitored, and was done weekly so as not to create gaps in the data; the majority of the most active systems located in Position~1 also fell within the field of view of Position~5. Time allotted for the project was increased to an average of \aprox5\ks\ per observation, thus providing better sensitivity to longer period pulsations.
  \item [Phase 4] (AO7\sd AO9): Weekly monitoring returned to Position~1, now renamed Position~A, with additional observations of the other positions (B, C and D, which are very close to 2, 3 and 4, respectively) being made once a month. These monthly pointings were \aprox15\ks, while the weekly ones were increased to \aprox6\sd7\ks.
  \item [Phase 5] (AO10): The time available for the monthly observations of alternate positions was invested into increasing the length of the weekly observations of Position~A to \aprox10\ks, this being the only monitored position.
  \item [Phase 6] (AO11): Having mainly monitored the central bar of the \smc, it was decided to move to another location at the northeastern tip of the bar, near position B. This new location, Position~X, was monitored for \aprox10\ks\ weekly for 18 weeks, but was abandoned due to lack of pulsar activity. Observations then changed to Position~D.
\end{description}

\subsection{Data reduction}

Cleaning of the raw light curves was achieved via a pipeline employing various \texttt{FTOOLS}\footnote{http://heasarc.gsfc.nasa.gov/ftools/} routines \citep{blackburn95}. Initial filtering required the data to come from observations offset from the target no more than 0.004\degree, with an elevation above the Earth's horizon of more than 5\degree, and not taken during times of high background. \textit{Good Xenon} mode data from the top anode layer only were used, within an energy range of 3--10\kev. Data were binned at 0.01\s\ intervals, while background files were generated in 16\s\ bins. For each time-step, the net flux was divided by the number of active \pcu s to give the \cpcus. Then, the light curves time tags were corrected to the Solar System barycenter, removing timing variations caused by the satellite's motion around the Earth and Sun. Finally, short light curves belonging to the same observation (which had been split up due to SAA passage, Earth occultation or flares) were pieced together into a long light curve spanning the whole observation.

\subsection{Collimator correction}

The collimator of each \pcu\ consists of a number of hexagonal cells joined in a honeycomb structure. The hexagonal tubes are not perfectly parallel with the result that all the \pcu s are slightly off center. Hence, the resulting collimator response pattern has an elliptical hexagon shape with the maximum throughput being slightly off center.

In order to account for differences in observed flux from a pulsar when observed at different pointing positions, a collimator correction was applied \textit{a posteriori} to each pulsar's count rate. A look-up table approach was used, where each pulsar had a collimator response calculated for each of the pointing positions used in the survey (see \tref{table_pointings}).

A pulsars with unknown position cannot be collimator corrected, so significant detections may not rise above the noise level in the long-term light curve unless the outbursts are very bright and/or the pulsar is located close to the center of the field of view. Type~I outbursts may only appear bright when the pulsar is close to the center of the field of view (as with SXP59.0 in Position~1/A, \fref{fig_sxp59.0}), in which case the collimator correction is small anyway. To overcome this limitation pulsars with unknown positions had coordinates ``guessed" for them on the basis of which observing positions they had been detected in throughout the mission; the coordinates assigned were approximately at the center of the region formed by the overlap of the observing positions in which it had been previously detected.

\subsection{Data analysis}

Once the cleaned light curve for an observation was obtained, it was run through a timing analysis package using the Lomb-Scargle method \citep{lomb76,scargle82} and numerically implemented following the prescription by \citet{press89}. Two different power spectra were generated for each light curve, each spanning different period spaces (from $P_{\rm min}$ to $P_{\rm max}$), at different timing resolutions ($\Delta f$), and pulsars were later searched for in their corresponding group according to their pulse period. The parameters used for each group are listed in \tref{table_LS_params}. The reason for creating two groups was to obtain densely sampled periodograms at long periods without creating excessively large files. The parameters were chosen such that both groups would contain approximately the same number of independent frequencies, $M$.

\input{tab2.tex}

Once the periodograms were calculated, it was necessary to establish an objective way to determine which peaks in the power spectra were real; this is covered in \aref{sec_appendix_a}.

\subsubsection{Pulsar Detection}

Pulsars were searched for within the data using an algorithm specifically created for the task. Each light curve's two power spectra were scanned to look for the pulsars in groups 1 and 2; the steps used were as follows:

\begin{list}{Step}{}
  \item 1: The highest peak (at a period $P$) above 90\% global significance in the power spectrum is found and, either identified as a known pulsar, or flagged as an ``unknown''.
  \item 2: The light curve is folded at period $P$ to produce a pulse profile.
  \item 3: Using the pulse profile as a template, the pulsations are then subtracted from the light curve and a power spectrum of the cleaned light curve is produced.
  \item 4: This power spectrum is subtracted from the previous one to create a pulsar-specific power spectrum (or \psps), which shows only the contribution of the individual pulsar to the power spectrum. This method allows one to see the possible harmonics that may have been lost in the noise or confused with the fundamental of another pulsar. The power of the fundamental and harmonic(s) peaks in the \psps\ is measured and a pulse amplitude estimated using \eref{eq_ls_totamp}. The significance of the detection is taken to be the local significance of a peak of power equal to that which the detected signal would have if the power in the harmonics were concentrated in the fundamental.
  \item 5: Repeat previous steps until all signals with peaks above 90\% significance have been removed.
  \item 6: To account for the remaining, dim pulsars, a stretch of the power spectrum centered on each pulsar's nominal frequency, and with a total width which is different for each pulsar (and depends on the pulsar's past period history), is searched for a peak. If no peak is found, then the significance of the detection is set to 0\% and the amplitude to that of the average power within the region. If there is a peak, then the significance of the detection is set to the local significance and the amplitude of the pulsar calculated from the power in the peak according to \eref{eq_ls_amp}. No harmonics are searched for in this case.
\end{list}

\subsubsection{Light Curve Generation and \porb\ Calculations}

Once the amplitude of the pulsed flux for every pulsar in every observation had been measured, a long-term light curve was created for each system showing its X-ray activity over the course of the survey. It is important to note that these light curves do not show the absolute flux, but rather the pulsed flux. Without knowing the pulsar's pulse fraction for each observation it is impossible to convert the pulsed flux amplitude into an absolute flux value. These light curves were then searched for periodicities, as the X-ray emission could show orbital modulations. Again, the Lomb-Scargle technique was employed, searching for periods in the appropriate range for each pulsar. In cases where Type~II outbursts contaminated the Lomb-Scargle periodogram, these were removed from the data. In general, all bright outbursts were initially removed and only added back into the light curve if they coincided with the calculated ephemerides and their inclusion increased the significance of the calculated orbitla period. It is important to note that the collimator response correction was only applied to the data points whose local significance upon detection was $\geq 99\%$, anything below this detection threshold was considered noise, and as such was not collimator corrected. \sref{sec_results} shows the results of these analyses.

\subsubsection{Orbital Profile Generation}

Based on the ideas of \citet{dejager88} and \citet{carstairsthesis}, we used the Phase-Independent Folding (PIF) technique to create the orbital profiles. The method is as follows: the folded light curve is obtained from $m$ sets of $n$-binned folded light curves. To begin with, the light curve is folded at the desired period and the time values converted to phase space (ranging from 0 to 1). This ``raw'' folded light curve is then binned into $n$ bins (of width $1/n$) in the standard way, with each bin starting at phase $a/n$ (with $a = 0,1,2,...n-1$). This step is repeated again, but this time the bins start at phase $a/n + 1/(n \times m)$. In this way we create $m$ folded light curves from the same data, each consisting of $n$ bins, and only differing in the starting phase of these bins\footnote{We define $\phi = 0$ at the time of the first data point in each observation.}. The general expression for the phase at which each bin begins is $a/n + m/(n \times m)$. Now each folded light curve is further divided into $l = n \times m$ sub-bins, such that in each light curve there will be $n$ groups of $m$ consecutive bins with the same flux value. The final folded light curve will have $l$ bins, each of which is the average of the bins from the $m$ sets of light curves. The error in each bin will be given by the standard error calculated from the $m$ values that were averaged for each $l$-bin. In the present work we have used $n \times m = 10 \times 5$.

The PIF method provides an efficient way of generating folded light curves from poorly sampled data as their shape will not depend on the starting point at which they are folded and, although only every \textit{m}\supers{th} bin will be independent, spurious flux values within bins will be evened out while real features will remain. Thus, we obtain the benefits of both wide bins (sufficient counts in each bin for good statistics) and narrow bins (sensitivity to narrow features in the profile).

\section{Results} \label{sec_results}

Following are the results obtained from the light curves of the observed \smc\ \xray\ pulsars. In the triple-panel plots (generally the \textit{(a)} plots), the top panel shows the \xray\ activity of each pulsar through the amplitude of the pulsed flux, with each solid line representing one observation. When an orbital period has been found, the ephemerides for the dates of maximum flux are over plotted as dashed vertical lines; where the orbital period is less than 25 days, only every other line is plotted for clarity. The middle panel shows the period at which the pulsar was detected, with the horizontal dashed line denoting the center of the pulsar's search range (note that only significant detections have their periods plotted). Finally, the bottom panel displays the significance of the neutron star's pulsations for each observation, with the three horizontal dashed lines marking, from bottom to top, the levels of 99, 99.99 and 99.9999\% local significance. We considered a pulsar had been detected when its local significance was $\geq 99\%$. In the \textit{(b)} plots we show the Lomb-Scargle periodogram of the pulsed flux light curve (top) and the light curve folded at the orbital period (bottom). When horizontal dashed lines are plotted on the periodogram, these represent different levels of significance, which are (from bottom to top) 99, 99.9, 99.99, and 99.999\%. The coordinates given for each system are the most accurate at time or writing. When a \bex\ system has a confirmed optical counterpart, the coordinates of the counterpart are used. When no optical counterpart is known, but the system has been detected by \chandra\ and/or \xmm, then it is the \xray\ coordinates that were given. For some of the pulsars without an exact position, the coordinates provided by scans with \rxte\ are given; these are the least precise and have errors of up to several arc minutes.

\subsection{SXP0.92}

\textbf{PSR J0045\minus7319\\
RA 00 45 35, dec \minus73 19 02}

\his First discovered as a radio pulsar by \citet{ables87} with a period of 0.926499\pmt0.000003\s\ using the Parkes 64\m\ radio telescope. \citet{kaspi93} observed Doppler shifts in the pulse period which where consistent with a 51\dy\ binary orbit with a companion star having mass $>4\Msun$. Optical observations of the field revealed a 16\supers{th} magnitude, 11\msunt, B1 main-sequence star, which is likely the companion \citep{bell94}.

\begin{figure}
   \centering
        \includegraphics[angle=90,width=0.95\linewidth]{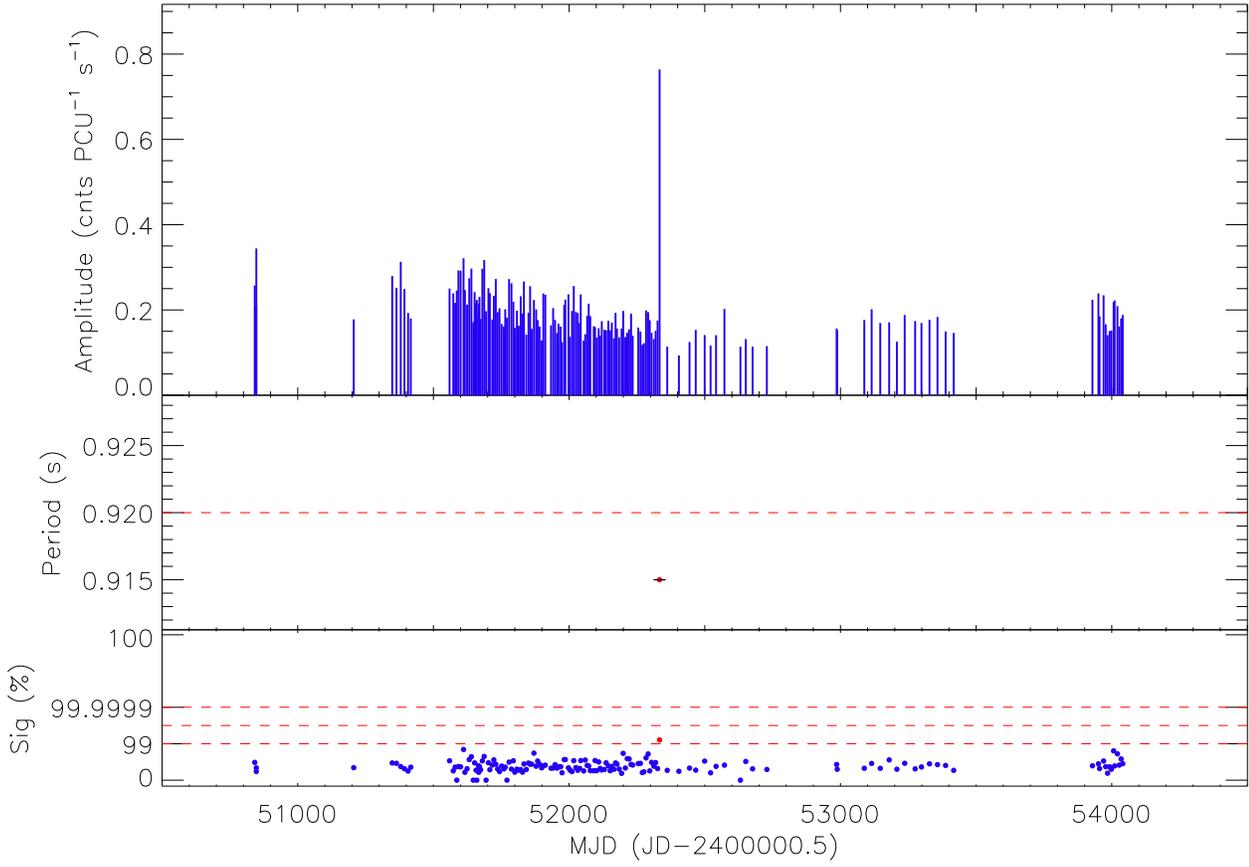}
   \caption{SXP0.92, \xray\ amplitude light curve.}
   \label{fig_sxp0.92}
\end{figure}

\sur SXP0.92 received \aprox2 years of uninterrupted coverage during AO5 and AO6 (see \fref{fig_sxp0.92}), during which time it was only once detected above 99\% local significance (MJD 52334). The power spectrum does not show any significant periods.

\subsection{SXP2.16}

\textbf{XTE SMC2165
RA 01 19 00, dec \minus73 12 27}

\his Discovered during the course of this survey by \citet{corbet2003iauc} at 2.1652\pmt0.0001\s.

\sur Due to its position near SMC~X-1 it has only been within the field of view on 3 occasions: MJD 51220, 51263 and 51310. It was on this last date that it was discovered (its only significant detection).

\subsection{SXP2.37}

\textbf{\smc~X-2\\
RA 00 54 34, dec \minus73 41 03}

\his \smc~X-2 was discovered in \sas\ observations of the \smc\ carried out by \citet{clark78}. Further outbursts were also observed by \heao\ and \rosat, but no pulsations were detected until it was observed in the present survey by \rxte\ during a long outburst lasting from January through to May, 2000 \citep{corbet2001,silasthesis}.

\begin{figure}
   \centering
        \includegraphics[angle=90,width=0.95\linewidth]{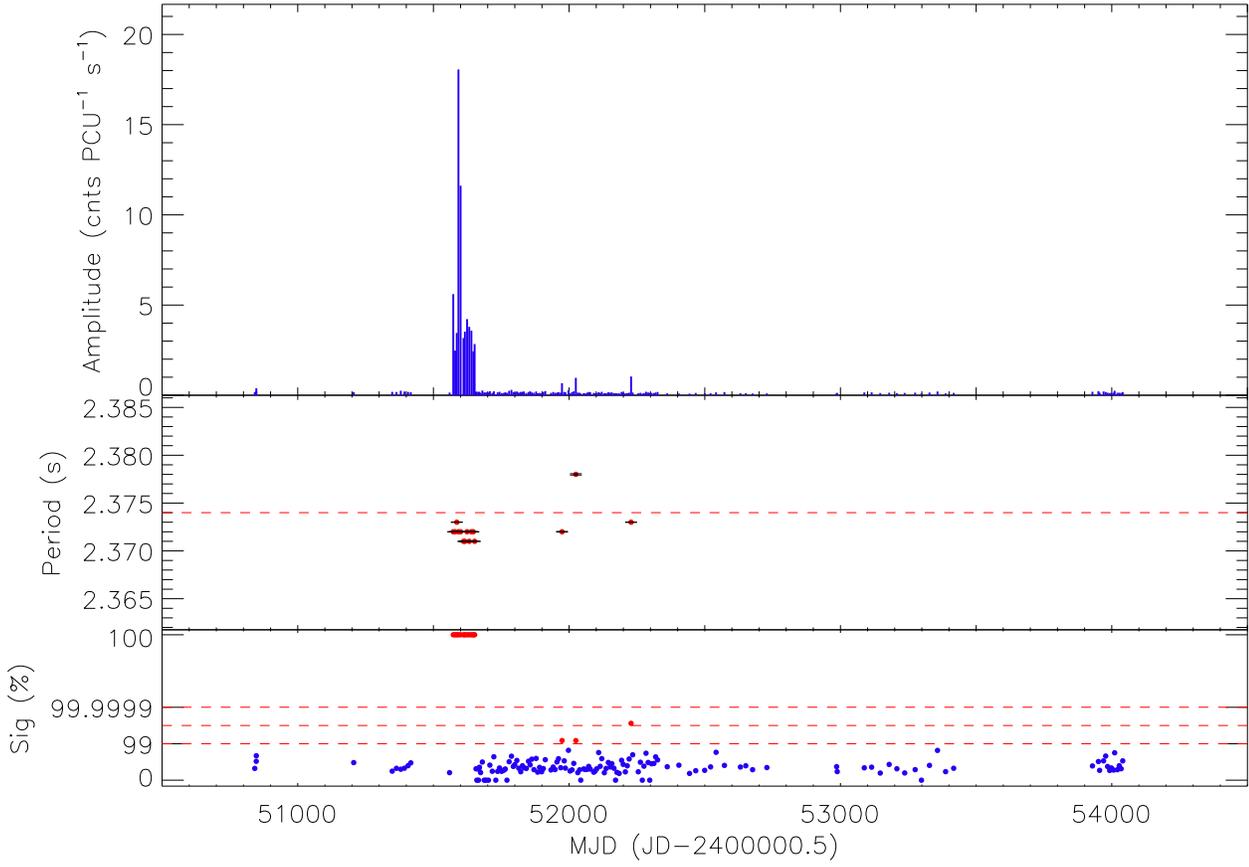}
   \caption{SXP2.37, \xray\ amplitude light curve.}
   \label{fig_sxp2.37}
\end{figure}

\sur After the aforementioned outburst, \smc~X-2 was only detected on 3 further occasions (MJD 51974, 52025 and 52228), but at a much lower flux of \fpul\ $<$ 1\cpcus (see \fref{fig_sxp2.37}). Lomb-Scargle analysis of the data following the major outburst shows no clear periods.

\subsection{SXP2.76}

\textbf{RX J0059.2\minus7138\\
RA 00 59 11.7, dec \minus71 38 48}

\his First reported by \citet{hughes94} from a \rosat\ observation showing 2.7632\s\ pulsations that varied greatly with energy. In the low-energy band of the \rosat\ PSPC (0.07\sd0.4\keV) the source appears almost unpulsed while in the high-energy band (1.0\sd2.4\keV) the flux is \aprox 50\% pulsed. Its optical counterpart was confirmed as a 14\supers{th} magnitude Be star by \citet{southwell96}. \citet{schmidtke2006} report a period of 82.1\dy\ with a maximum at MJD 52188.9 from analysis of \ogle~III data.

\begin{figure}
   \centering
      \includegraphics[angle=90,width=0.95\linewidth]{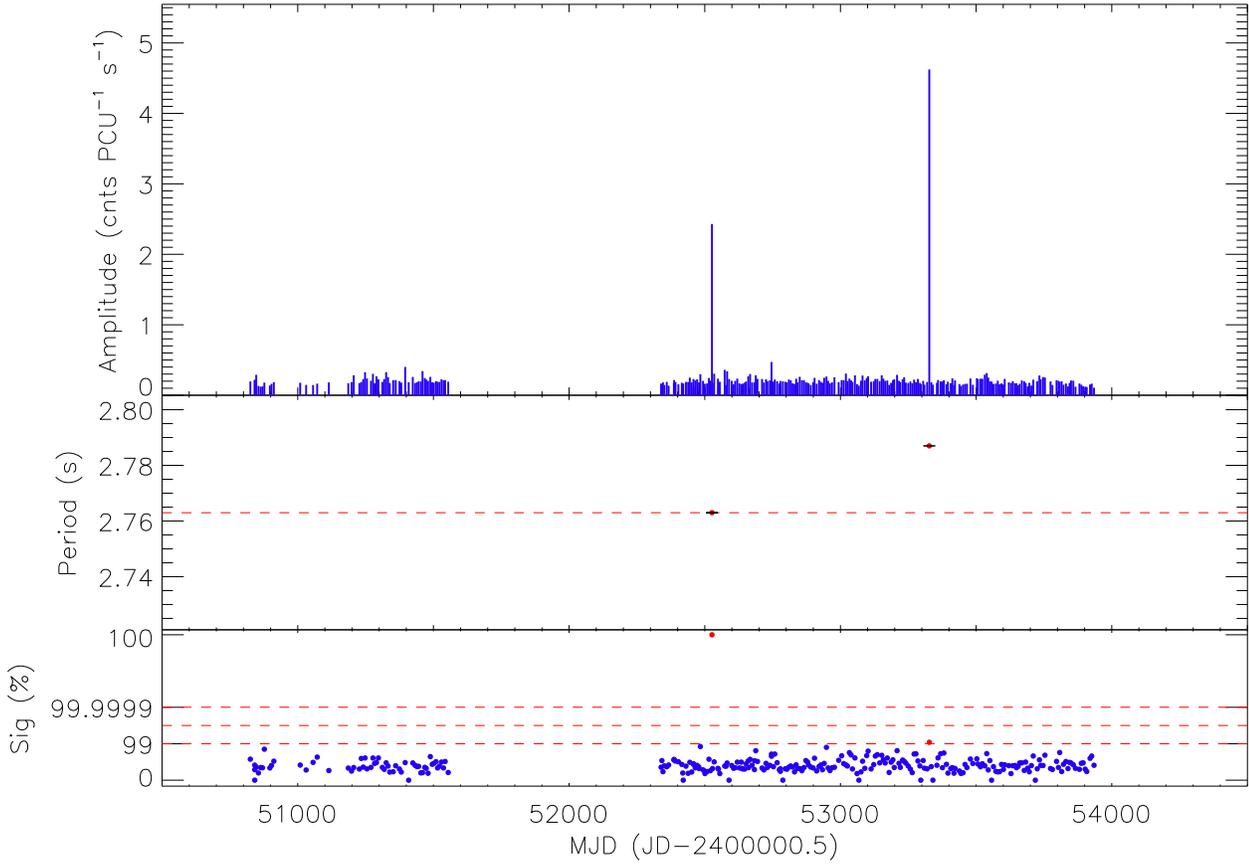}
   \caption{SXP2.76, \xray\ amplitude light curve.}
   \label{fig_sxp2.76}
\end{figure}

\sur This source is near the edge of the field of view of Position 1/A so we would expect to detect only the brighter outbursts. Only 2 detections were made (MJD 52527 and 53327) and no orbital period can be extracted from the \xray\ data (see \fref{fig_sxp2.76}). We find no power at the reported optical period and note that the two \xray\ detections occured \aprox10 days after and \aprox11 days before the optical ephemeris's predicted maximum.

\subsection{SXP3.34}

\textbf{AX J0105\minus722, RX J0105.3\minus7210 \\
RA 01 05 02, dec \minus72 11 00}

\his Was reported as an \asca\ source with 3.34300\pmt0.00003\s\ pulsations by \citet{yokogawa98a}. Its optical counterpart is proposed to be [MA93] 1506 \citep{coe2005}, who also find an 11.09\dy\ modulation in \macho\ data. Although this could be the orbital period of the system (as it falls within the expected range on the Corbet diagram \citep{corbet84}), \citet{schmidtke2005} report the find of a strong 1.099\dy\ period in \macho\ data and that the 11.09\dy\ value is an alias of this main period. They attribute the modulation to non-radial pulsation in the Be star.

\sur There have been no significant detections of the pulsar during this survey and timing analysis of the light curve reveals no clear periodicities.

\subsection{SXP4.78}

\textbf{XTE J0052\minus723 \\
RA 00 52 06.6, dec \minus72 20 44}

\his Was discovered by the present survey in late December 2000 and reported in \citet{laycock2003b}, where [MA93] 537 was proposed as the optical counterpart. Another possible counterpart is suggested in \citet{coe2005} as the star AzV129 is found to have a 23.9\pmt0.1\dy\ period in both \macho\ colours, which would agree with the expected orbital period inferred from the Corbet diagram.

\begin{figure}
   \centering
        \includegraphics[angle=90,width=0.95\linewidth]{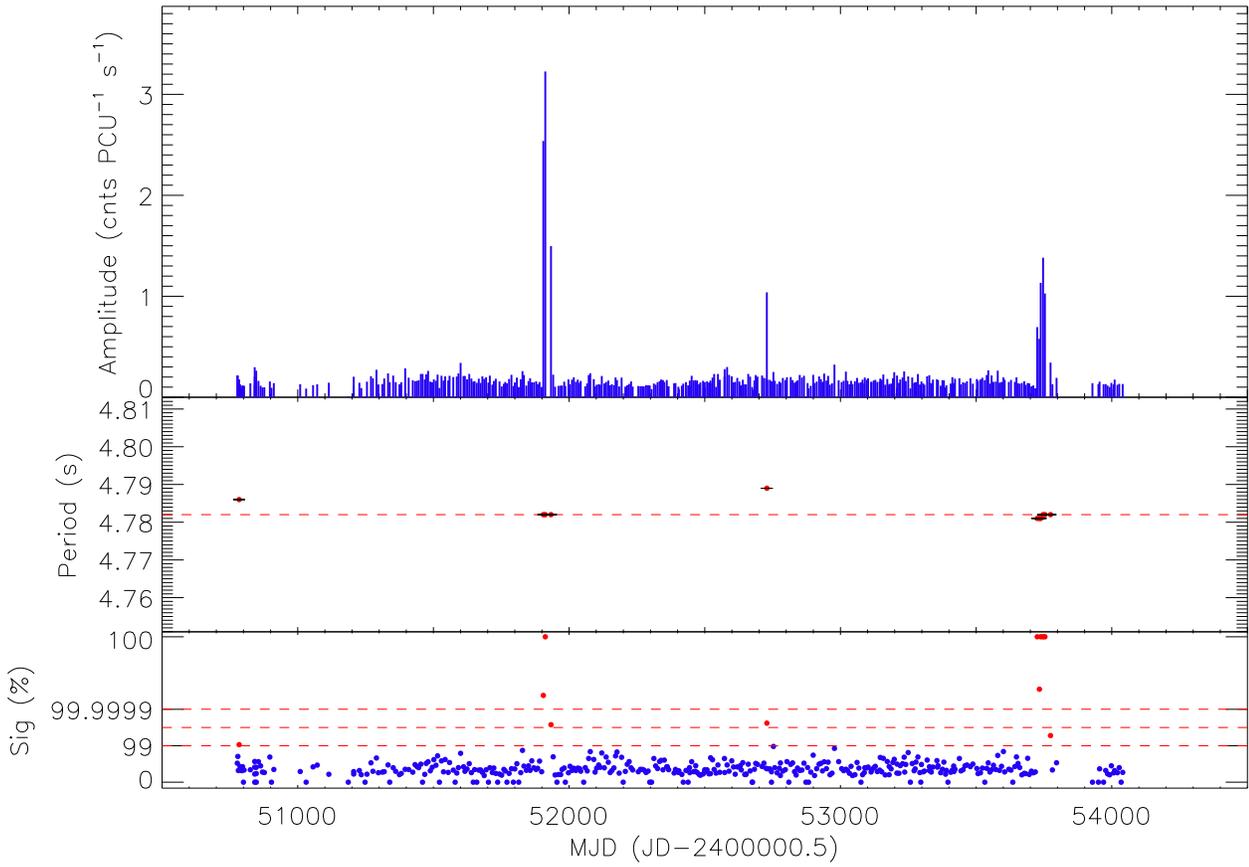}
   \caption{SXP4.78, \xray\ amplitude light curve.}
   \label{fig_sxp4.78}
\end{figure}

\sur SXP4.78 was detected on one occasion after its initial outburst (MJD 52729) at a much weaker \fpul $\simeq$ 0.8\cpcus. It then began a relatively long, bright (peaking at \aprox1.2\cpcus) outburst on December 21, 2005 (MJD 53725) that lasted \aprox7 weeks (see \fref{fig_sxp4.78}). Despite the long outburst, no orbital modulation is apparent. Timing analysis using all the data finds no periods, while analysis of the data outside the two bright outbursts suggests a weak period at 34.1\dy

A 1\ks\ observation with \chandra\ was carried out on March 3\rd\ 2006 in an attempt to establish the pulsar's coordinates. Unfortunately, the outburst had ended and no source was detected within the \rxte\ error box provided by \citet{laycock2003b}.

\subsection{SXP6.85}

\textbf{XTE J0103\minus728 \\
RA 01 02 53.1, dec = \minus72 44 33}

\his First detected in April 2003 with a pulse period of 6.8482\pmt0.0007\s\ \citep{corbet2003atel}. It was detected in outburst with \xmm\ on October 2\nd\ 2006, which provided a more accurate position \citep{haberl2007atel}. This allowed its optical counterpart to be identified as a Be star with V = 14.59, with a possible 24.82\dy\ periodicity \citep{schmidtkecowley2007}.

\begin{figure}
   \centering
        \includegraphics[angle=90,width=0.95\linewidth]{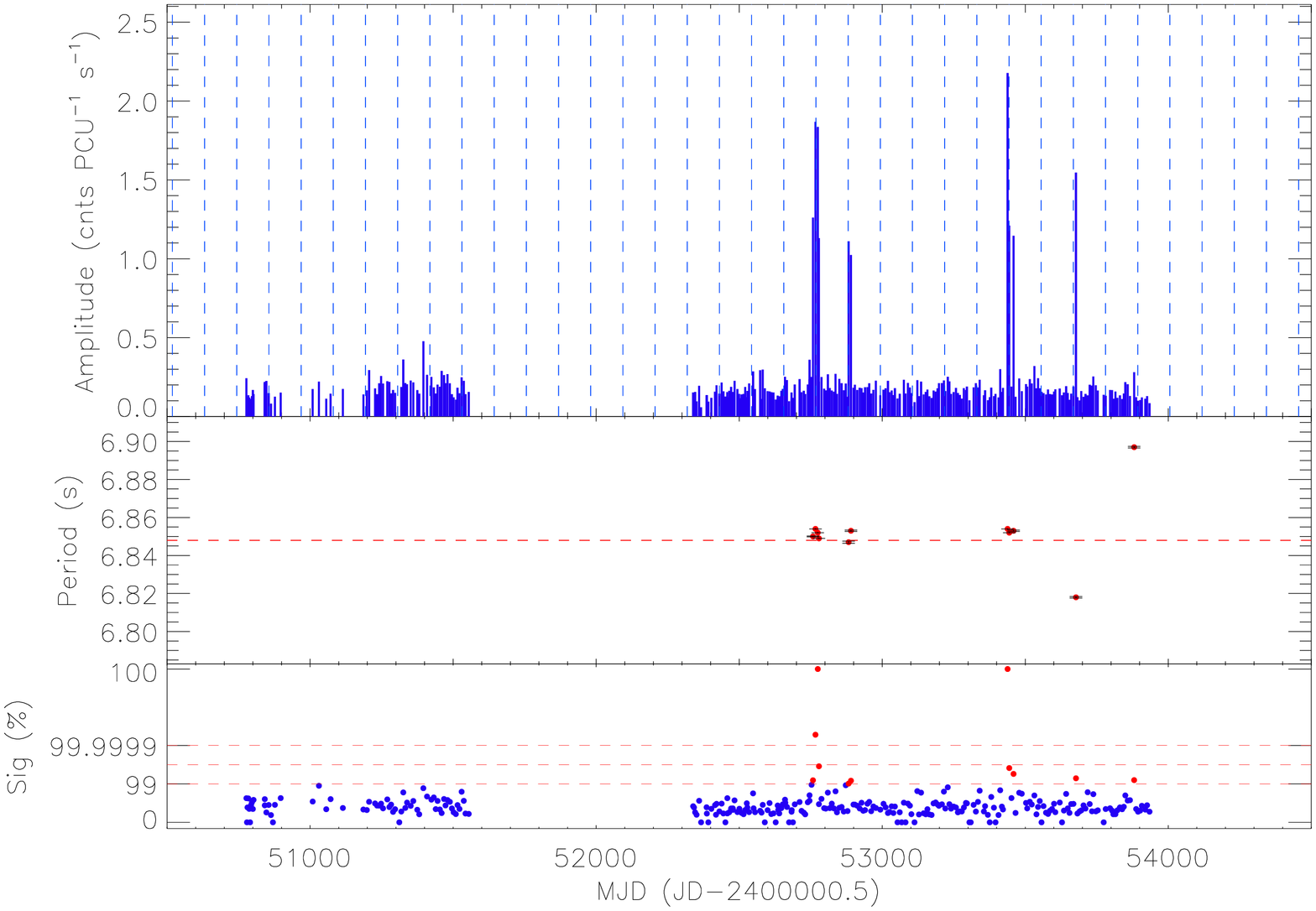}
        \includegraphics[angle=90,width=0.95\linewidth]{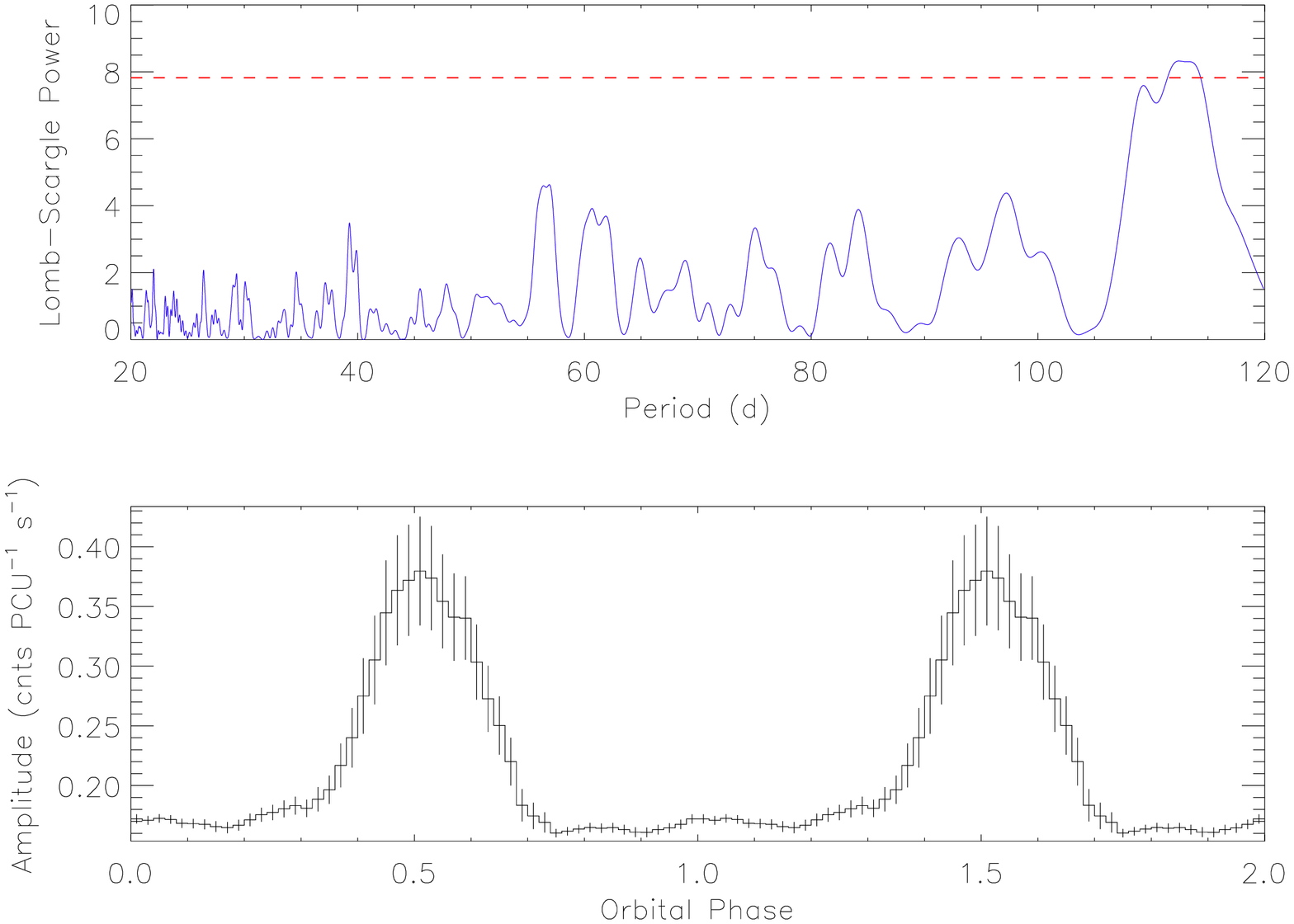}
   \caption{SXP6.85. a) \textit{Top}: \xray\ amplitude light curve. b) \textit{Middle}: Lomb-Scargle power spectrum; \textit{bottom}: light curve folded at 112.5\dy.}
   \label{fig_sxp6.85}
\end{figure}

\sur It has been detected on 3 other occasions (circa MJD 52885, 53440 and 53677) at varying fluxes (\fpul\ $\approx$ 0.8\sd1.8\cpcus), but high above the noise level (see \fref{fig_sxp6.85}). Lomb-Scargle analysis of the data outside the 4 bright detections shows no significant period. When including these outbursts, we find a significant period of 112.5\dy; this is the minimum time lapse between any two outbursts and probably drives the result. Given its pulse period, and based on the spin-orbit relation, we do not discard the possibility that the real orbital period be $\frac{1}{2}$ or $\frac{1}{3}$ of this value. The current \xray\ ephemeris is MJD 52318.5\pmt7.9 + n\x112.5\pmt0.5\dy. We note that the outburst detected on October 2\nd\ 2006 (MJD 54010) is consistent with the proposed ephemeris.

\subsection{SXP7.78 }

\textbf{\smc~X-3 \\
RA 00 52 05.8, dec \minus72 26 03.2}

\his Originally detected by \citet{clark78}, it was not identified with the previously detected \rxte\ 7.78\s\ source until 2004 \citep{edge2004atel}. \citet{corbet2003head,corbet2004proc} proposed an orbital period from a series of recurrent \xray\ outbursts (in the present survey) of 45.1\pmt0.4\dy. An optical modulation in \macho\ data was reported by \citet{cowley2004} (44.86\dy) and \citet{coe2005} (44.6\pmt0.2\dy); \citet{williamthesis} also found a strong 44.8\pmt0.2\dy\ modulation in the \ogle\ counterpart, present even when there was no significant \xray\ activity.

\begin{figure}
   \centering
        \includegraphics[angle=90,width=0.95\linewidth]{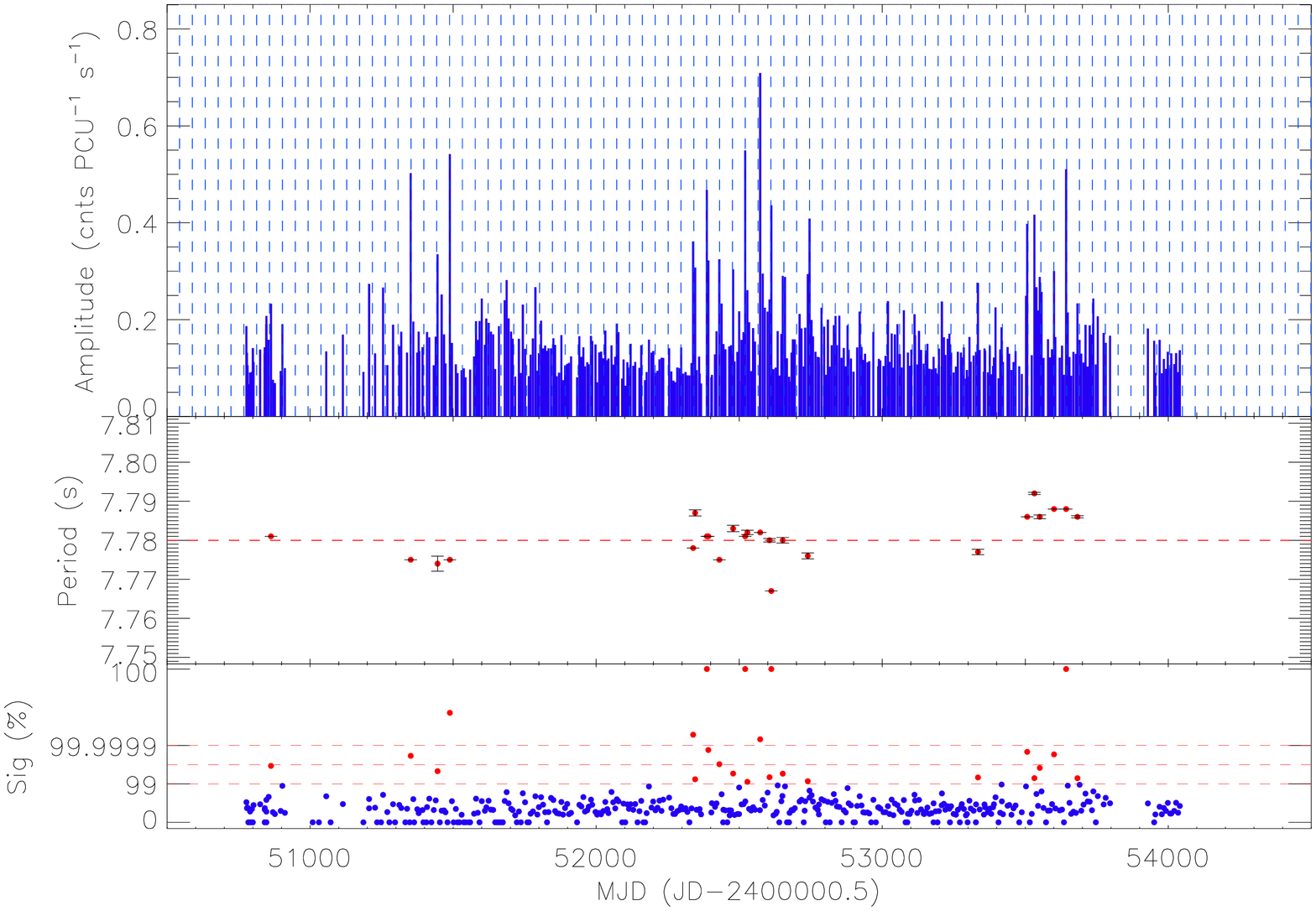}
        \includegraphics[angle=90,width=0.95\linewidth]{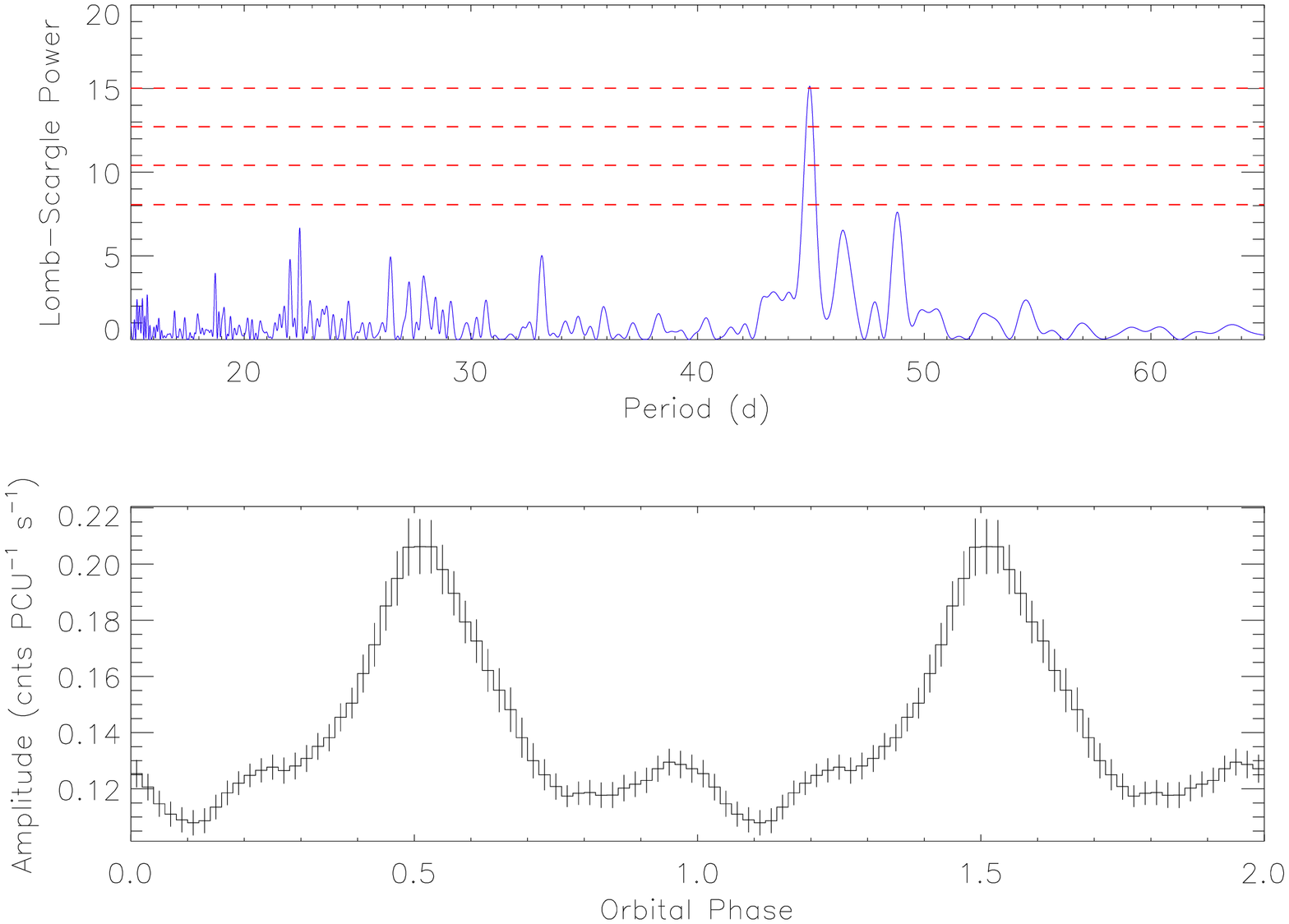}
   \caption{SXP7.78. a) \textit{Top}: \xray\ amplitude light curve. b) \textit{Middle}: Lomb-Scargle power spectrum; \textit{bottom}: light curve folded at 44.92\dy.}
   \label{fig_sxp7.78}
\end{figure}

\sur It has displayed in the past 10 years 3 distinct periods of outbursts (\fpul\ $\approx$ 0.3\cpcus), lasting \aprox200\sd400\dy\ (see \fref{fig_sxp7.78}(a)). Timing analysis of the complete light curve reveals a clear period at 44.92\dy; the ephemeris we derive is MJD 52250.9\pmt1.4 + n\x44.92\pmt0.06\dy. The folded light curve may show evidence of detections at apastron. During the longer of the outburst episodes (circa MJD 52500) the pulsar shows some spin up, about \pdoteq{3.7}{-10}, implying \lumxge{2.3}{37} (\ble{3.6}{12}). This pulsar is unique in that, despite the spin up observed during each of the individual outburst episodes, the overall spin evolution seems to show a long-term spin down.

\subsection{SXP8.02}

\textbf{CXOU J010042.8\minus721132 \\
RA 01 00 41.8, dec \minus72 11 36}

\his Proposed as the first anomalous \xray\ pulsar (AXP) in the \smc\ by \citet{lamb2002}, they found the source to have displayed little variability in the past 20 years. It is characterised by a very soft spectrum and low luminosity (\aprox1.5\x10\supersm{35}\ergps).

\begin{figure}
   \centering
        \includegraphics[angle=90,width=0.95\linewidth]{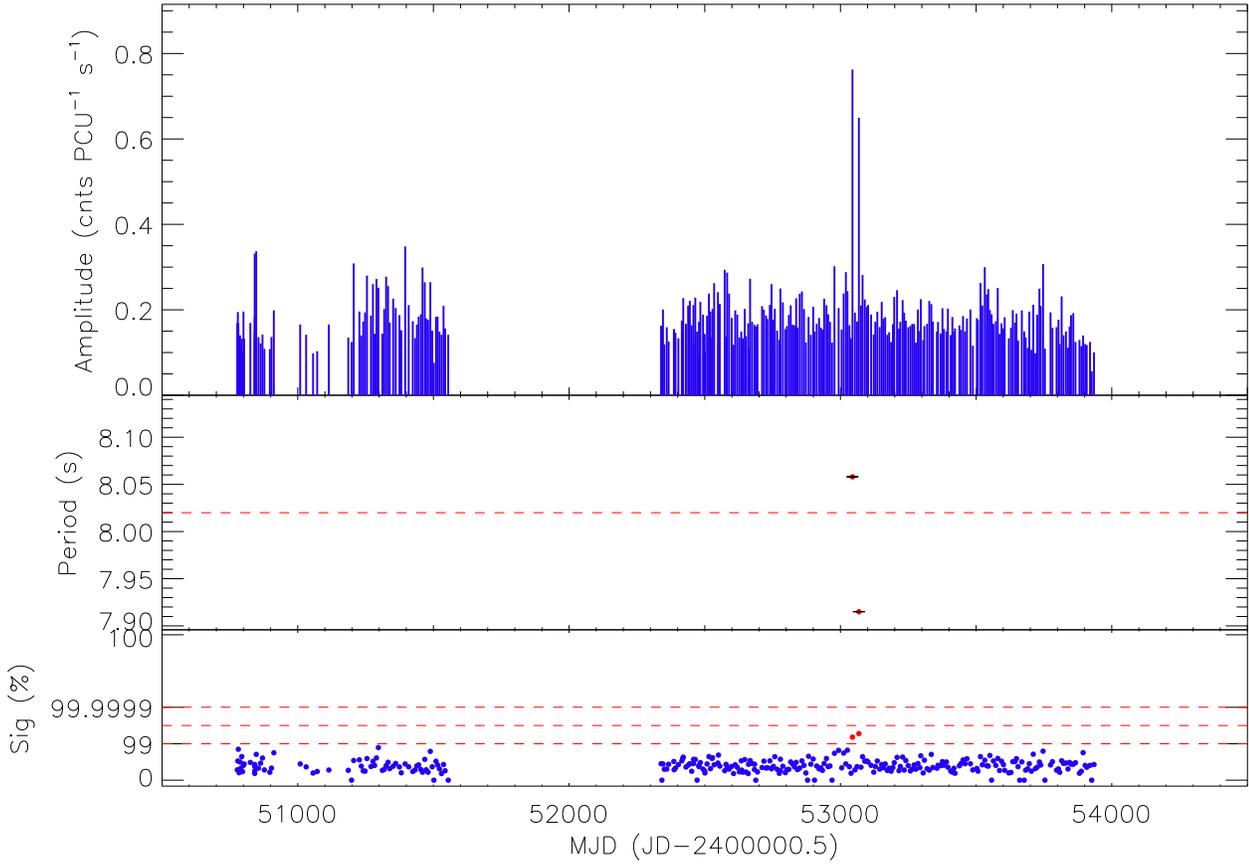}
   \caption{SXP8.02, \xray\ amplitude light curve.}
   \label{fig_sxp8.02}
\end{figure}

\sur During AO5 and AO6 it was outside the field of view of our observations. Coverage from AO7 onwards has been good and timing analysis of these dates shows a possible periodicity at \aprox23.2\dy, which is likely driven by the only two clear detections, separated by that time range (see \fref{fig_sxp8.02}). This period disappears when removing these two detections from the data. If SXP8.02 is truly an anomalous pulsar, it is not expected to show periodicity in its \xray\ emission. The two significant detections were observed at around \lumxeq{5.0}{36} (unabsorbed, assuming a 50\% pulse fraction and a power law spectrum of $\gamma = 1$).

\subsection{SXP8.88}

\textbf{RX J0051.8\minus7231, 1E0050.1\minus7247, 1WGA J0051.8\minus7231 \\
RA 00 51 52.0, dec \minus72 31 51.7}

\his \citet{israel97} detected this source several times between 1979 and 1993. A number of optical counterparts have been proposed, and \citet{haberl2000} believe [MA93] 506 to be the correct one. From a number of outbursts during July 2003\sd May 2004 \citet{corbet2004atel} derived an orbital ephemeris of MJD 52850\pmt2 + n\x28.0\pmt0.3\dy. \citet{schmidtkecowley2006} find an optical period of 33.4\dy\ in \ogle\ and \macho\ data. Note: This system was incorrectly named SXP8.80 in \citet{coe2005}.

\begin{figure}
   \centering
        \includegraphics[angle=90,width=0.95\linewidth]{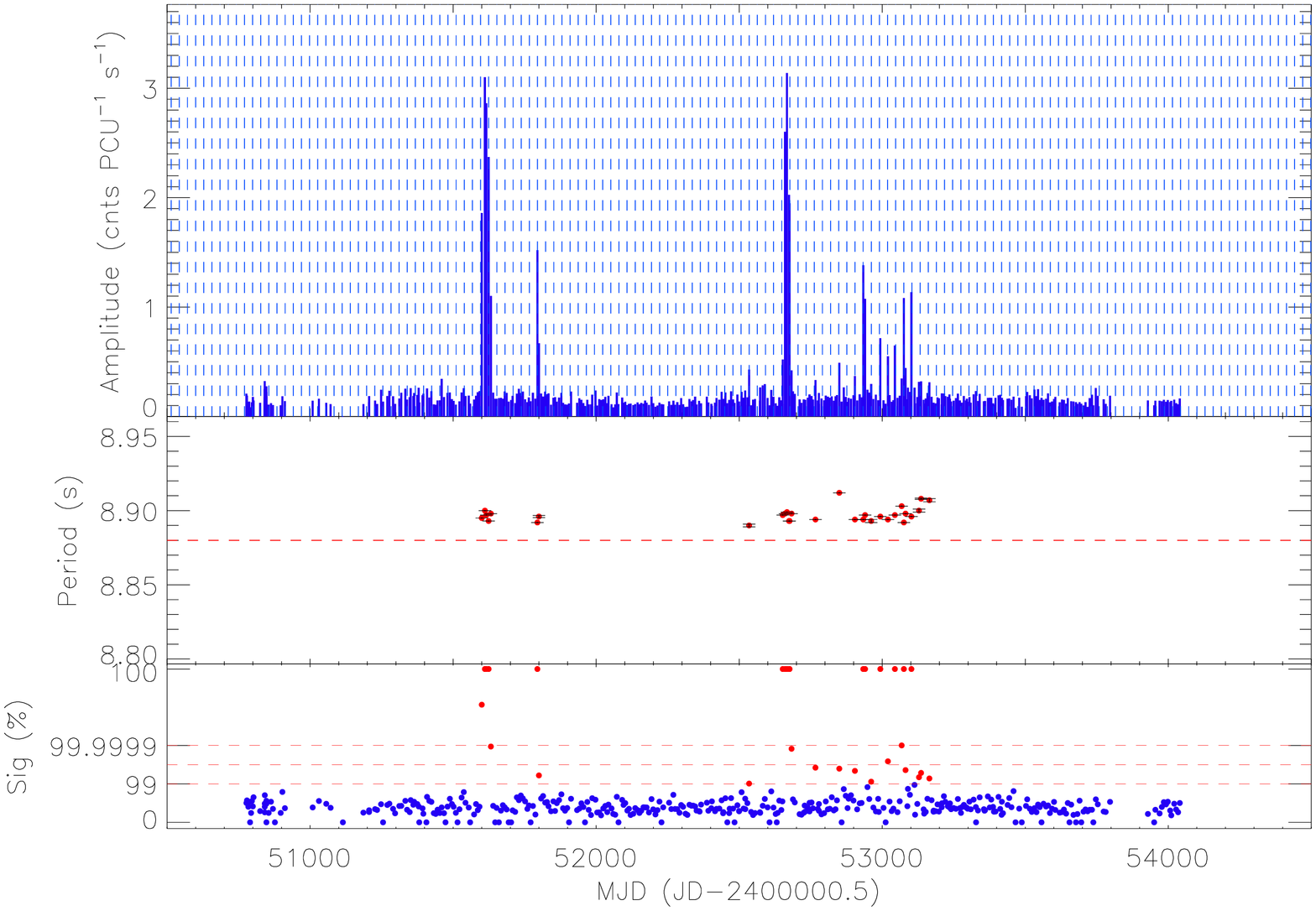}
        \includegraphics[angle=90,width=0.95\linewidth]{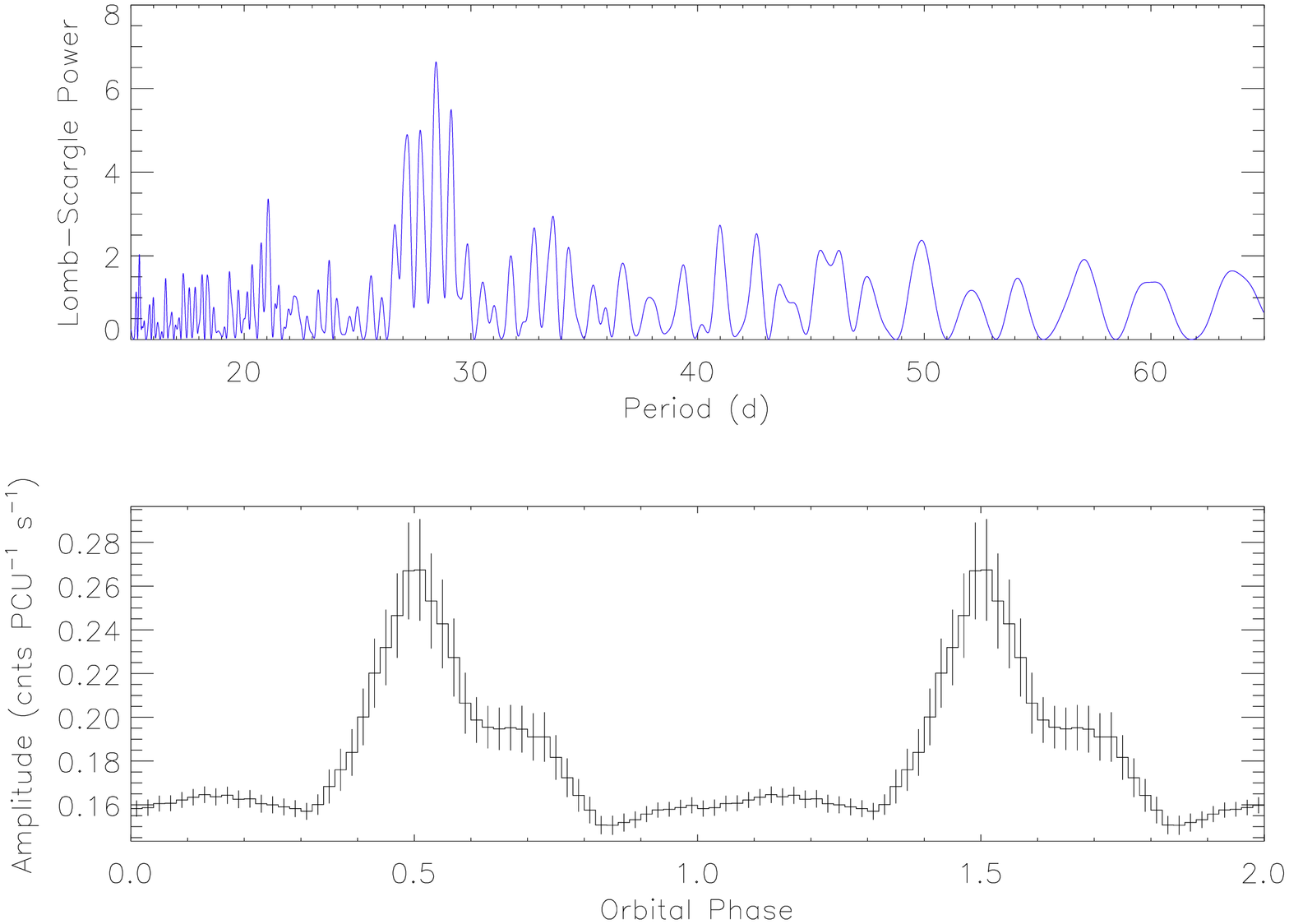}
   \caption[SXP8.88.]{SXP8.88.  a) \textit{Top}: \xray\ amplitude light curve. b) \textit{Middle}: Lomb-Scargle power spectrum; \textit{bottom}: light curve folded at 28.47\dy. Only the data outside the two large outbursts were used to obtain the power spectrum and folded profile of b).}
   \label{fig_sxp8.88}
\end{figure}

\sur Two Type~II outbursts together with the aforementioned series of Type~I outbursts have been detected (see \fref{fig_sxp8.88}). A period of \aprox28\dy\ is found using Lomb-Scargle analysis on the data not containing the Type~II outbursts, similar to the value from \citet{corbet2004atel}. The new ephemeris from the survey data is MJD 52392.2\pmt0.9 + n\x28.47\pmt0.04\dy. We note that the two Type~II outbursts began roughly at the times of expected maximum flux, lasted about 1 orbit, and peaked 0.5\porb\ later. This could be explained by a number of scenarios: a) The neutron star forms an accretion disc around periastron, which it consumes throughout the orbit; b) an accretion disc is formed around periastron which is accreted onto the neutron star along the orbit at the same time as wind accretion (which would peak around apastron) is taking place; c) the Be star ejects matter forcefully enough that it moves outwards as an annulus and the neutron star is able to capture material along its orbit as it moves away from the Be star, but capture decreases as soon as it passes apastron.

None of these scenarios completely explain the behaviour of the \xray\ emission during the Type~II outbursts. If a) is correct, then why would accretion become greatest at apastron on both occasions. Scenario c) suffers a similar drawback as it would require the annulus to move at similar speeds on both outbursts in order for the outburst to peak precisely at apastron. Scenario b) would require a very dense wind in order for its accretion to be comparable to disc accretion. Simultaneous observations of this system in the \xray\ and the optical during an outburst would help clarify this behaviour.

\subsection{SXP9.13}

\textbf{AX J0049\minus732, RX J0049.2\minus7311 \\
RA 00 49 13.6, dec \minus73 11 39}

\his Discovered during an \asca\ observation in November 1997 (MJD 50765), pulsations were detected at 9.1321\pmt0.0004\s\ and \lumxeq{3.3}{35} \citep{imanishi1998iauc}. There has been much debate as to which is the correct optical counterpart to this source, \citet{schmidtke2004} and \citet{filipovic2000} identify it with the \rosat\ source RX J0049.5\minus7310 but \citet{coe2005} conclude that it is a \halpha\ source coincident with RX J0049.2\minus7311. Timing analysis of the \ogle\ data shows a peak at 40.17\dy\ and its probable harmonic at 20.08\dy\ \citep{williamthesis}. It should be noted that \citet{schmidtke2004} find a 91.5\dy\ or possibly an \aprox187\dy\ period for RX J0049.5\minus7310. It was detected by \asca\ on one further occasion on MJD 51645 at \lumxeq{1.9}{35} \citep{yokogawa2003}.

\begin{figure}
   \centering
        \includegraphics[angle=90,width=0.95\linewidth]{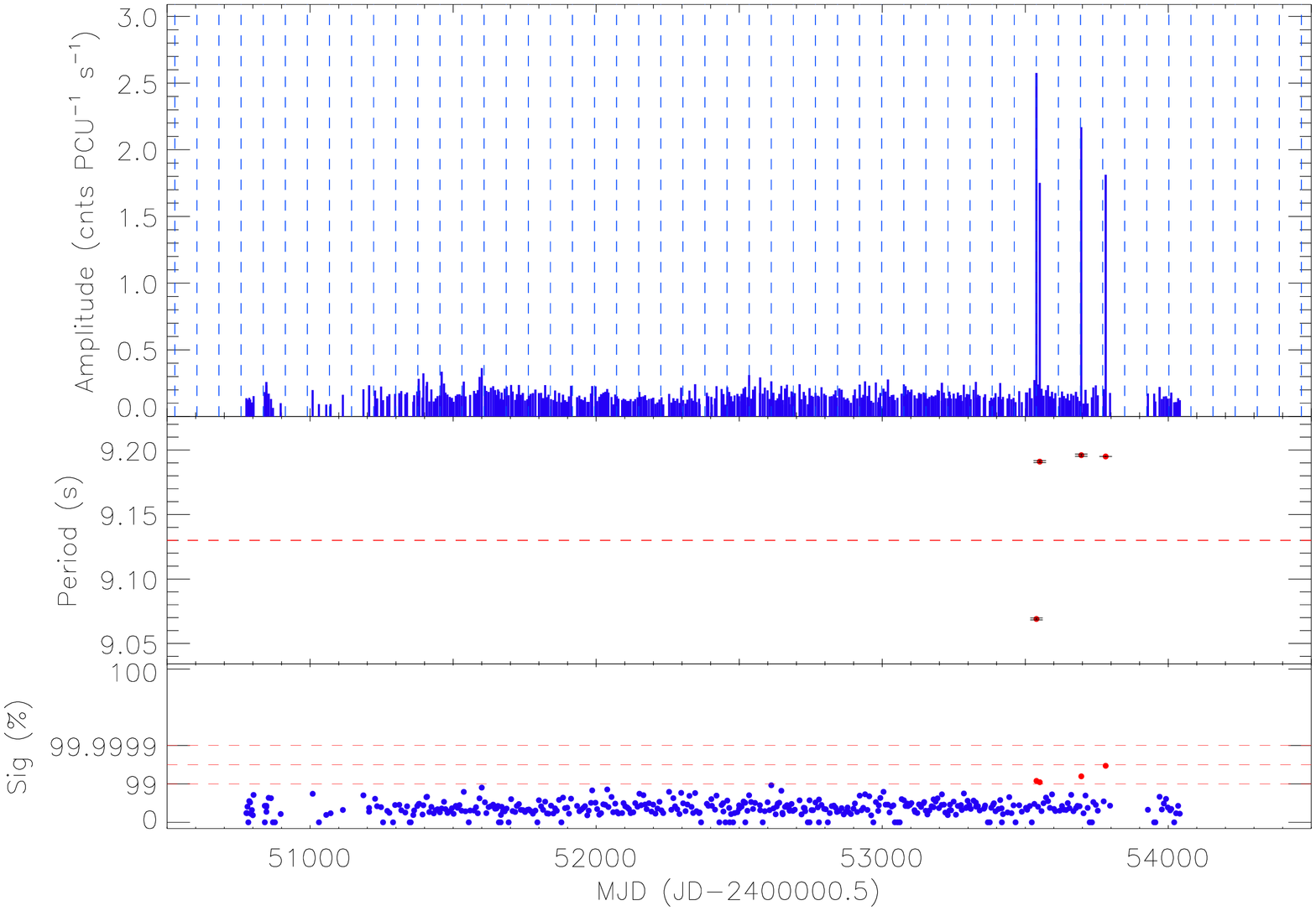}
        \includegraphics[angle=90,width=0.95\linewidth]{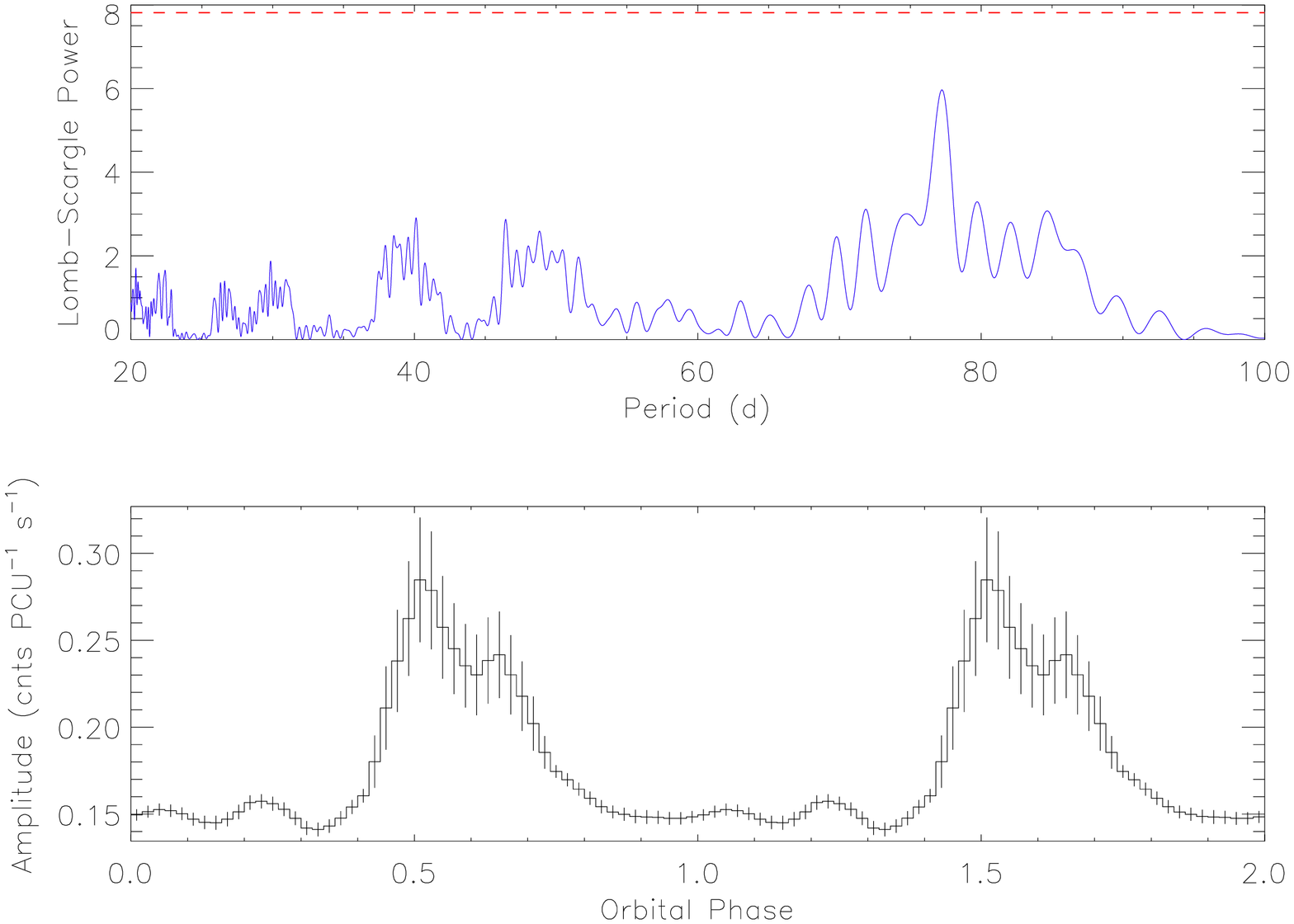}
   \caption{SXP9.13. a) \textit{Top}: \xray\ amplitude light curve. b) \textit{Middle}: Lomb-Scargle power spectrum; \textit{bottom}: light curve folded at 77.2\dy.}
   \label{fig_sxp9.13}
\end{figure}

\sur Although coverage of this source has been very complete, it has only been detected in outburst 3 times (circa MJD 53545, 53700 and 53780). Lomb-Scargle analysis of data prior to this date shows power at a period of 77.4\dy; when analysing the whole data set, the most significant peak is at 77.2\dy\ (\fref{fig_sxp9.13}(b)). In both cases the ephemerides agree with the 3 outbursts (\fref{fig_sxp9.13}(a)), which strongly suggests this could be the orbital period of the system. However, it should be noted that this value is different from the reported optical periods. The ephemeris we derive is MJD 52380.5\pmt2.3 + n\x77.2\pmt0.3\dy. The above reported \asca\ detections did not find the system in outburst and cannot be used to ratify this ephemeris.

\subsection{SXP15.3}

\textbf{RX J0052.1\minus7319 \\
RA 00 52 14, dec \minus73 19 19}

\his \citet{lamb99iauc} found 15.3\s\ pulsations in \rosat\ and \batse\ data from 1996. They estimate the luminosity to be \aprox10\supersm{37}\ergps\ with a pulse fraction of \aprox27\%. \citet{williamthesis} finds an ephemeris of MJD 50376.1 + n\x75.1\pmt0.5\dy\ describes the modulation in the \macho\ and \ogle\ light curves. It should be noted that this ephemeris is likely driven by the large 1996 \xray\ outburst which is clearly visible in the \ogle\ data; this Type~II outburst peaked around MJD 50375.

\begin{figure}
   \centering
        \includegraphics[angle=90,width=0.95\linewidth]{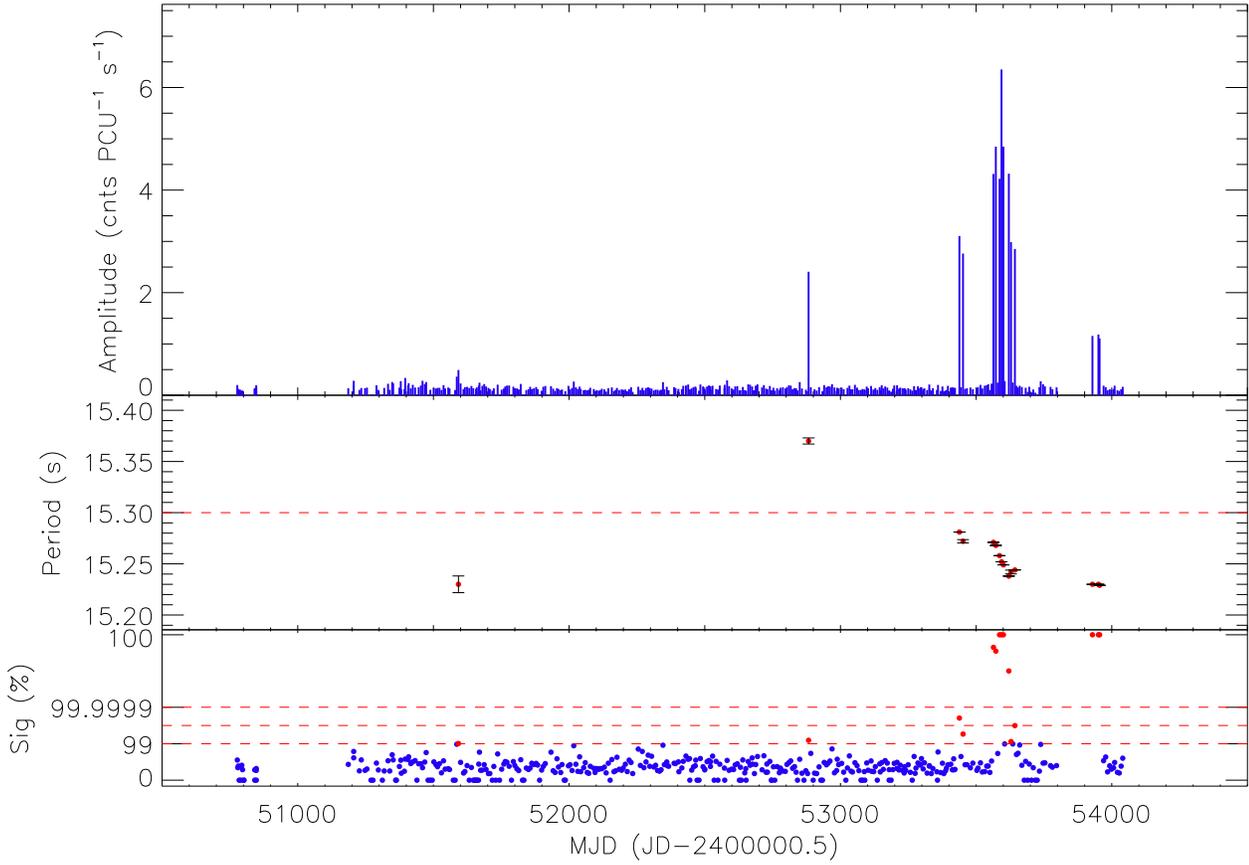}
   \caption{SXP15.3, \xray\ amplitude light curve.}
   \label{fig_sxp15.3}
\end{figure}

\sur There was a very minor detection of SXP15.3 in February 2000 (MJD 51592), and a more significant one in August 2003 (MJD 52883) at \fpul\ $\simeq$ 1.6\cpcus\ (see \fref{fig_sxp15.3}). Two more bright detections separated by 13 days occurred in March 2005 (circa MJD 53445); then, on July 12\supers{th} (MJD 53564) a very bright outburst began, lasting until October 17\supers{th} (MJD 53661, \aprox100 days long). During this time the pulsed flux oscillated between \aprox2.0 and \aprox5.6\cpcus. A clear spin up is detected up until September 15\supers{th}, when the period started increasing. The maximum period change was $\Delta P = 3.3\times 10^{-2}$ over the course of 56.68\dy, giving a \pdoteq{6.74}{-9}. We derive the expected luminosity (see \aref{sec_appendix_b} for method) from such a level of spin up (if it were all intrinsic with no orbital contribution) to be \lumxge{8.6}{37} (\ble{1.5}{13}).

It is likely this outburst lasted for more than one orbital cycle (expected to be \aprox30\sd50\dy\ from the Corbet diagram) so some orbital modulation might be visible in the period data. An attempt was made to fit these data to an orbital model with constant global spin up, and also with a piece-wise approach where the spin up varies throughout the outburst (following the method outlined in \citet{wilson2003}). Although no definite value was found, variations in the period curve suggest a period of \aprox28\dy.

Timing analysis of the data outside the long outburst revealed no periodicities. Furthermore, the optical ephemeris proposed by \citet{williamthesis} does not agree with any of the detections for this source.

\subsection{SXP16.6}

\textbf{XTE J0050\minus731 \\
No position available}

\his Discovered with a deep 121\ks\ observation taken for this survey in September 2000. It was initially misidentified as RX J0051.8\minus7310 but later disproved \citet{yokogawa2002}, and so SXP16.6 remains unassociated with any known source, although it is often still mistakenly referred to as RX J0051.8\minus7310.

\begin{figure}
   \centering
        \includegraphics[angle=90,width=0.95\linewidth]{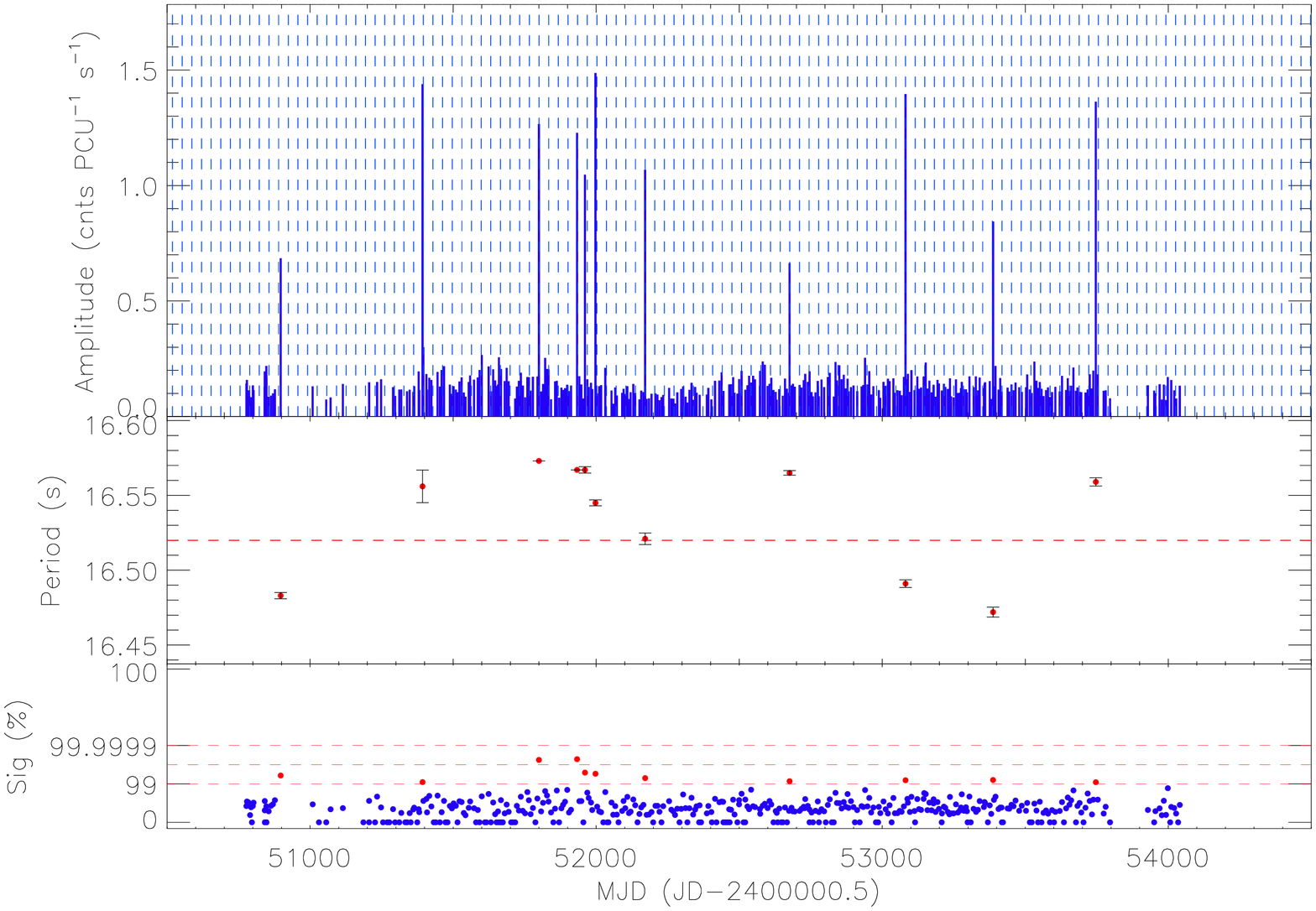}
        \includegraphics[angle=90,width=0.95\linewidth]{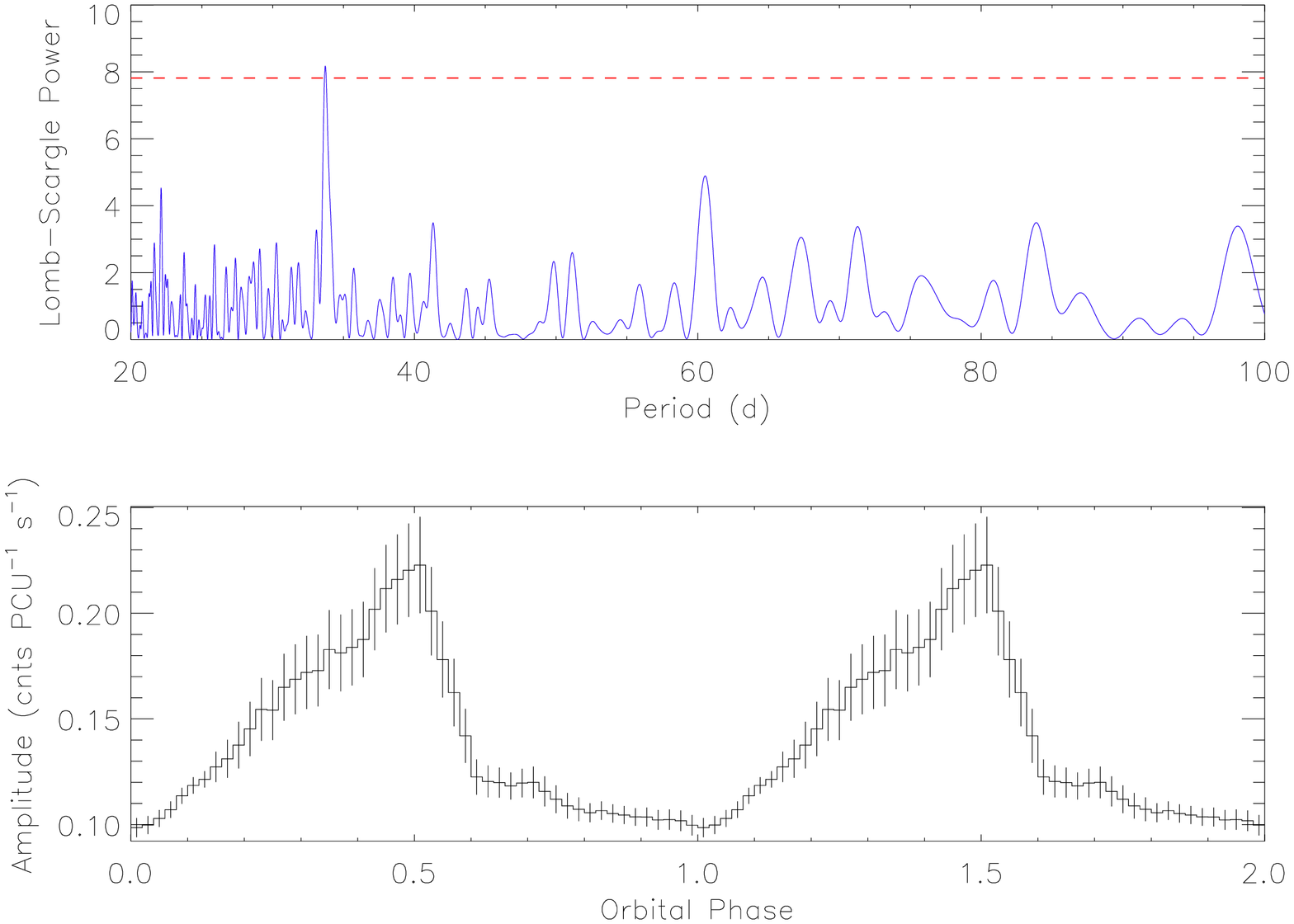}
   \caption{SXP16.6. a) \textit{Top}: \xray\ amplitude light curve. b) \textit{Middle}: Lomb-Scargle power spectrum; \textit{bottom}: light curve folded at 33.72\dy.}
   \label{fig_sxp16.6}
\end{figure}

\sur There have been a large number of detections of SXP16.6, and Lomb-Scargle analysis finds a strong modulation at 33.72\dy\ (see \fref{fig_sxp16.6}(b)), which we propose as the orbital period of the system. The ephemeris is MJD 52373.5\pmt1.0 + n\x33.72\pmt0.05\dy.

\subsection{SXP18.3}

\textbf{XTE J0055\minus727 \\
RA 00 50, dec \minus72 42}

\his Reported by \citet{corbet2003atelb} from observations in November/December 2003. An approximate position was established from scans with \rxte 's \pca\ \citep{corbet2003atelb}. No optical counterpart has yet been identified. A number of further outbursts between May\sd October 2004 provide an ephemeris of MJD 53145.7\pmt1.3 + n\x34.6\pmt0.4\dy\ from O-C (observed vs. calculated) orbital calculations \citep{corbet2004atelb}.

\begin{figure}
   \centering
        \includegraphics[angle=90,width=0.95\linewidth]{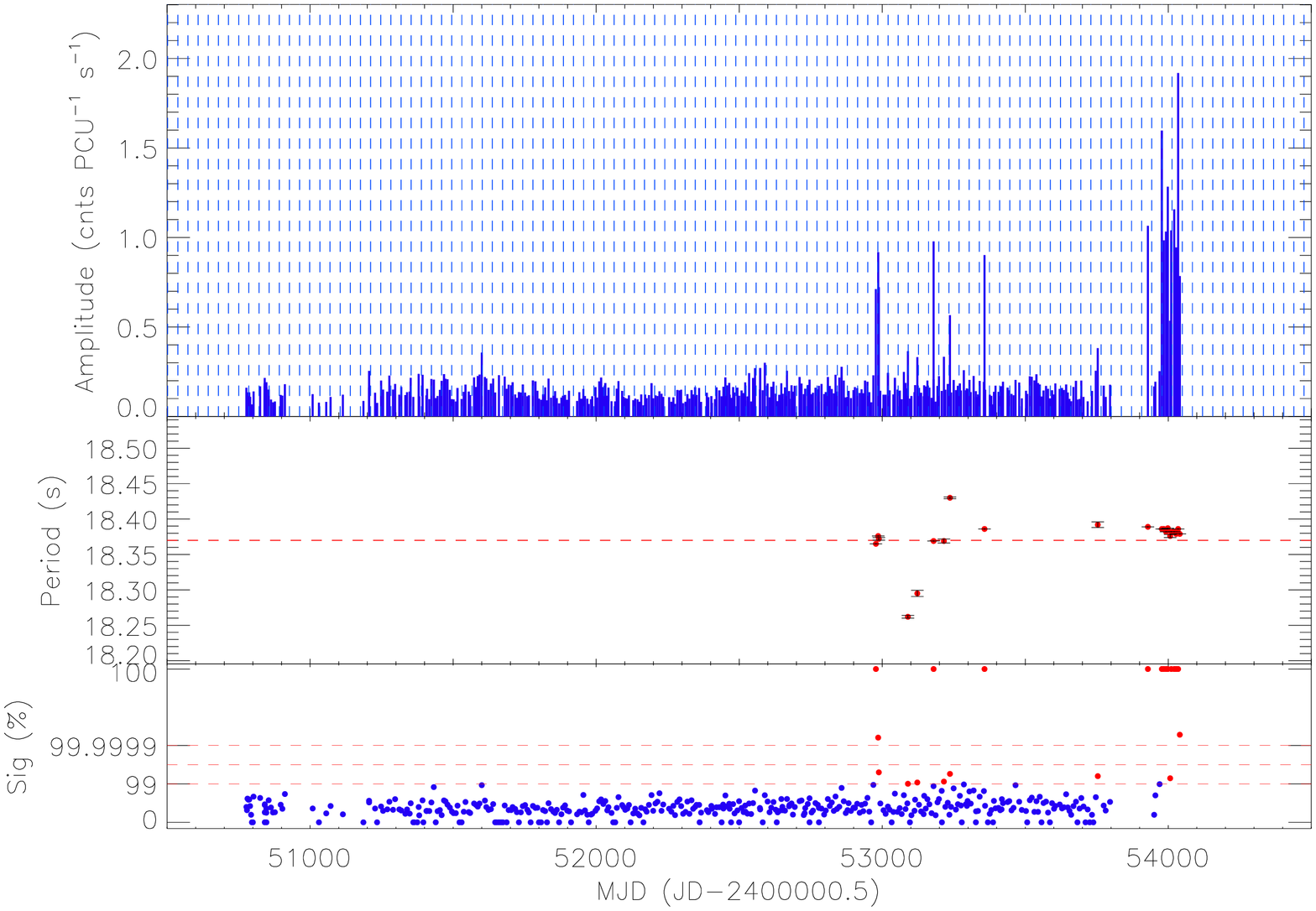}
        \includegraphics[angle=90,width=0.95\linewidth]{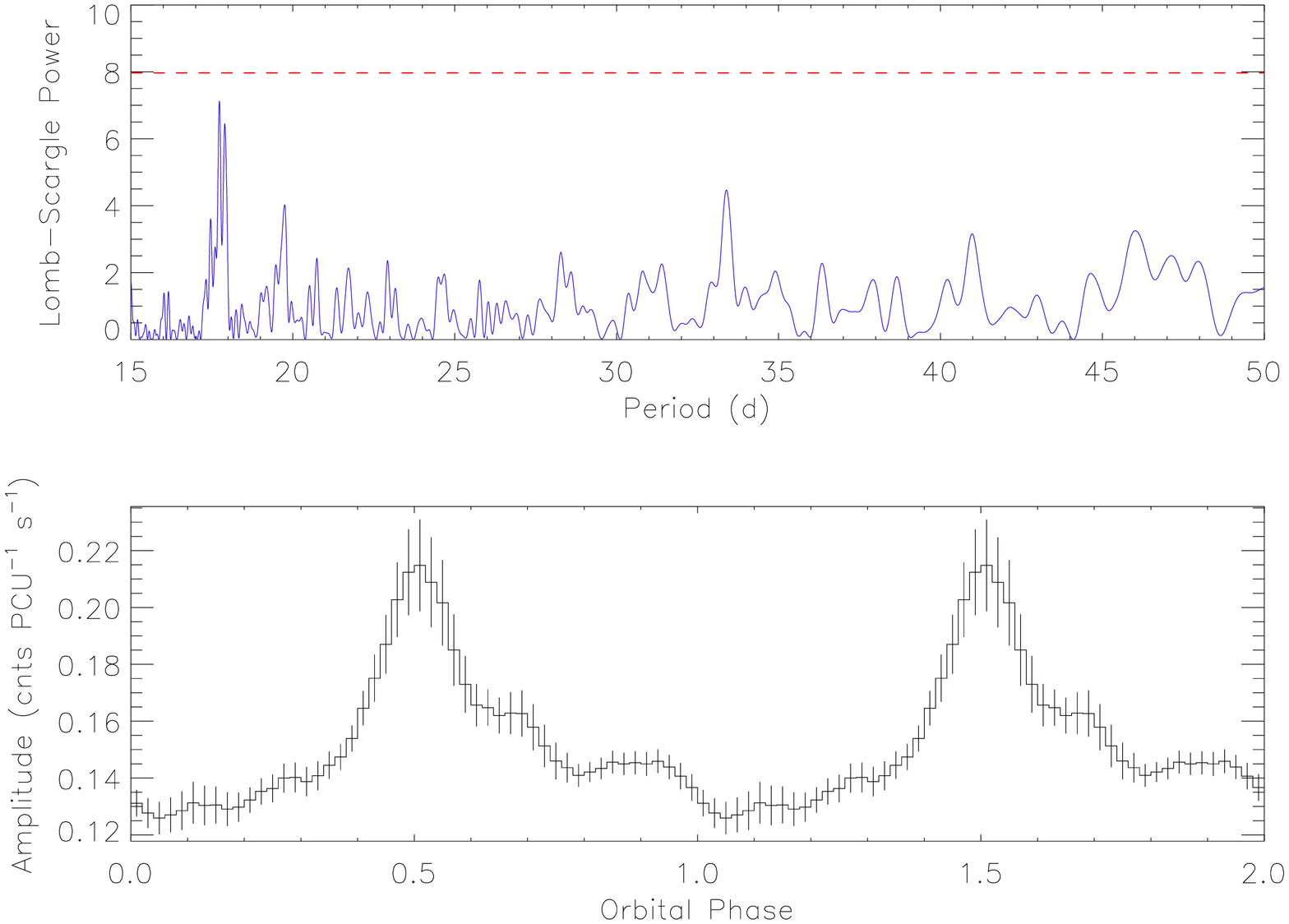}
   \caption{SXP18.3. a) \textit{Top}: \xray\ amplitude light curve (for clarity, only every other ephemeris line has been plotted). b) \textit{Middle}: Lomb-Scargle power spectrum; \textit{bottom}: light curve folded at 17.73\dy (only data prior to MJD 53900 was used in both cases).}
   \label{fig_sxp18.3}
\end{figure}

\sur A long, bright outburst began around MJD \aprox53925 (July 2006), which was ongoing as of January 2007 (see \fref{fig_sxp18.3}(a)). Lomb-Scargle analysis of data prior to this outburst finds a peak at 17.73\dy, which is \aprox$\frac{1}{2}$ the period proposed by \citet{corbet2004atelb}. We find no significant power at this period and note that some of the significant detections do not agree with it. Thus, the new ephemeris for this system is MJD 52275.6\pmt0.9 + n\x17.73\pmt0.01\dy.

\subsection{SXP22.1}

\textbf{RX J0117.6\minus7330 \\
RA 01 17 40.5, dec \minus73 30 52.0}

\his Discovered with \rosat\ as an \xray\ transient in September 1992 \citep{clark96iauc}, its companion was identified as a 14.2 magnitude OB star by \citet{charles96iauc}. It was observed with \rosat\ again in October, but no pulsations were detected \citep{clark97} The companion's classification as a Be (B1-2) star was confirmed by \citet{coe98}. 22\s\ pulsations were finally detected in \rosat\ and \batse\ data by \citet{macomb99}.

\sur There are only 3 observations of this pulsar's region and it was not significantly detected in any of them.

\subsection{SXP31.0}

\textbf{XTE J0111.2\minus7317, AX J0111.1\minus7316 \\
RA 01 11 09.0, dec \minus73 16 46.0}

\his Was simultaneously discovered by \rxte\ and \batse\ during a giant outburst that began late October 1998 \citep{chakrabarty98a,wilson98iauc}. \citet{schmidtke2006} report a 90.4\dy\ periodicity in \ogle~III data.

\sur Being in the same field as SMC~X-1, it was only observed 3 times, the first during the end of the aforementioned giant outburst. The two detections are \aprox43 days apart and a spin up of ~0.4\s\ is measured, which implies a luminosity of \lumxge{2.7}{38}. As \citet{laycock2005}, in an in depth spectral analysis of these observations, derive a value of \lumxge{4.6}{37}, we can assume we are seeing Doppler shifted periods due to orbital motion. No further information can be derived from these observations.

\subsection{SXP34.1}

\textbf{RX J0055.4\minus7210 \\
RA 00 55 26.9, dec \minus72 10 59.9}

\his Discovered in archival Chandra data at 34.08\pmt0.03\s\ and lying 0.6\asec\ from a known \rosat\ source \citep{edge2004atelb,edge2004}.

\begin{figure}
   \centering
      \includegraphics[angle=90,width=0.95\linewidth]{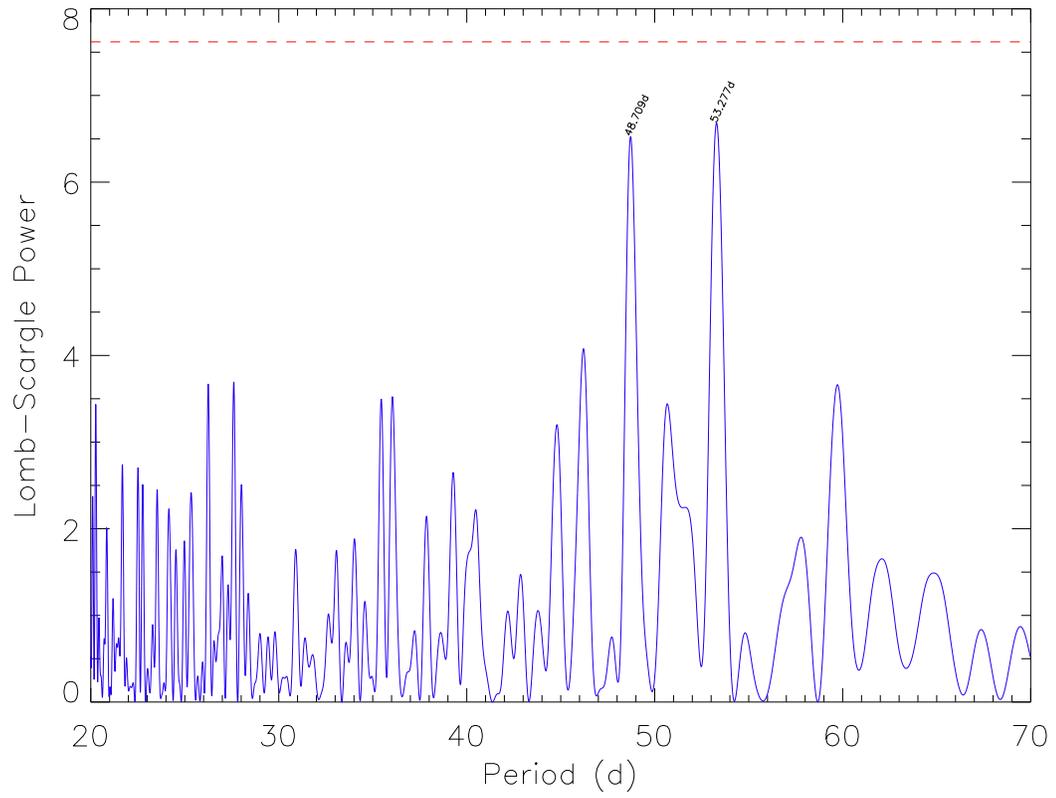}
   \caption{SXP34.1, \xray\ amplitude light curve.}
   \label{fig_sxp34.1}
\end{figure}

\sur Only two significant detections are seen (MJD 50777 and 53690) and no clear periods can be found with timing analysis (see \fref{fig_sxp34.1}).

\subsection{SXP46.6} \label{sxp46.6}

\textbf{XTE J0053\minus724, 1WGA 0053.8\minus7226 \\
RA 00 53 58.5, dec \minus72 26 35}

\his Discovered in the first observation of this survey (November 25\supers{th} 1998) with a period of 46.63\pmt0.04\s\ while it was undergoing a long outburst \citep{corbet98iauc}. \citet{laycock2005} derive a period of 139\pmt6\dy\ from 6 \xray\ outbursts in the earlier part of this survey. Two candidates for the optical counterpart were proposed by \citet{buckley2001}. \citet{schmidtke2007} confirm it is star B and find two periods in \ogle\ data: 69.2\pmt0.3\dy\ and 138.4\pmt0.9\dy; they suggest the possibility that the orbital period is the shorter or the two values.

\begin{figure}
   \centering
        \includegraphics[angle=90,width=0.95\linewidth]{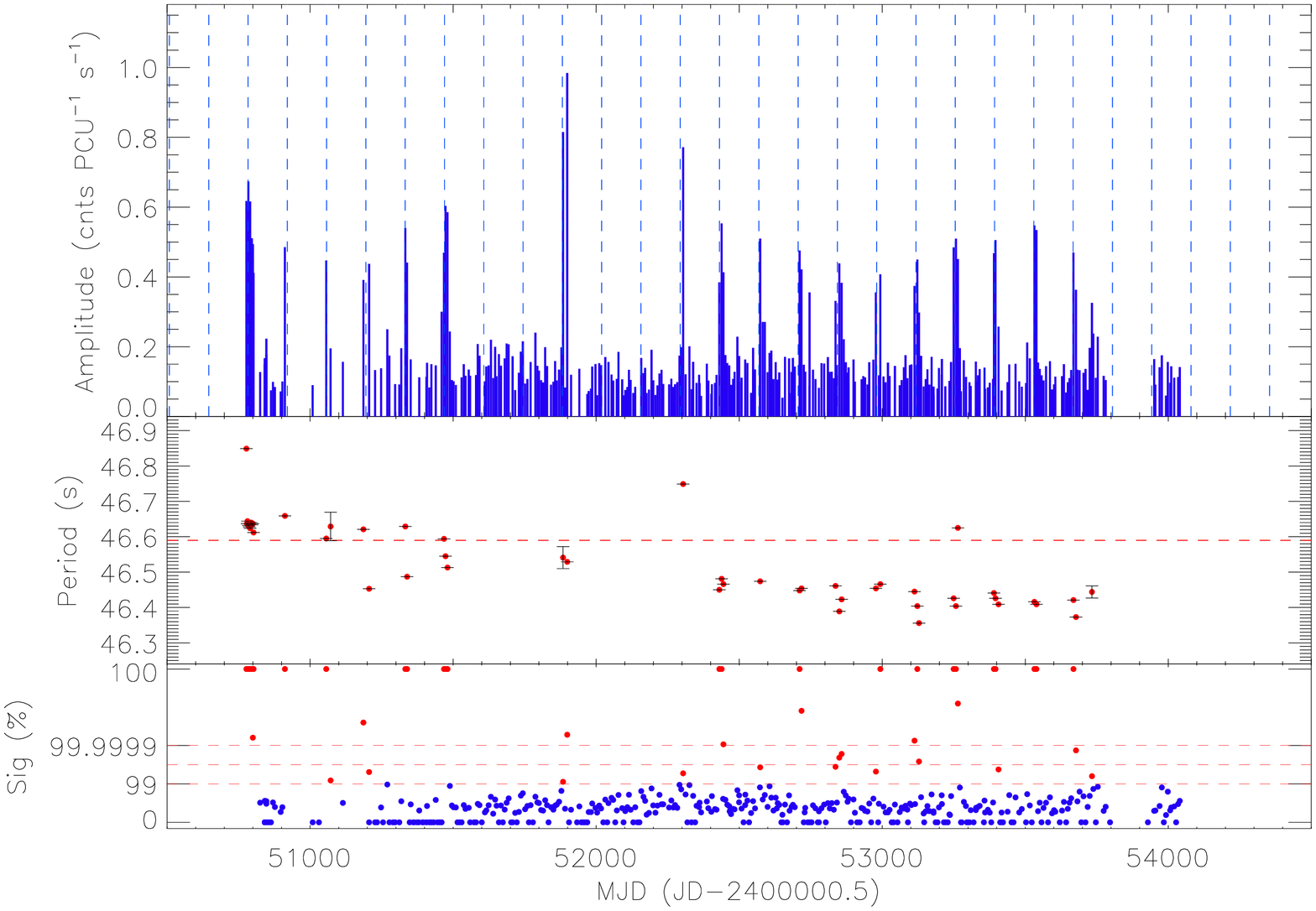}
        \includegraphics[angle=90,width=0.95\linewidth]{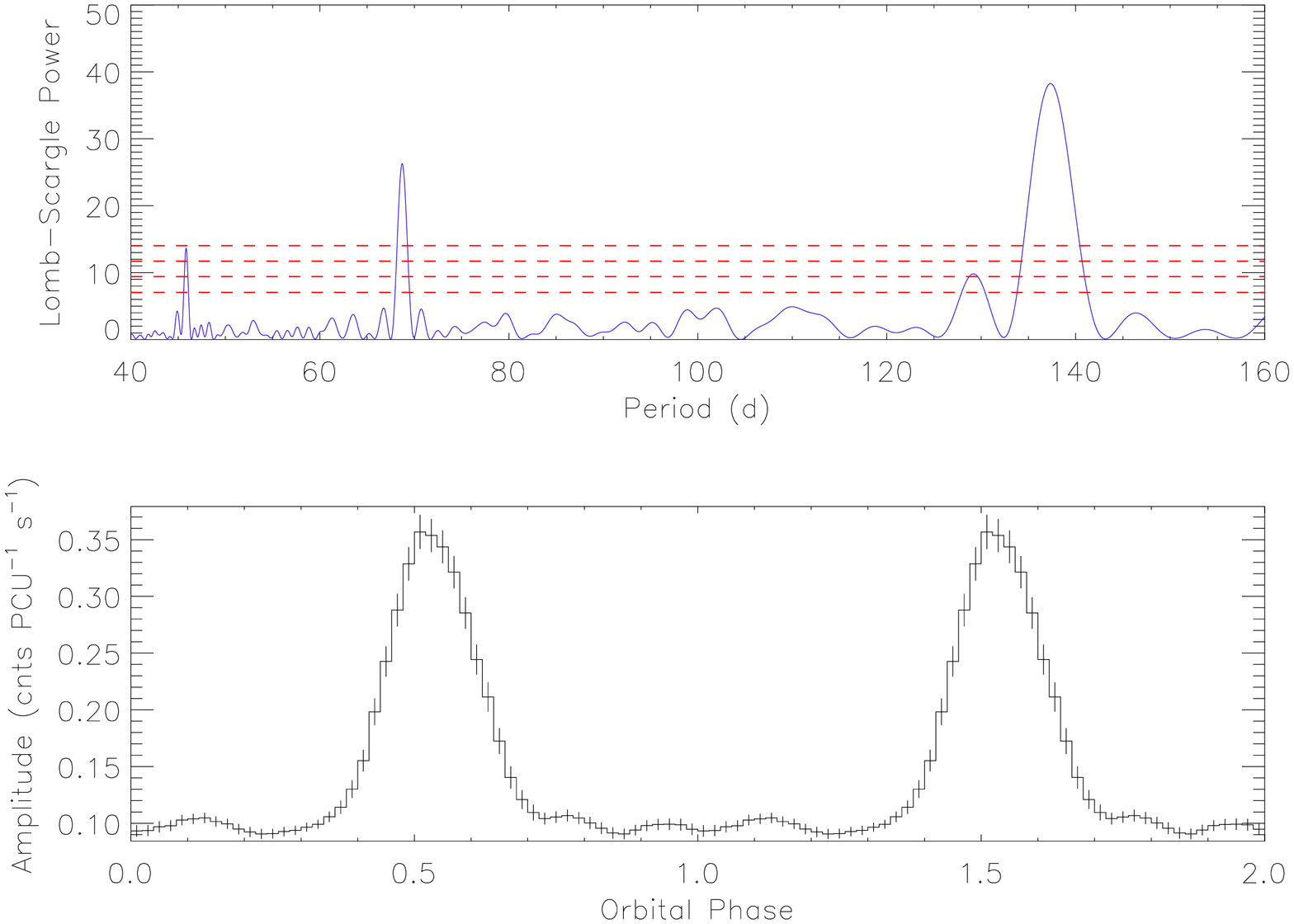}
   \caption{SXP46.6. a) \textit{Top}: \xray\ amplitude light curve. b) \textit{Middle}: Lomb-Scargle power spectrum; \textit{bottom}: light curve folded at 137.36\dy.}
   \label{fig_sxp46.6}
\end{figure}

\sur The source was thought to be inactive after January 2002. In the meantime a new \smc\ pulsar with a 46.4\s\ period was announced \citep{corbet2002iauc}. Lomb-Scargle analysis of both pulsars revealed the same orbital periods and very similar ephemeris, suggesting they were the same pulsar which has been slowly spinning up \citep{galache2005atel}. The estimated luminosity required for a spin up of \pdoteq{1.05}{-9} during MJD 50800\sd51900 is \lumxge{9.9}{35} (\ble{6.0}{12}), and for a \pdoteq{4.68}{-10} during MJD, \lumxge{4.4}{35} (\ble{4.0}{12}). The ephemeris of the outbursts is now best described by MJD 52293.9\pmt1.4 + n\x137.36\pmt0.35\dy\ (see \fref{fig_sxp46.6}(a)). Although the folded light curve shows a small increase in flux half a phase away from maximum, we do not believe this is evidence that the orbital period is half the calculated value. Given the clear, regular outbursts experienced by this system throughout the survey we believe the true orbital period is \aprox137\dy.

\subsection{SXP51.0}

\textbf{SMC 51 \\
No position available}

\his Was erroneously proposed as a new 25.5\s\ pulsar in \citet{lamb2002b} from a deep 121\ks\ observation. \citet{silasthesis} identifies the 25.5\s\ peaks in the power spectrum as harmonics of SXP51.0's true pulse period. No position is available; it lies within Position 4, and most likely in the overlap between Positions 1/A and 4.

\begin{figure}
   \centering
        \includegraphics[angle=90,width=0.95\linewidth]{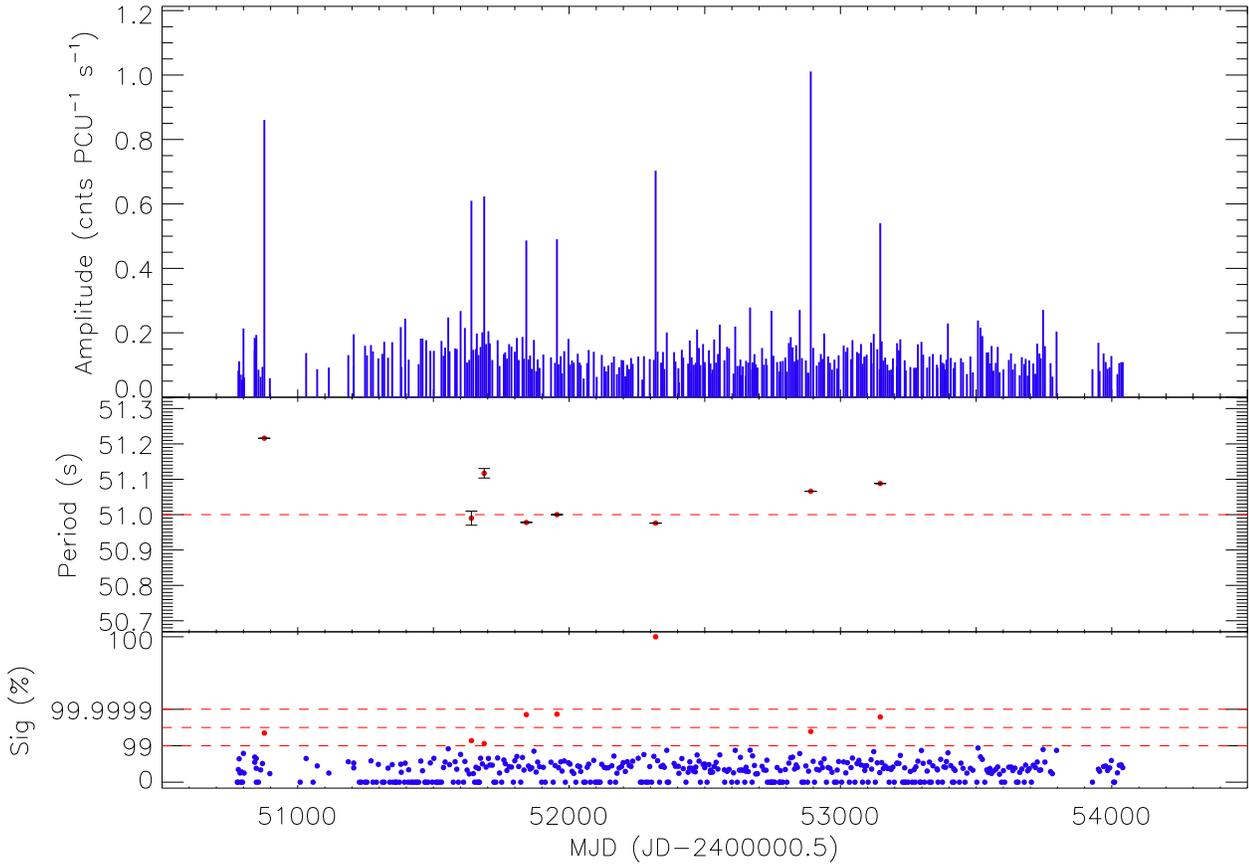}
   \caption{SXP51.0, \xray\ amplitude light curve.}
   \label{fig_sxp51.0}
\end{figure}

\sur Despite numerous significant detections (see \fref{fig_sxp51.0}), timing analysis finds no strong periodicities in the light curve.

\subsection{SXP59.0}

\textbf{XTE J0055\minus724, RX J0054.9\minus7226, 1WGA J0054.9\minus7226 \\
RA 00 54 56.6, dec \minus72 26 50}

\his Discovered in \rxte\ observations of the vicinity of \smc~X-3 \citep{marshall98iauc}, it showed 4 bright, very similar, outbursts during January 1998\sd September 1999 from which \citet{laycock2005} derived an orbital period of 123\pmt 1\dy. The optical counterpart was established by \citet{stevens99}; \cite{schmidtkecowley2005} found a 60.2\pmt0.8\dy\ period from timing analysis of \ogle\ and \macho\ data which they propose as the orbital period.

\begin{figure}
   \centering
        \includegraphics[angle=90,width=0.95\linewidth]{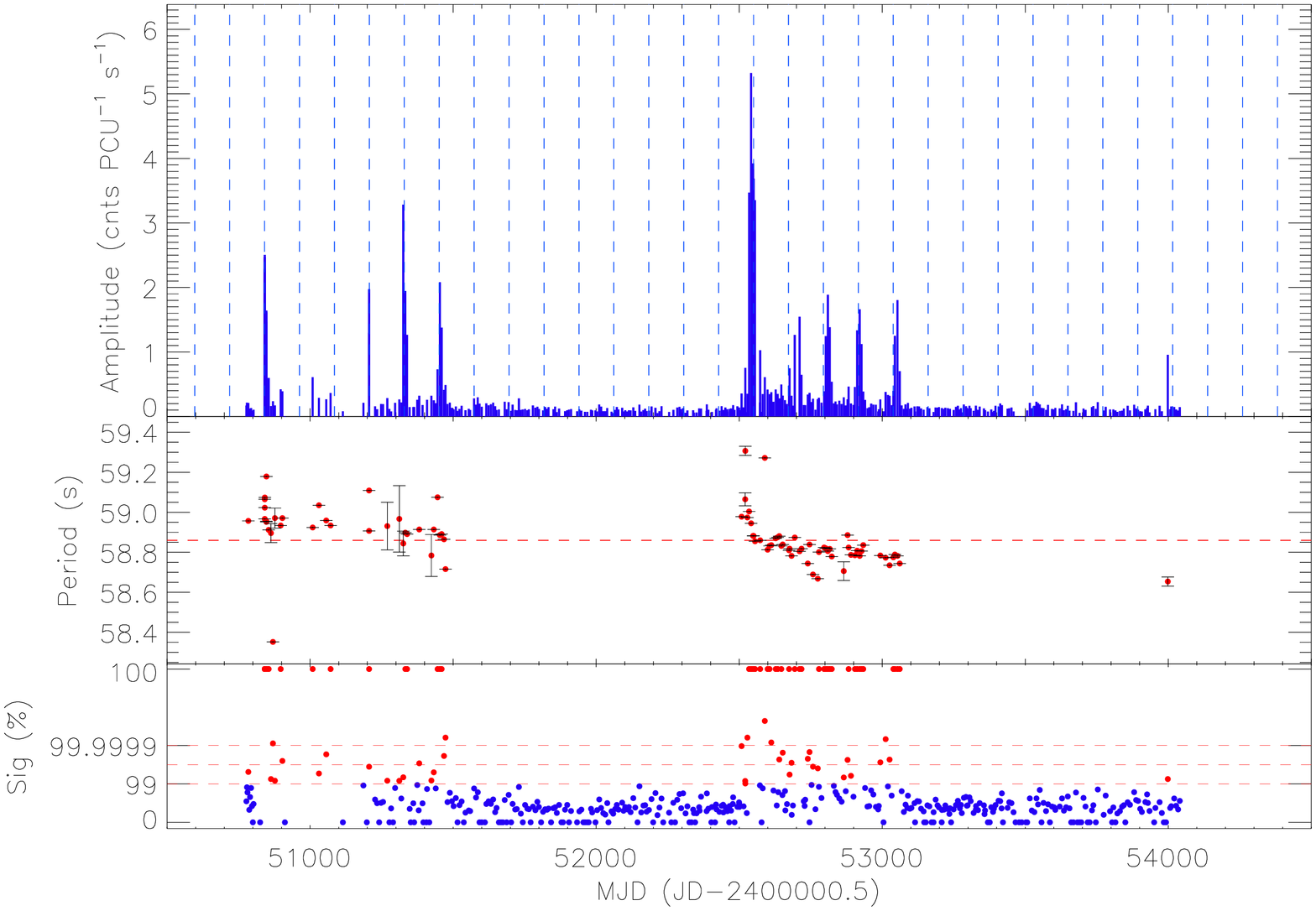}
        \includegraphics[angle=90,width=0.95\linewidth]{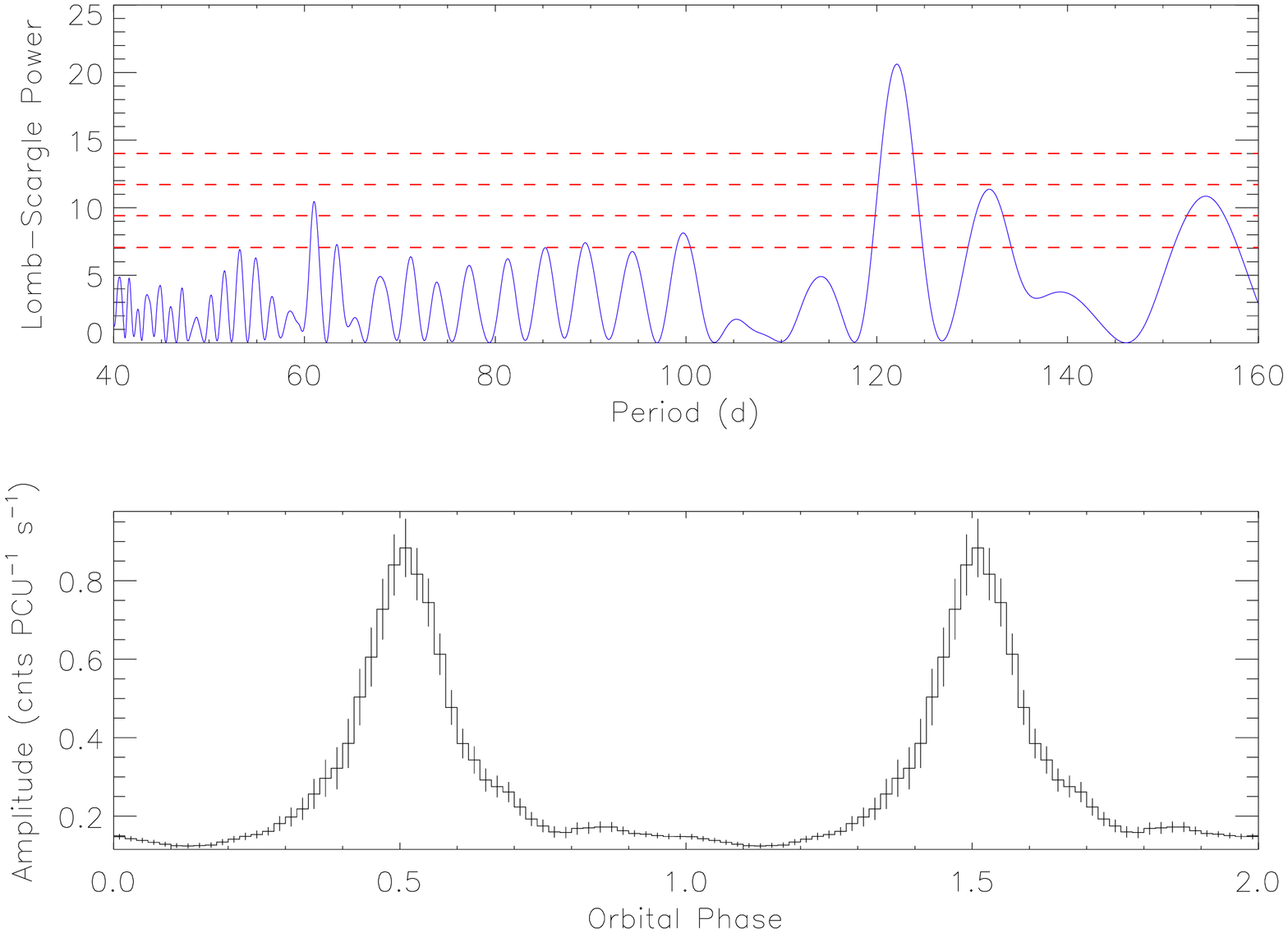}
   \caption{SXP59.0. a) \textit{Top}: \xray\ amplitude light curve. b) \textit{Middle}: Lomb-Scargle power spectrum; \textit{bottom}: light curve folded at 122.10\dy.}
   \label{fig_sxp59.0}
\end{figure}

\sur SXP59.0 remained undetected from September 1999 until a bright outburst in mid 2002 (circa MJD 52520) kicked off a series of 5 outbursts (see \fref{fig_sxp59.0}(a)). From the whole data range we extract an ephemeris of MJD 52306.1\pmt3.7 + n\x122.10\pmt0.38\dy. We note that this period is twice the optical period but we detect no significant flux half a phase from maximum that would suggest the orbital period is not 122 days.

During the 1998\sd99 outbursts a spin up was detected of \pdoteq{4.7}{-9} and again throughout the 5 outbursts of 2002\sd2003, with a value of \pdoteq{5.9}{-9}. The luminosities associated with these spin ups are, respectively, \lumxge{2.6}{36} (\ble{1.3}{13}) and \lumxge{3.3}{36} (\ble{1.4}{13}). In between both groups of outbursts SXP59.0 was observed to have spun down during the \aprox1100 days it was undetected. The average spin down was \pdoteq{-4.2}{-9}, which would be associated with an estimated \lumxge{2.9}{36}. Although SXP59.0 was further away from the center of the field of view during AO5\sd AO6 (MJD 51600\sd52300), it should have been picked up as a number of the detections during the second outburst season were made when \rxte\ was pointing at Position D (essentially the same coordinates as Position 5). In view of this, the spin down mechanism for SXP59.0 must be something other than reverse accretion torque and is likely due to the propeller effect.

\subsection{SXP65.8}

\textbf{CXOU J010712.6-723533 \\
RA 01 07 12.63, dec \minus72 35 33.8}

\his Discovered as part of a \chandra\ survey of the \smc\ wing that is reported in \citet{mcgowan2007}. They detected a source at \lumxeq{3}{36} (37\pmt5\% pulse fraction) with pulsations at 65.78\pmt0.13\s. Its position was found to coincide with the emission line star [MA93] 1619, a $V = 16.6$ Be star. A 110.6\dy\ period has been observed in the \macho\ data for this object \citep{schmidtkecowley2007b}.

\sur Due to its location outside the heavily observed positions, this pulsar has very little data. We find no significant detections and no periodic modulation is apparent in the \xray\ light curve.

\subsection{SXP74.7}

\textbf{RX J0049.1\minus7250, AX J0049\minus729 \\
RA 00 49 04, dec \minus72 50 54}

\his Discovered in the first observation of this survey with a period of 74.8\pmt0.4\s\ \citep{corbet98iauc}. \citet{kahabka1998iauc} identified it with the \rosat\ source RX J0049.1\minus7250 and \citet{stevens99} found a single Be star within the \rosat\ error radius which they proposed as the optical counterpart. Only 3 \xray\ outbursts where observed in the early stages of this survey (before MJD 52300), from which \citet{laycock2005} derived a possible orbital period of 642\pmt59\dy\ based on the separation between the outbursts. \citet{schmidtkecowley2005} and \citet{williamthesis} find a 33.4\dy\ periodicity in \ogle\ data.

\begin{figure}
   \centering
        \includegraphics[angle=90,width=0.95\linewidth]{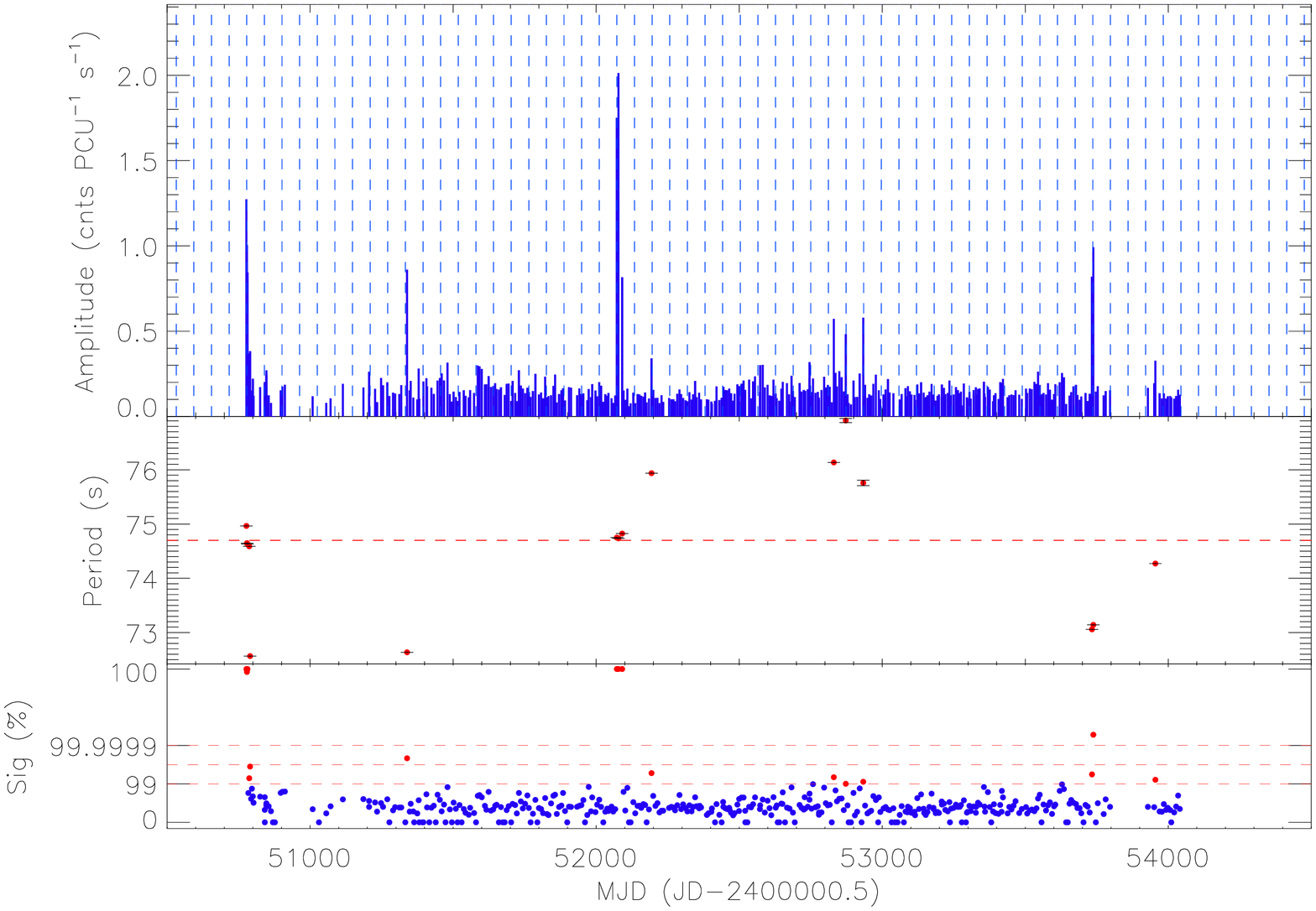}
        \includegraphics[angle=90,width=0.95\linewidth]{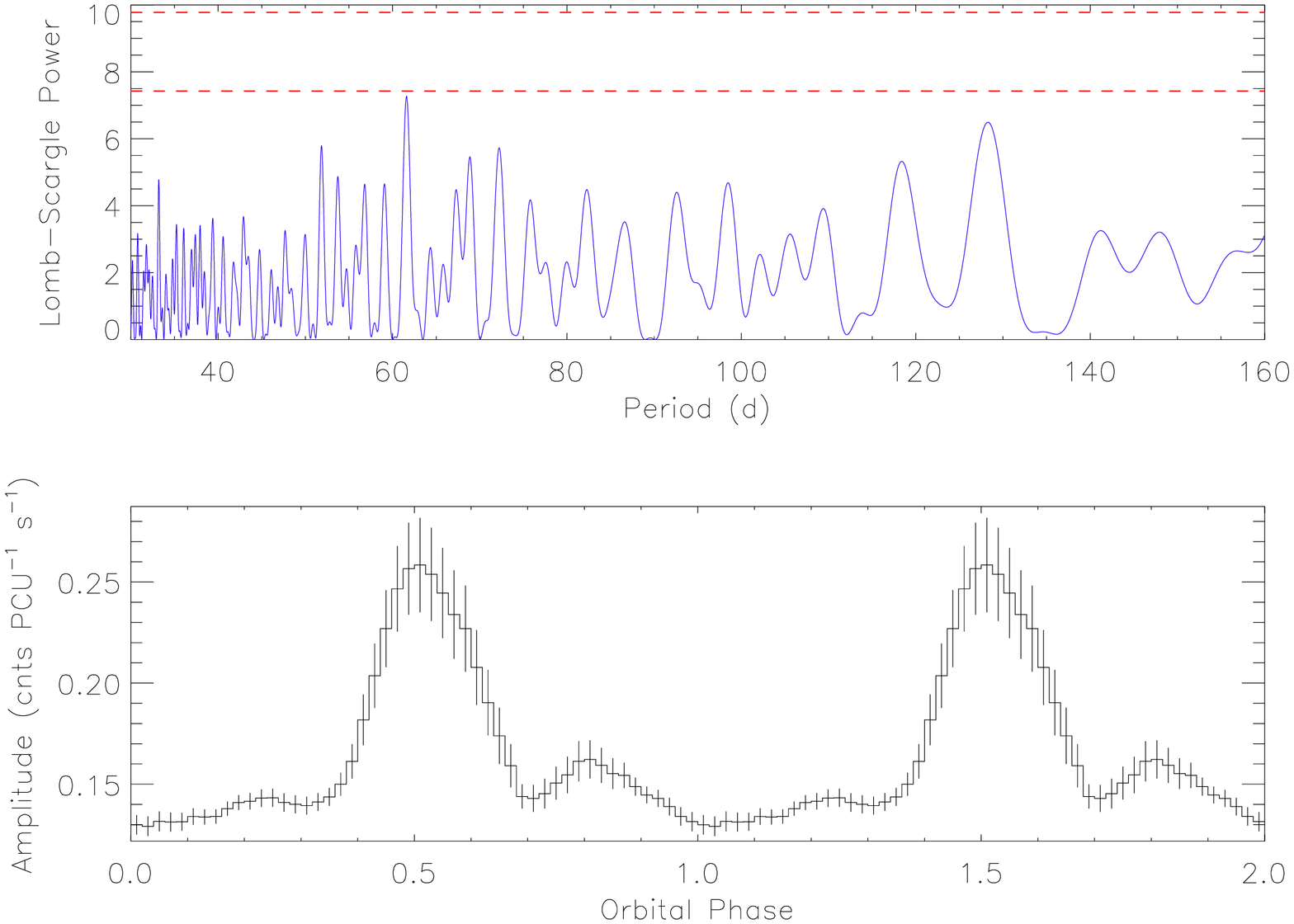}
   \caption{SXP74.7. a) \textit{Top}: \xray\ amplitude light curve. b) \textit{Middle}: Lomb-Scargle power spectrum; \textit{bottom}: light curve folded at 61.6\dy.}
   \label{fig_sxp74.7}
\end{figure}

\sur Lomb-Scargle analysis of the data finds a period at \aprox62\dy, which although not strong, shows a convincing profile (see \fref{fig_sxp74.7}(b)); its ephemeris is MJD 52319.0\pmt3.1 + n\x61.6\pmt0.2\dy. There is a weak peak in the power spectrum at 33.2\dy, close to the reported optical period, but folding the light curve at this period does not produce a clear modulation.

\subsection{SXP82.4}

\textbf{XTE J0052\minus725 \\
RA 00 52 09, dec \minus72 38 03}

\his First observed by \rxte\ in this survey \citep{corbet2002iauc}, its position was determined from archival \chandra\ observations \citep{edge2004atelc}. \ogle\ data show a strong modulation at \aprox380\dy\ (\pcom{Edge 2006}).

\begin{figure}
   \centering
        \includegraphics[angle=90,width=0.95\linewidth]{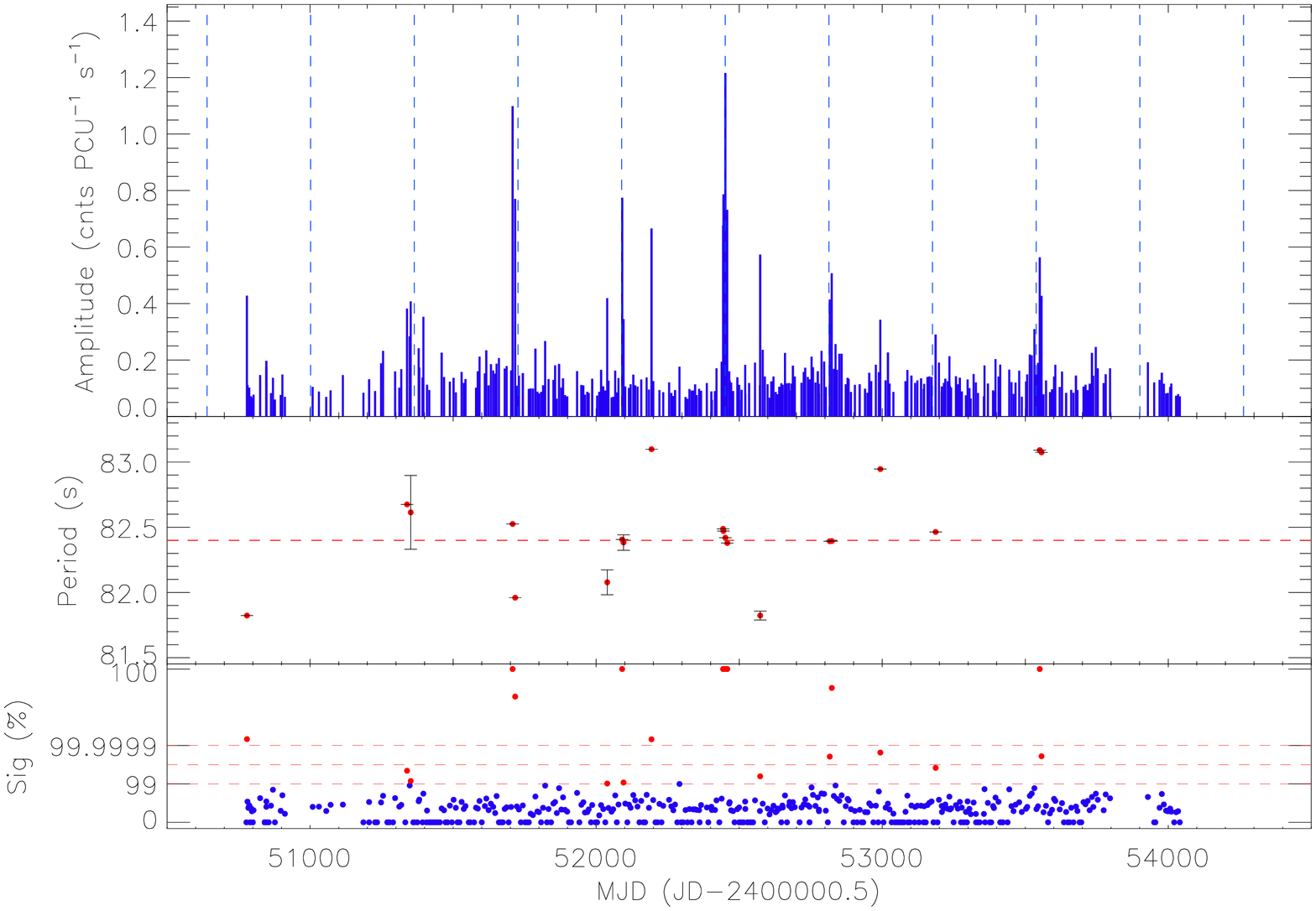}
        \includegraphics[angle=90,width=0.95\linewidth]{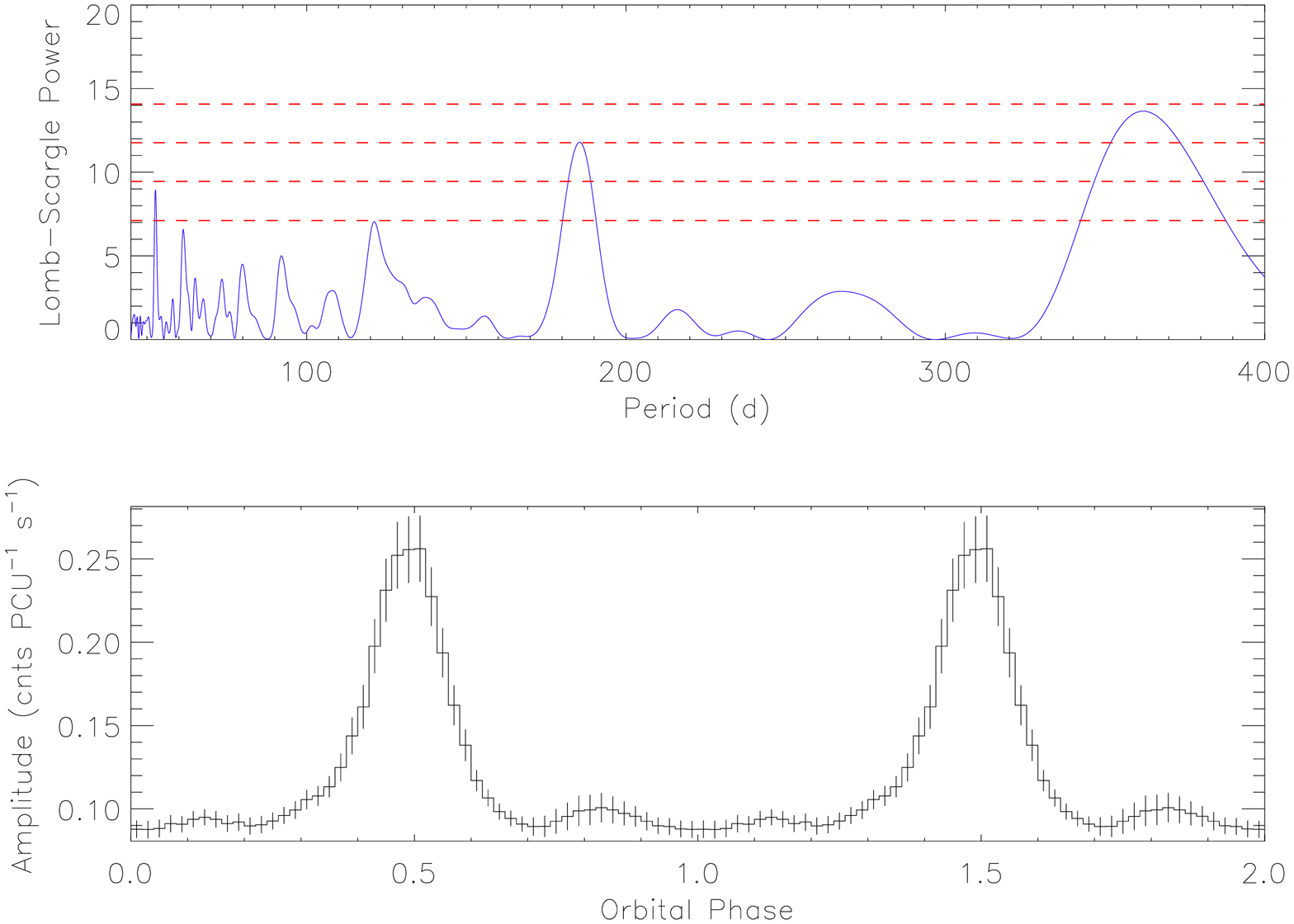}
   \caption{SXP82.4. a) \textit{Top}: \xray\ amplitude light curve. b) \textit{Middle}: Lomb-Scargle power spectrum; \textit{bottom}: light curve folded at 362.3\dy.}
   \label{fig_sxp82.4}
\end{figure}

\sur The Lomb-Scargle periodogram for the light curve shows a very significant peak at \aprox362\dy\ with a number of harmonics (see \fref{fig_sxp82.4}(b)); the ephemeris derived is MJD 52089.0\pmt3.6 + n\x362.3\pmt4.1\dy. Although this period is longer than would be expected given its \ps\ position in the Corbet diagram, its similarity to the optical period would seem to confirm it is the actual orbital period of the system. We note that in the aforementioned \chandra\ observation of this puslar on MJD~52459, it was detected at \lumxeq{3.4}{36} \citep{edge2004}; this date is \aprox8 days after our predicted periastron passage and the luminosity exhibited is consistent with a Type~I outburst (which was also detected with \rxte\, see \fref{fig_sxp82.4}(a)).

\subsection{SXP89.0}

\textbf{XTE SMC pulsar \\
No position available}

\his Reported by \citet{corbet2004proc} from observations in March 2002, it is located within the field of view of Position~1/A.

\begin{figure}
   \centering
        \includegraphics[angle=90,width=0.95\linewidth]{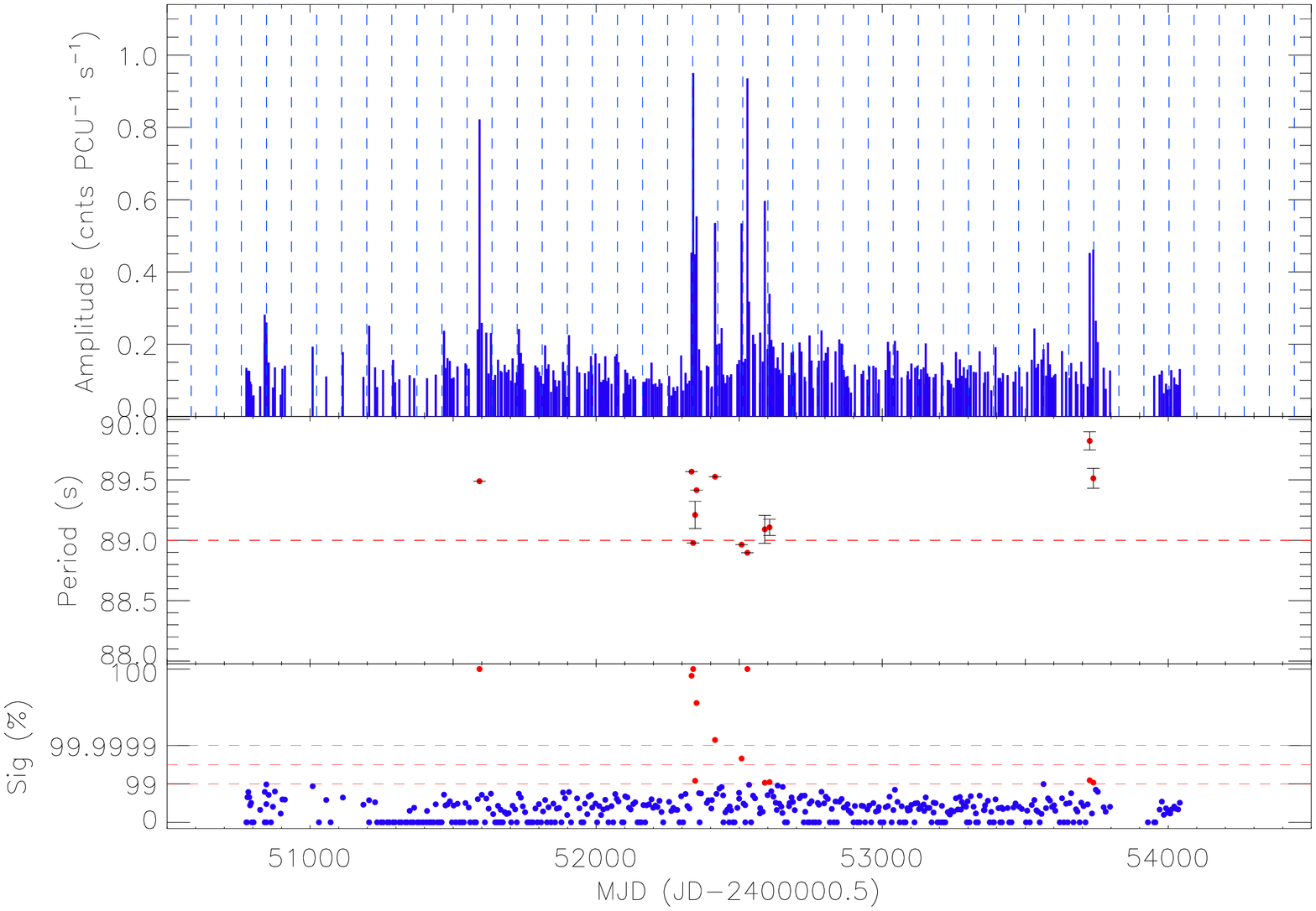}
        \includegraphics[angle=90,width=0.95\linewidth]{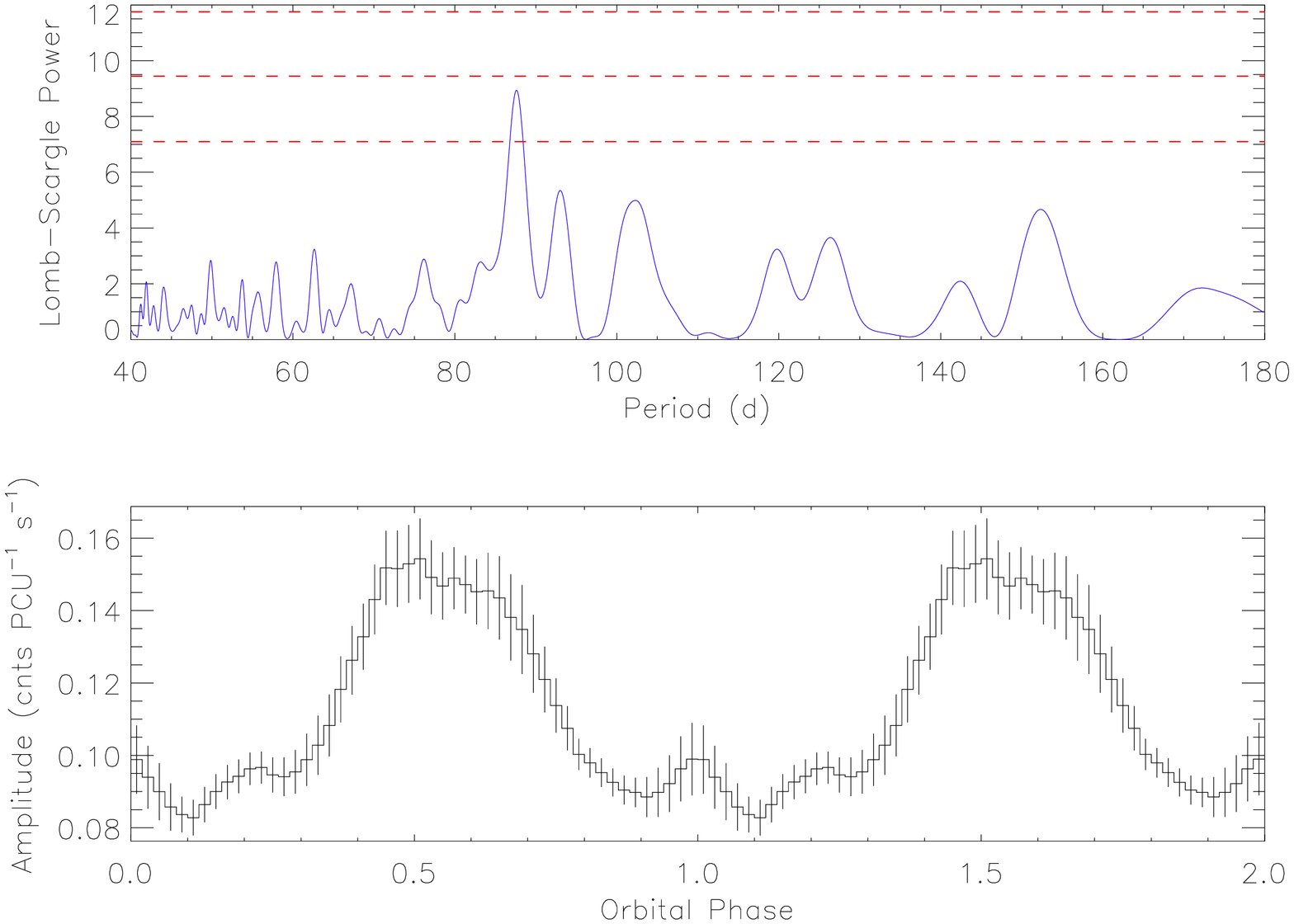}
   \caption{SXP89.0. a) \textit{Top}: \xray\ amplitude light curve. b) \textit{Middle}: Lomb-Scargle power spectrum; \textit{bottom}: light curve folded at 87.6\dy.}
   \label{fig_sxp89.0}
\end{figure}

\sur The first outburst is a single detection in February 2000 (MJD 51592), 2 years before the official discovery; 4 others occurred 2 years later in a short space of time--within \aprox260 days (see \fref{fig_sxp89.0}(a)). They follow a high-low-high-low brightness pattern, with very similar high/low countrates. The separation between them is \aprox88\dy\ which could be expected to be the orbital period. In fact, timing analysis finds a strong period at 87.6\dy, with an ephemeris of MJD 52337.5\pmt6.1 + n\x87.6\pmt0.3\dy.

\subsection{SXP91.1}

\textbf{AX J0051\minus722, RX J0051.3\minus7216 \\
RA 00 50 55, dec \minus72 13 38}

\his Discovered in the first observation in this survey with a period of 92\pmt1.5\s\ \citep{marshall97iauc}, further analysis improved this measurement to 91.12\pmt0.05\s\ \citep{corbet98iauc}. An orbital period of 115\pmt5\dy\ was derived by \citet{laycock2005} from early survey data (before MJD 52200). \citet{stevens99} identified the optical counterpart; \citet{schmidtke2004} find an 88.25\dy\ period in their analysis of \macho\ data for this star.

\begin{figure}
   \centering
        \includegraphics[angle=90,width=0.95\linewidth]{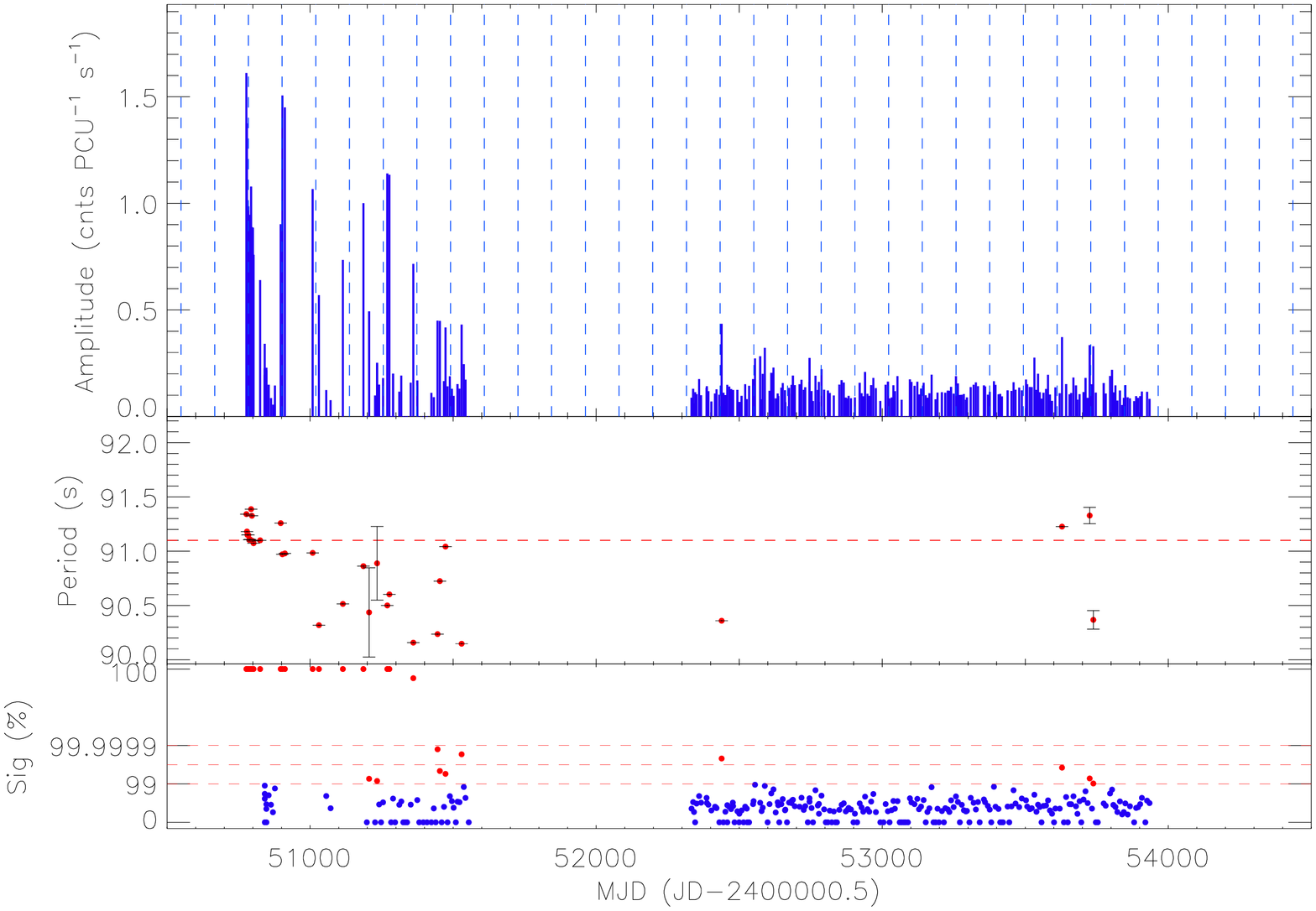}
        \includegraphics[angle=90,width=0.95\linewidth]{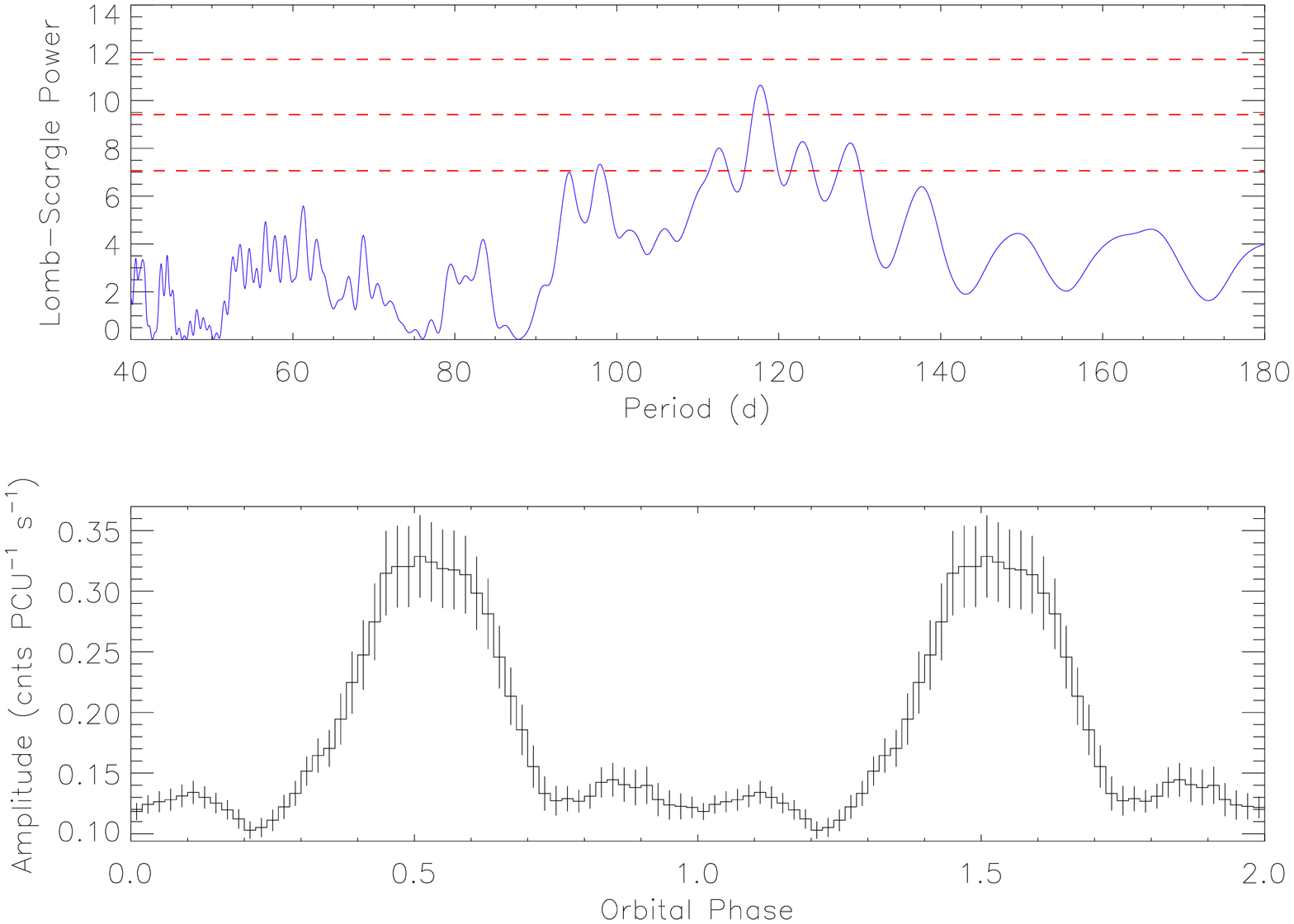}
   \caption{SXP91.1. a) \textit{Top}: \xray\ amplitude light curve. b) \textit{Middle}: Lomb-Scargle power spectrum; \textit{bottom}: light curve folded at 117.8\dy.}
   \label{fig_sxp91.1}
\end{figure}

\sur \citet{corbet2004proc} reported the discovery of a new 89\s\ pulsar from \xte\ observations in March 2002, it was located within the field of view of Position~1/A. After studying the long-term lightcurves of these two pulsars, we believe SXP89.0 is actually SXP91.1 after having spun up (see \fref{fig_sxp91.1}(a)). The Lomb-Scargle periodogram of the whole light curve returns a period of \aprox101\dy, which is slightly shorter than the period found by \citet{laycock2005}. Timing analysis of the light curve post MJD 52300 shows no clear periods. The ephemeris we derive is MJD 52197.9\pmt8.2 + n\x117.8\pmt0.5\dy. The average spin up during MJD 50750\sd51550 is calculated to be \pdoteq{1.8}{-8}, with a luminosity of \lumxge{3.6}{36} (\ble{2.5}{13}).

\begin{figure}
   \centering
        \includegraphics[angle=90,width=0.95\linewidth]{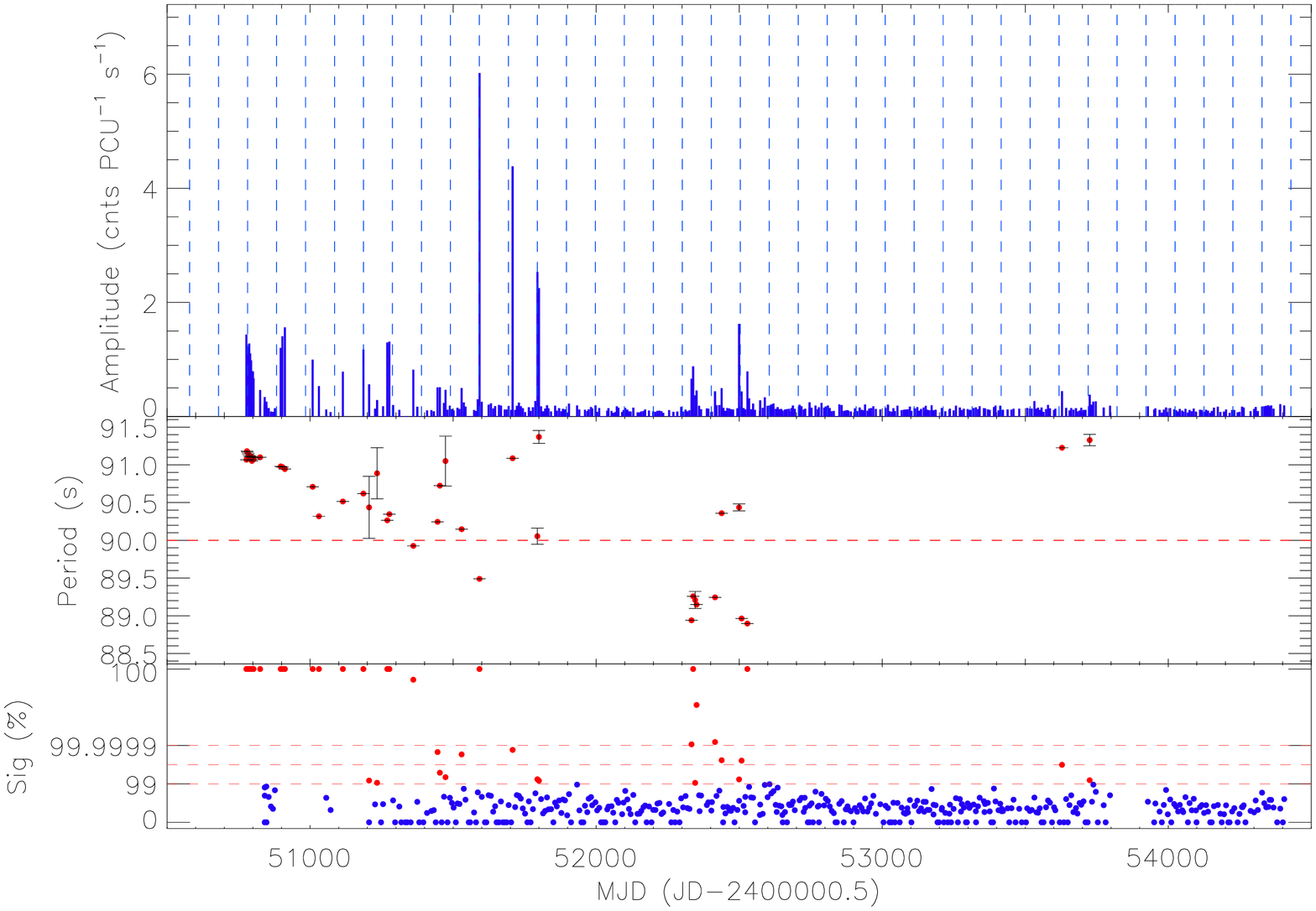}
        \includegraphics[angle=90,width=0.95\linewidth]{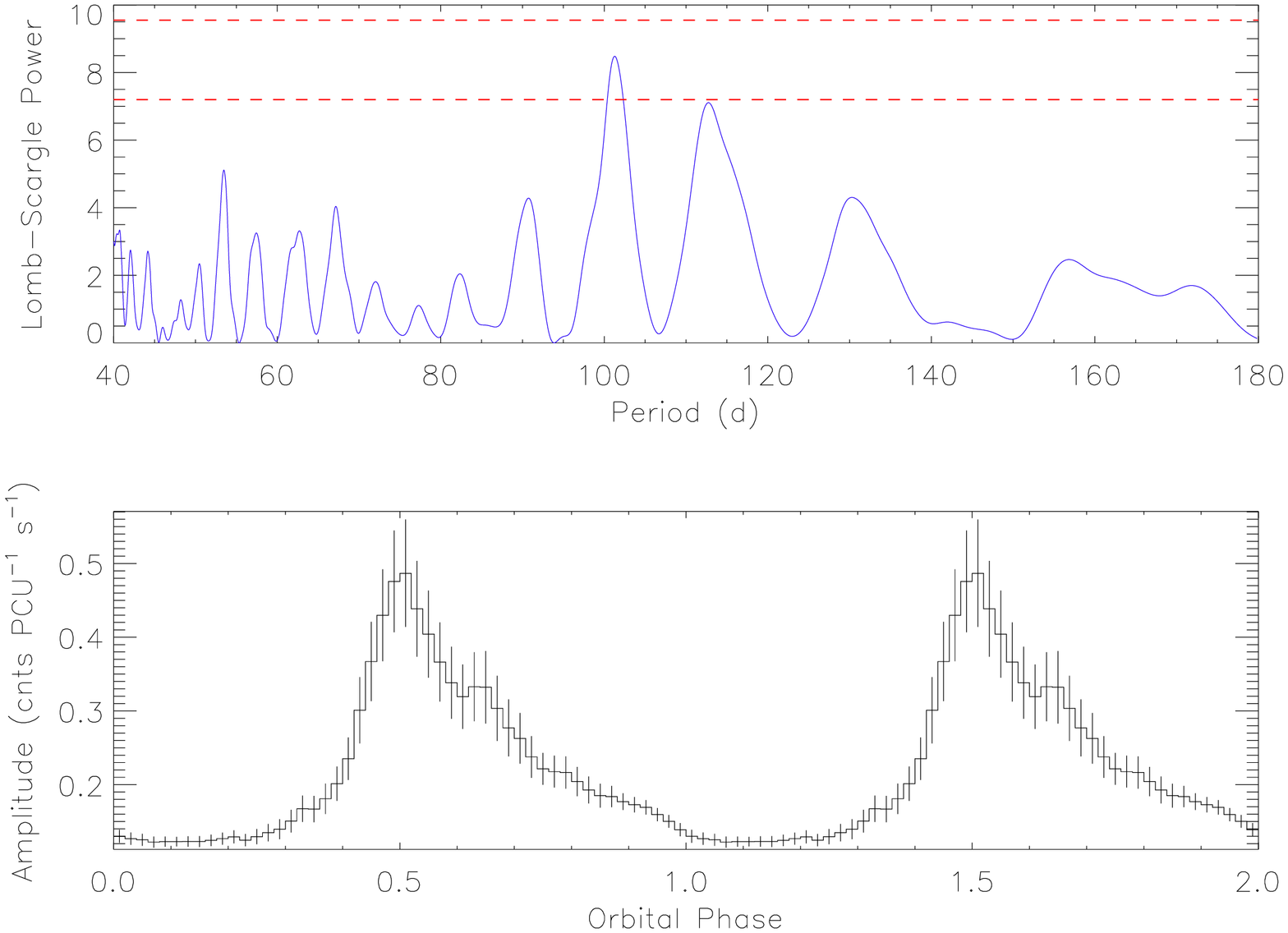}
   \caption{Consolidated light curve of SXP89.0 and SXP91.1. a) \textit{Top}: \xray\ amplitude light curve. b) \textit{Middle}: Lomb-Scargle power spectrum; \textit{bottom}: light curve folded at 101.3\dy.}
   \label{fig_sxp90.0}
\end{figure}

Given the proximity in spin period between SXP91.1 and SXP89.0, and the fact that the optical orbital period of SXP91.1 is so similar to the \xray\ period of SXP89.0, it is possible that they may be one and the same pulsar. In \fref{fig_sxp90.0}(a) we show the consolidated light curve for both pulsars showing the possibility that SXP91.1 spun up sufficiently to be later detected as a separate pulsar. We are hesitant to claim they are one single system because the orbital ephemeris of this consolidated data set does not coincide with either system's ephemeris: MJD 52301.2\pmt3.0 + n\x101.3\pmt0.4\dy. However, the orbital profile (see \fref{fig_sxp90.0}(b)) is convincing and appears similar to that of SXP8.88 (\fref{fig_sxp8.88}(b)) or SXP18.3 (\fref{fig_sxp18.3}(b)).

\subsection{SXP95.2}

\textbf{SMC 95 \\
RA 00 52, dec \minus72 45}

\his Was discovered in March 1999 in data from this survey \citep{laycock2002}; an approximate position was obtained with \pca\ scans over the source and has a large uncertainty. The \xray\ orbital period suggested from the two bright outbursts before MJD 52300 is 283\pmt8\dy\ \citep{laycock2005}.

\begin{figure}
   \centering
        \includegraphics[angle=90,width=0.95\linewidth]{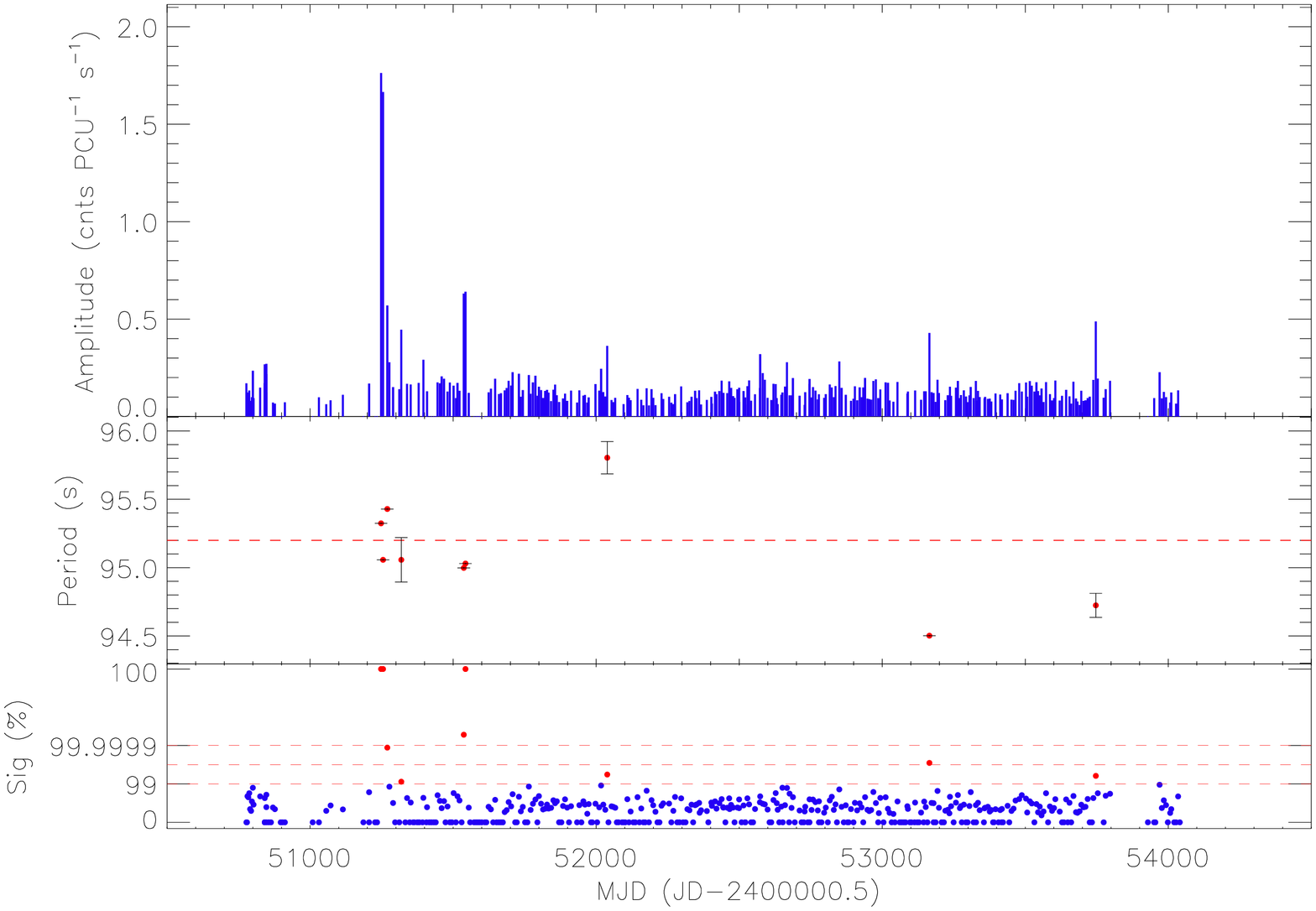}
   \caption{SXP95.2, \xray\ amplitude light curve.}
   \label{fig_sxp95.2}
\end{figure}

\sur Only 3 other marginal detections are available in the data (see \fref{fig_sxp95.2}). Lomb-Scargle analysis does not return any clear period, either including or excluding the two bright outbursts, although the former has its highest power at 71.3\dy, an ephemeris that is consistent with the source's significant detections and may reflect its orbital period.

\subsection{SXP101}

\textbf{RX J0057.3\minus7325, AX J0057.4\minus7325 \\
RA 00 57 26.8, dec \minus73 25 02}

\his Discovered in an \asca\ observation at a period of 101.45\pmt0.07\s\ \citep{yokogawa2000b}, and identified also as a \rosat\ source. Two optical counterparts were suggested by \citet{edge2003}; from a \chandra\ observation \citet{mcgowan2007} pinpoint the counterpart as a $V = 14.9$ star that exhibits a 21.9\dy\ periodicity in both \ogle~III and \macho\ data. \citet{schmidtkecowley2007b} find the same period.

\begin{figure}
   \centering
        \includegraphics[angle=90,width=0.95\linewidth]{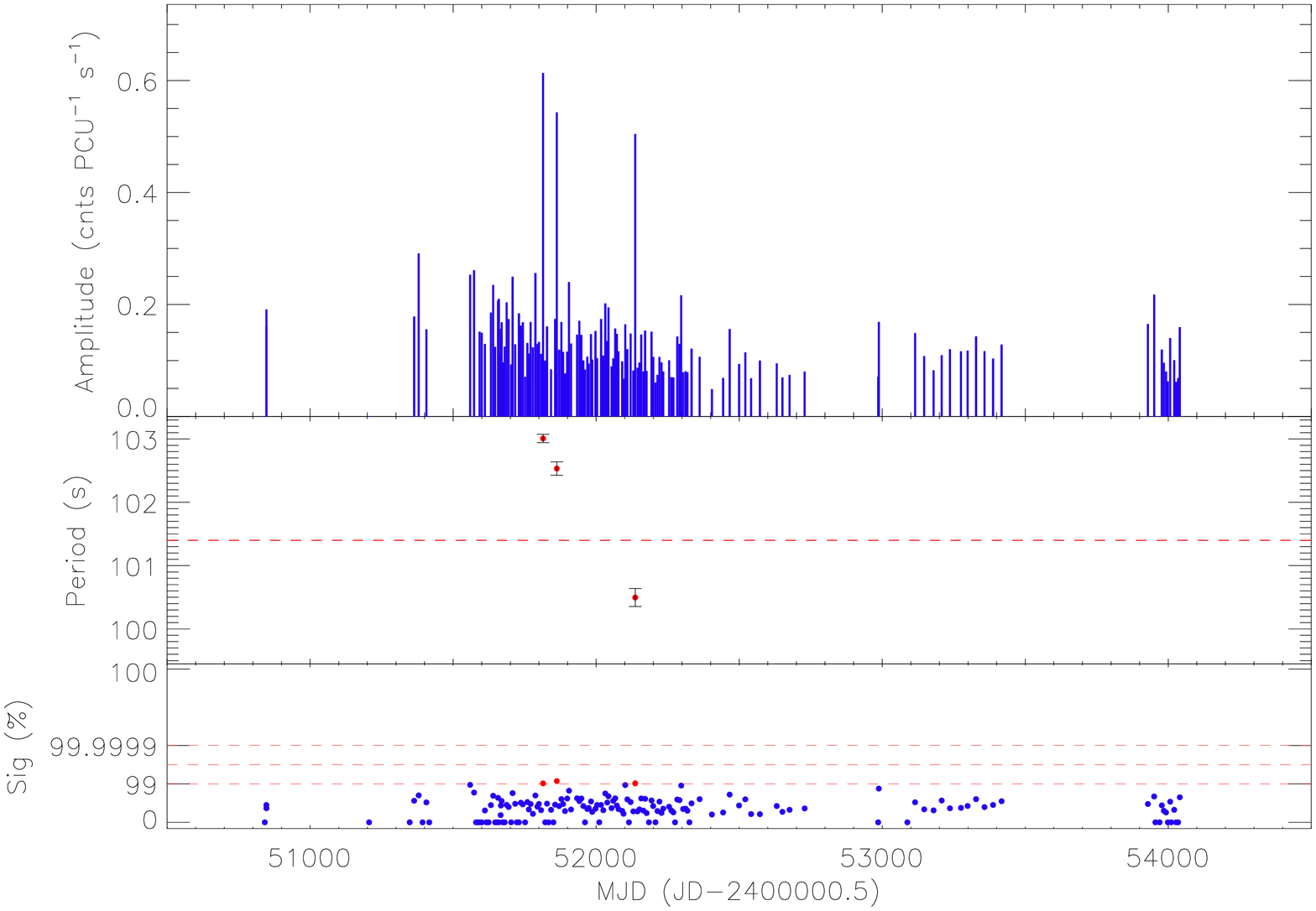}
   \caption{SXP101, \xray\ amplitude light curve.}
   \label{fig_sxp101}
\end{figure}

\sur This pulsar lies in the SE edge of the wing and was in the field of view of Positions 4 and 5, so coverage is only continuous throughout AO5 and AO6 (see \fref{fig_sxp101}); during this time only 3 outbursts of low brightness were observed (MJD 51814, 51863 and 52137). Timing analysis returns no significant periods, although the highest peak in the periodogram is at 25.2\dy, similar to the optical period.

\subsection{SXP138}

\textbf{CXOU J005323.8\minus722715 \\
RA 00 53 23.8, dec \minus72 27 15.0}

\his Discovered in archival \chandra\ data \citep{edge2004atelc}, the optical counterpart is [MA93] 667 \citep{williamthesis}. The \macho\ light curves for the companion star reveal peaks at \aprox125.1\dy\ intervals (stronger in the red band). \citet{schmidtkecowley2006} report finding a weak periodicity in the 122\sd123\s\ region in \macho\ data.

\begin{figure}
   \centering
        \includegraphics[angle=90,width=0.95\linewidth]{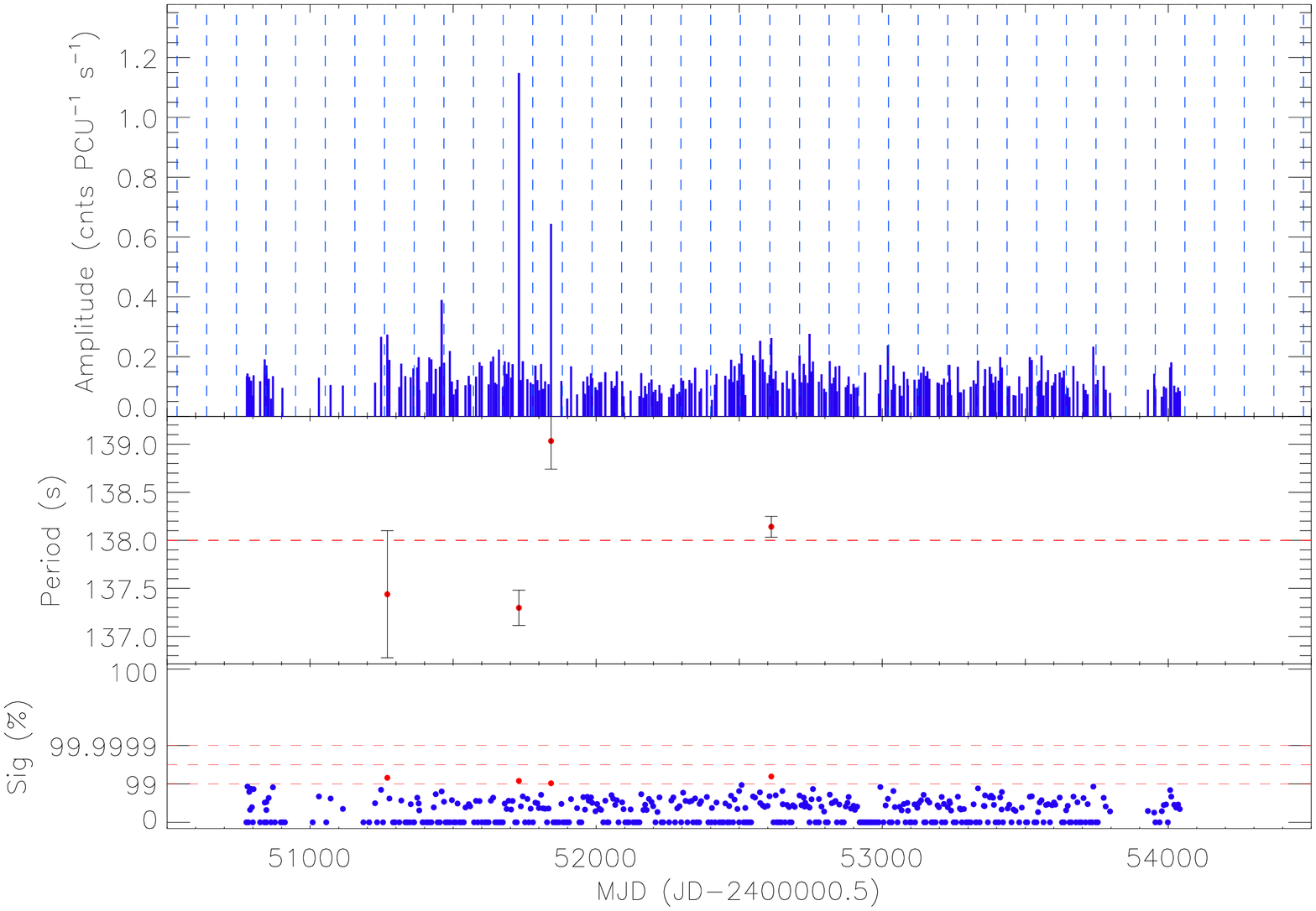}
        \includegraphics[angle=90,width=0.95\linewidth]{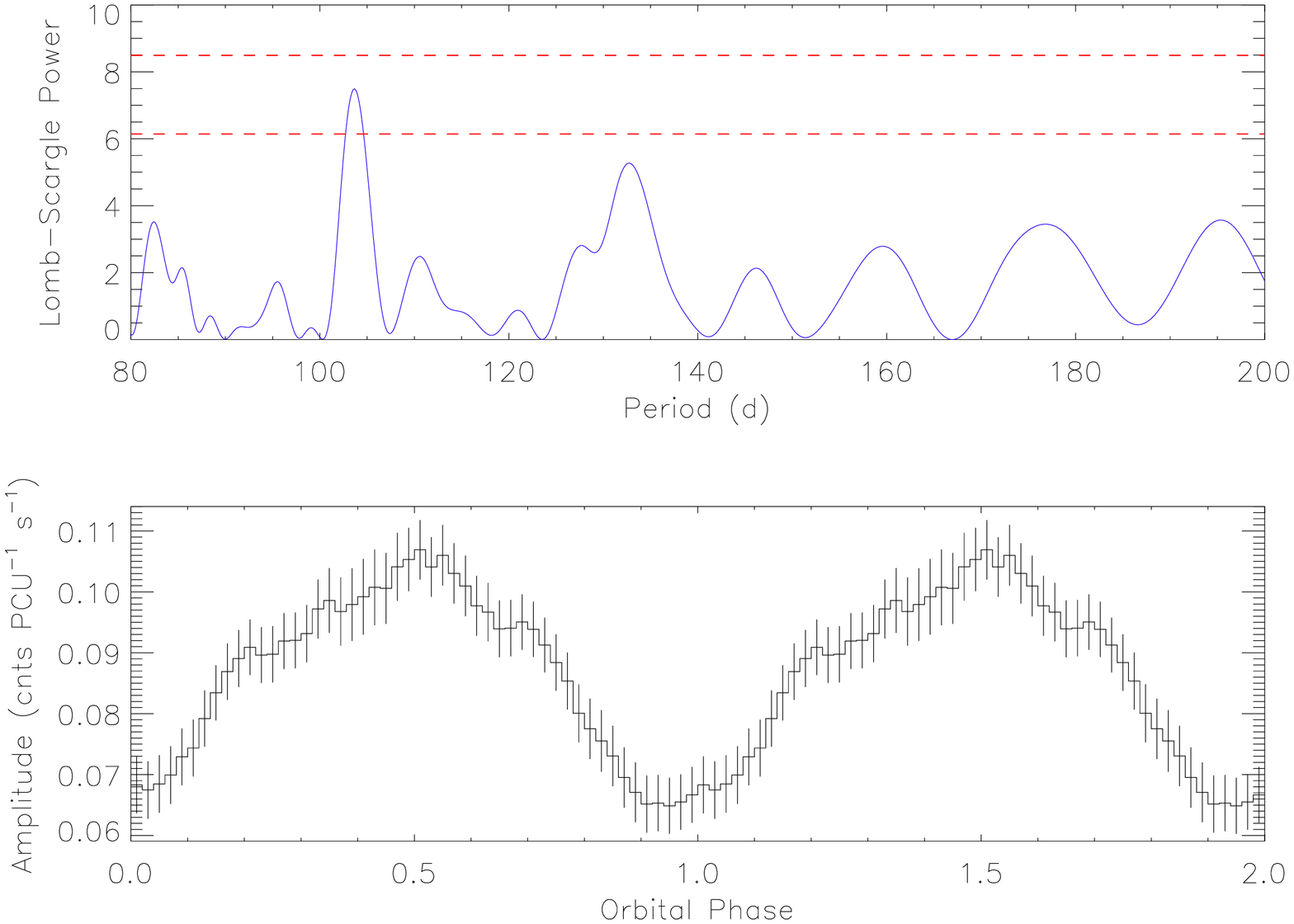}
   \label{fig_sxp138}
   \caption{SXP138. a) \textit{Top}: \xray\ amplitude light curve. b) \textit{Middle}: Lomb-Scargle power spectrum; \textit{bottom}: light curve folded at 103.6\dy.}
\end{figure}

\sur \xray\ data from this survey show two brighter detections \aprox112\dy\ apart (see \fref{fig_sxp138}(a)). Lomb-Scargle analysis finds no significant periods; however, if these two detections are removed from the data, a period of 103.6\dy\ appears in the power spectrum. Its ephemerides (shown by the vertical, dashed lines in \fref{fig_sxp138}(a)) do not correspond with the two detections, and the profile (see bottom panel of \fref{fig_sxp138}(b)) looks almost sinusoidal, unlike the profiles exhibited by other systems in the \smc. The maxima for this period occur at MJD 52400.2\pmt5.2 + n\x103.6\pmt0.5\dy.

\subsection{SXP140}

\textbf{XMMU J005605.2\minus722200, 2E0054.4\minus7237 \\
RA 00 56 05.7, dec \minus72 22 00}

\his Discovered in \xmm\ observations by \citet{sasaki2003}. The optical counterpart is believed to be [MA93] 904 \citep{haberlpietsch}. \citet{schmidtkecowley2006} find a 197\pmt5\dy\ period in \macho\ data.

\begin{figure}
   \centering
        \includegraphics[angle=90,width=0.95\linewidth]{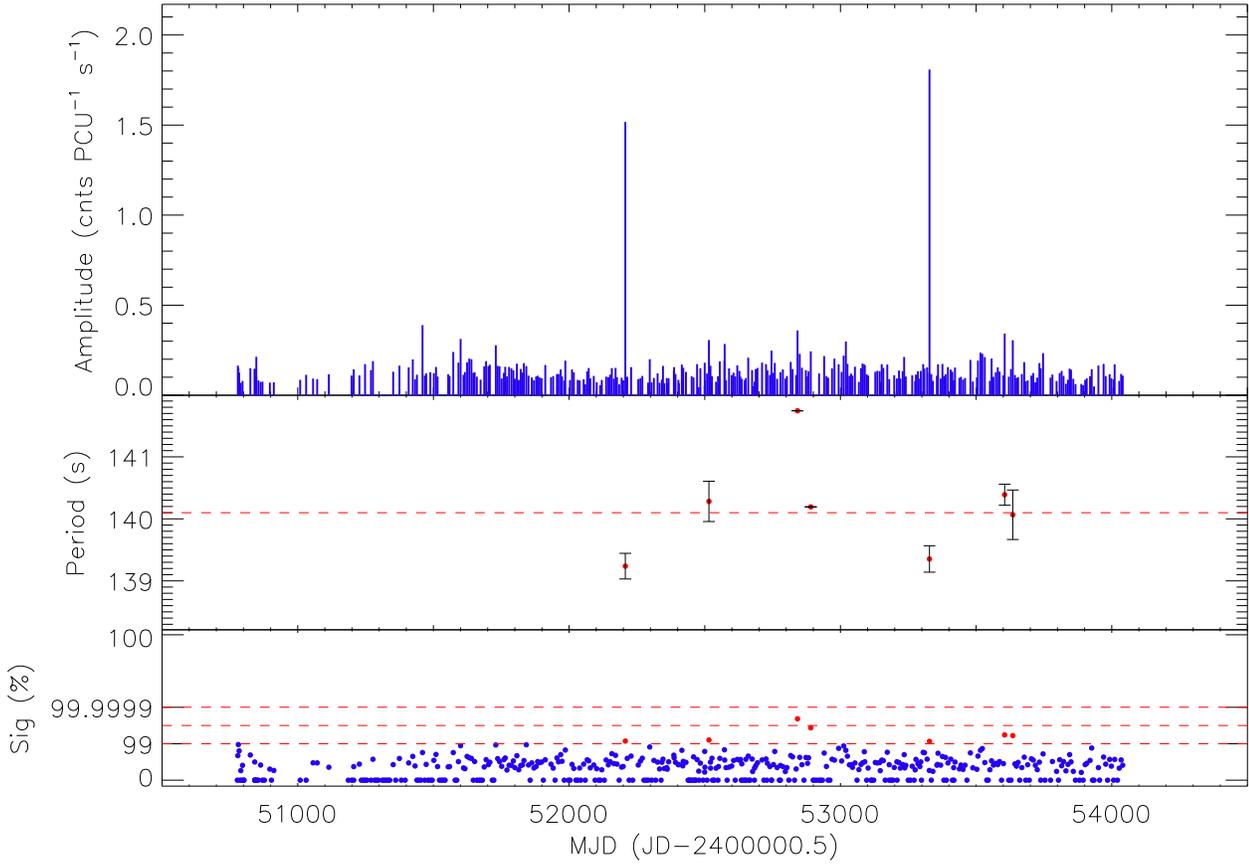}
   \caption{SXP140, \xray\ amplitude light curve.}
   \label{fig_sxp140}
\end{figure}

\sur None of the detections has been longer than 1 week, with only 2 of them showing significant brightness (see \fref{fig_sxp140}). Timing analysis returns no clear periods, although the periodogram of the data, excluding the two bright detections, does show some power at \aprox197\dy. As a number of different periods have similar power, this may only be coincidence. The folded light curve at 197\dy\ does not show a typical orbital profile.

\subsection{SXP144}

\textbf{XTE SMC pulsar \\
No position available}

\his Detected in observations from this survey in April 2003 by \citet{corbet2003atel}, who later reported an ephemeris of MJD 52779.2\pmt2.9 + n\x61.2\pmt1.6\dy\ \citep{corbet2003atelc}.

\begin{figure}
   \centering
        \includegraphics[angle=90,width=0.95\linewidth]{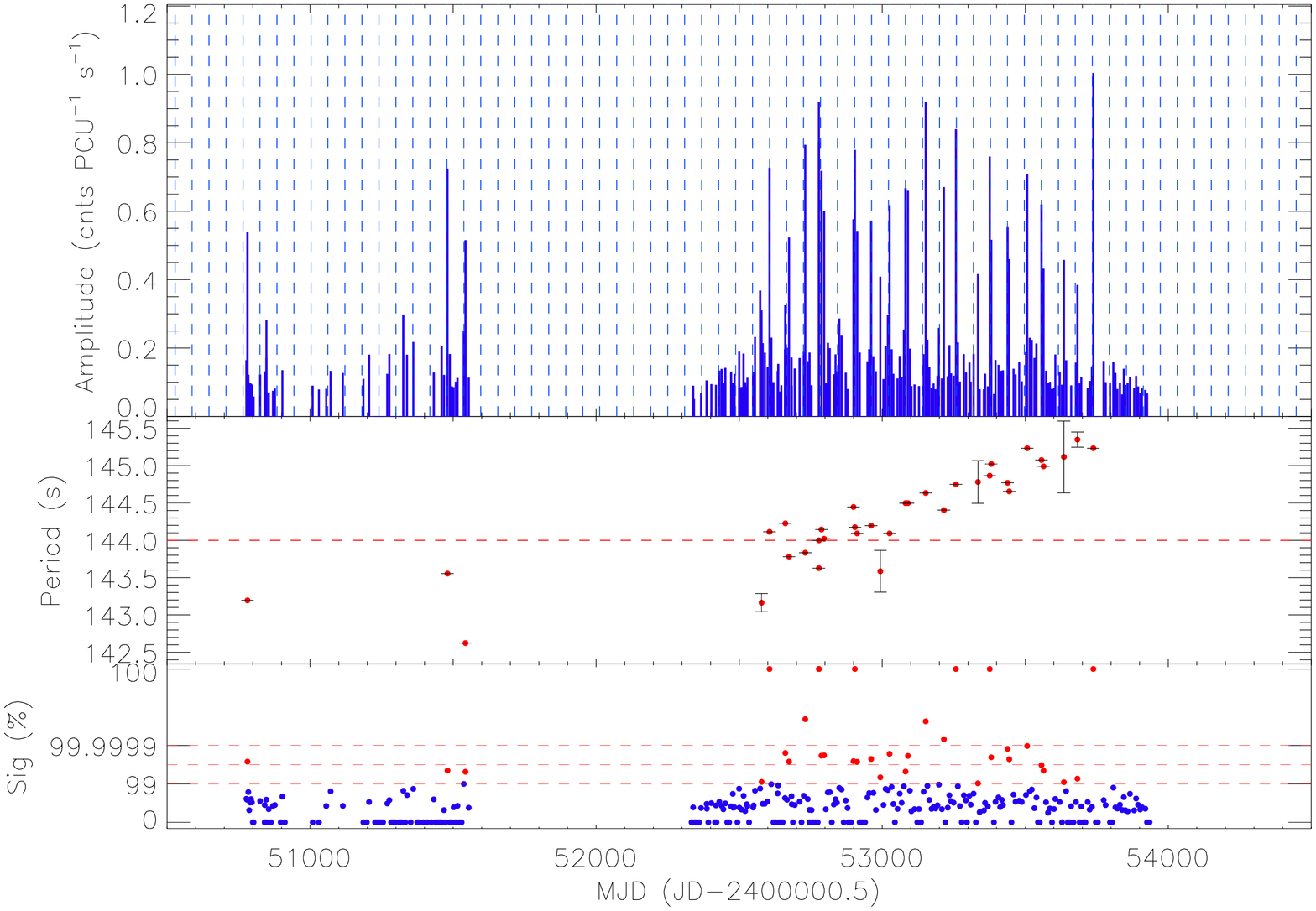}
        \includegraphics[angle=90,width=0.95\linewidth]{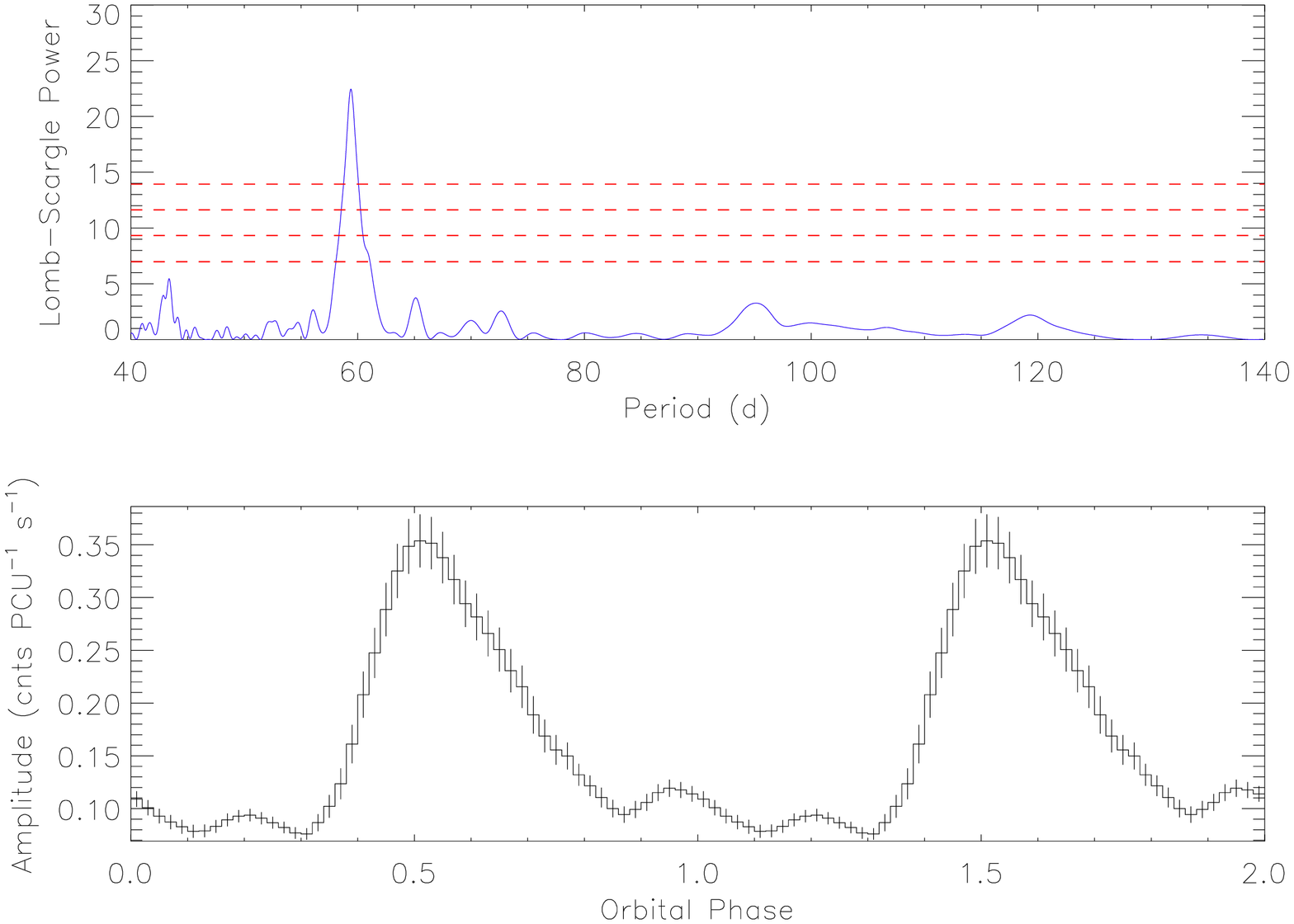}
   \caption{SXP144. a) \textit{Top}: \xray\ amplitude light curve. b) \textit{Middle}: Lomb-Scargle power spectrum; \textit{bottom}: light curve folded at 59.38\dy.}
   \label{fig_sxp144}
\end{figure}

\sur Although there are a few minor detections before the initial discovery, April 2003 (\aprox MJD 52750) saw the beginning of a regular pattern of outbursts which continued until February 2006 (\aprox MJD 53800, see \fref{fig_sxp144}(a)). The neutron star has displayed an extremely linear and constant spin \textit{down} during this time, with an average \pdoteq{1.6}{-8}, from which we derive a \lumxge{1.1}{36} (\ble{2.4}{13}). The improved outburst ephemeris from Lomb-Scargle analysis is MJD 52368.9\pmt1.8 + n\x59.38\pmt0.09\dy. This period is shorter than might have have expected from the pulse-orbit relationship, but what is most unusual about this pulsar is that it is spinning \textit{down}, moving it even further away from the Be group on the Corbet diagram.

\subsection{SXP152}

\textbf{CXOU J005750.3\minus720756 \\
RA 00 57 49, dec \minus72 07 59}

\his \citet{haberl2000} suggested that this object is a Be binary pulsar based on the \halpha -emitting object [MA93] 1038, although \rosat\ observations of this source had not detected any pulsations. These were found in a long \chandra\ observation by \citet{macomb2003} at a period of 152.098\pmt0.016\s\ (they report a very high pulse fraction of 64\pmt3\% and a \lumxeq{2.6}{35}), and in an \xmm\ observation by \citet{sasaki2003} at 152.34\pmt0.05\s.

\begin{figure}
   \centering
        \includegraphics[angle=90,width=0.95\linewidth]{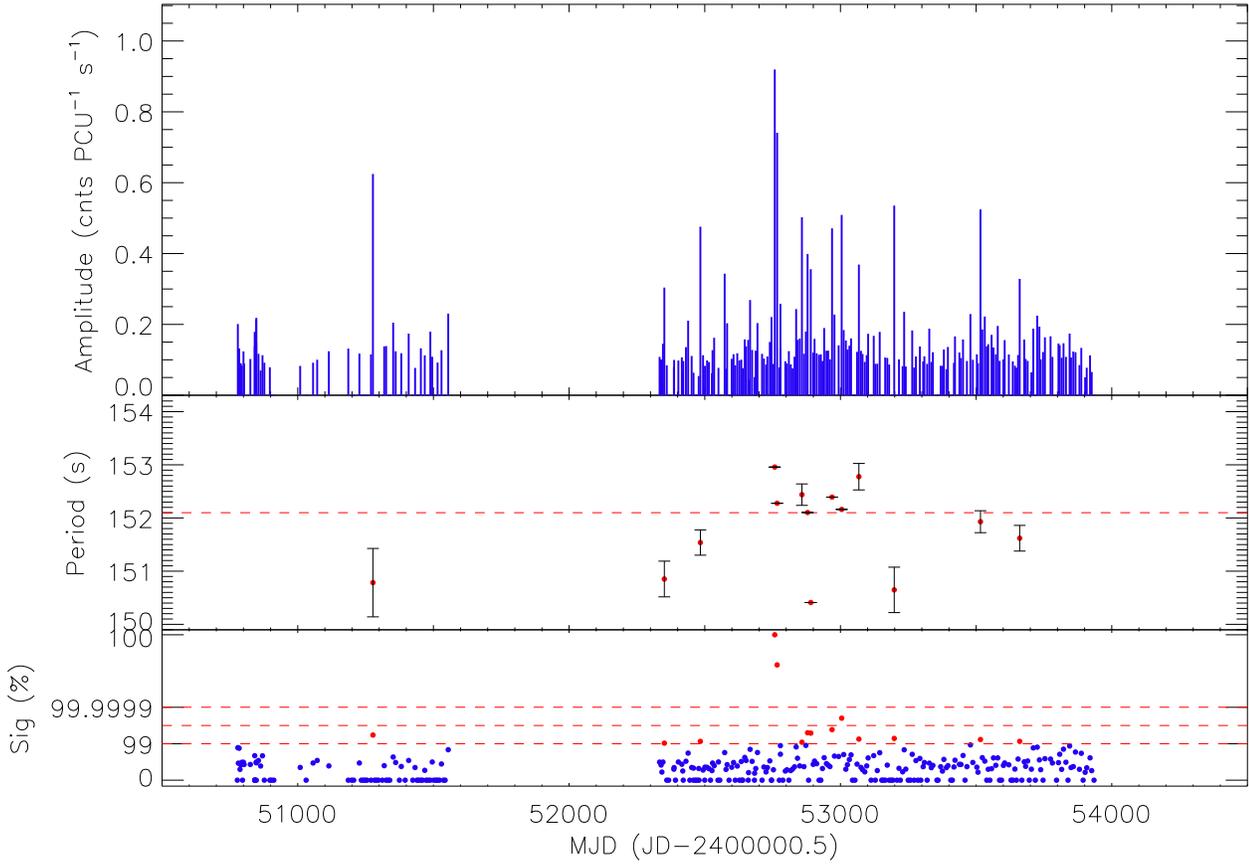}
   \caption{SXP152, \xray\ amplitude light curve.}
   \label{fig_sxp152}
\end{figure}

\sur The observational history is shown in \fref{fig_sxp152}. The periodogram of the light curve shows no clear orbital period, although the highest power peak is at \aprox107\dy, which would agree with the expected value from the Corbet diagram. The lack of periodic outbursts, despite its clear \xray\ activity, may point towards a low eccentricity system. This would limit accretion onto the neutron star to times when the Be star ejecta are dense enough, and would be independent of the orbital phase. Analysis of the optical light curve of the companion Be star is needed.

\subsection{SXP169} \label{sxp169}

\textbf{XTE J0054\minus720, AX J0052.9\minus7158, RX J0052.9\minus7158 \\
RA 00 52 54.0, dec \minus71 58 08.0}

\his First detected by \rxte\ in December 1998 at a period of 169.30\s\ \citep{lochneriauc98}. \citet{laycock2005} suggested a possible orbital period of 200\pmt40\dy. \citet{galache2005atel} reported an orbital period of 68.6\pmt0.2\dy\ based on data from this survey while \citet{schmidtke2006} found a period of 67.6\pmt0.3\dy\ in \ogle~-III data.

\begin{figure}
   \centering
        \includegraphics[angle=90,width=0.95\linewidth]{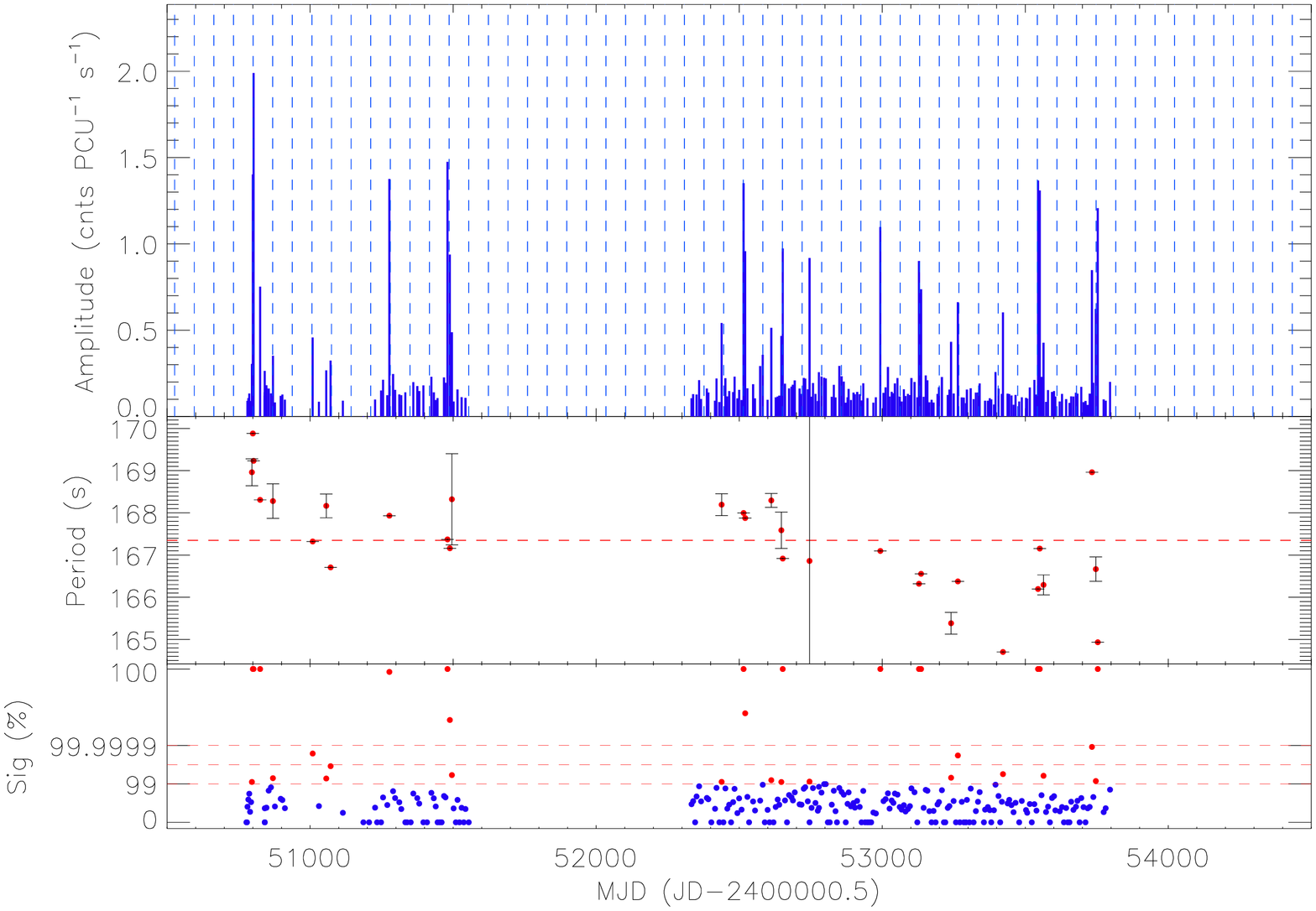}
        \includegraphics[angle=90,width=0.95\linewidth]{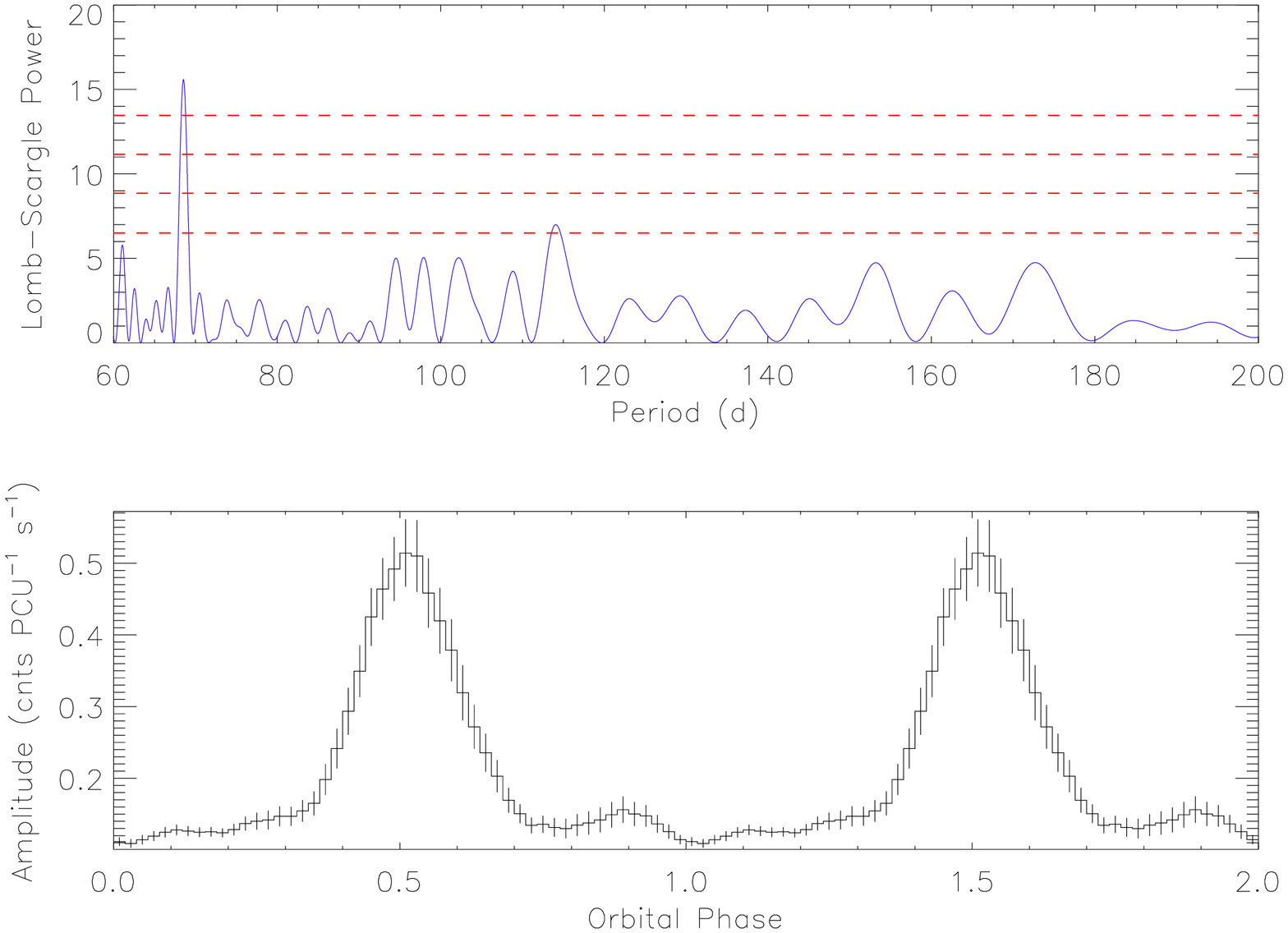}
   \caption{SXP169. a) \textit{Top}: \xray\ amplitude light curve. b) \textit{Middle}: Lomb-Scargle power spectrum; \textit{bottom}: light curve folded at 68.54\dy.}
   \label{fig_sxp169}
\end{figure}

\sur \citet{corbet2004proc} announced a new \smc\ pulsar at 164.7\s, with an unknown position. After comparing the long term light curves and the ephemerides from timing analysis it became apparent that SXP165 and SXP169 were the same source. A consolidated light curve is presented here (\fref{fig_sxp169}(a)), where the spin up of SXP169 throughout the survey is apparent. The estimated spin up during MJD 50800\sd51500 is \pdoteq{2.5}{-8}, implying \lumxge{1.2}{36} (\ble{3.0}{13}); for the remaining data the spin up is \pdoteq{2.0}{-8}, implying \lumxge{9.6}{35} (\ble{2.6}{13}). Lomb-Scargle analysis provides a clear period, and the outbursts are described by the ephemeris MJD 52240.1\pmt2.1 + n\x68.54\pmt0.15\dy, in agreement with the reported optical period.

\subsection{SXP172}

\textbf{AX J0051.6\minus7311, RX J0051.9\minus7311 \\
RA 00 51 52, dec \minus73 10 35}

\his Found in an \asca\ observation \citep{torii2000iauc}, it was identified with the \rosat\ source RX J0051.9\minus7311, which has a Be optical counterpart \citep{cowley97}. \citet{laycock2005} suggest a possible orbital period of \aprox67\dy\ based on the \xray\ activity up until MJD 52350. \citet{schmidtkecowley2006} report an optical period of 69.9\pmt0.6\dy\ in \ogle~II data. This pulsar has been detected on 17 occasions by \einstein, \rosat\ and \asca, but never above \lumxeq{7.8}{35} \citep{yokogawa2000e}.

\begin{figure}
   \centering
        \includegraphics[angle=90,width=0.95\linewidth]{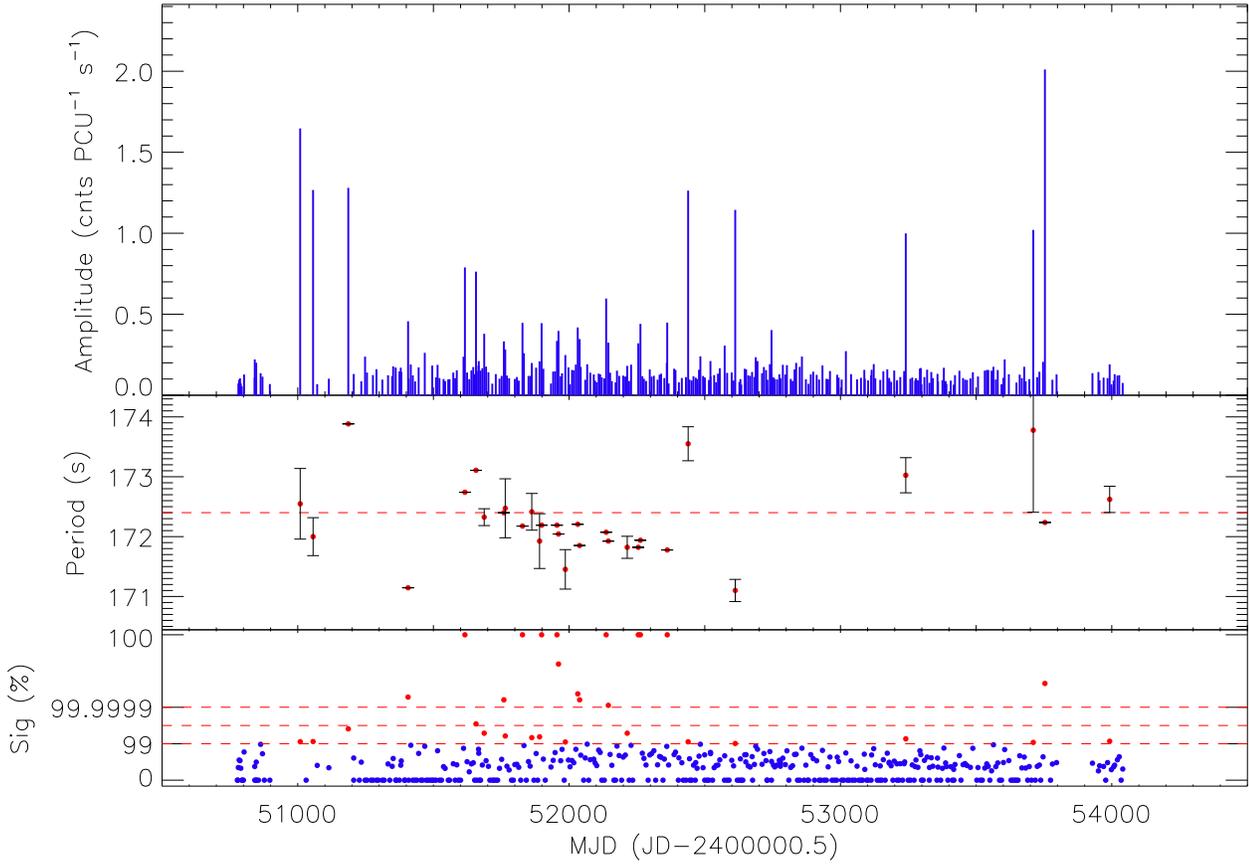}
   \caption{SXP172, \xray\ amplitude light curve.}
   \label{fig_sxp172}
\end{figure}

\sur SXP172 underwent a phase of intense, semi-regular, activity during MJD 51600\sd52400 (see \fref{fig_sxp172}). Lomb-Scargle analysis was carried using only data from this period, and also the whole data set, with no conclusive outcome. However, it is clear that the series of outbursts during the aforementioned dates are separated by \aprox70 days (consistent with the optical period). It is possible we are seeing contamination from another pulsar of similar pulse period, maybe located within a different pointing position. As past missions did not detect this pulsar in outburst, it is hard to confirm an \xray\ ephemeris.

\subsection{SXP202}

\textbf{XMMU J005920.8\minus722316 \\
RA 00 59 20.8, dec \minus72 23 16}

\his Detected in a number of archival \xmm\ observations and reported in \citet{majid2004}; the authors found an early B type star at the \xray\ coordinates and classified it as a HMXB.

\begin{figure}
   \centering
        \includegraphics[angle=90,width=0.95\linewidth]{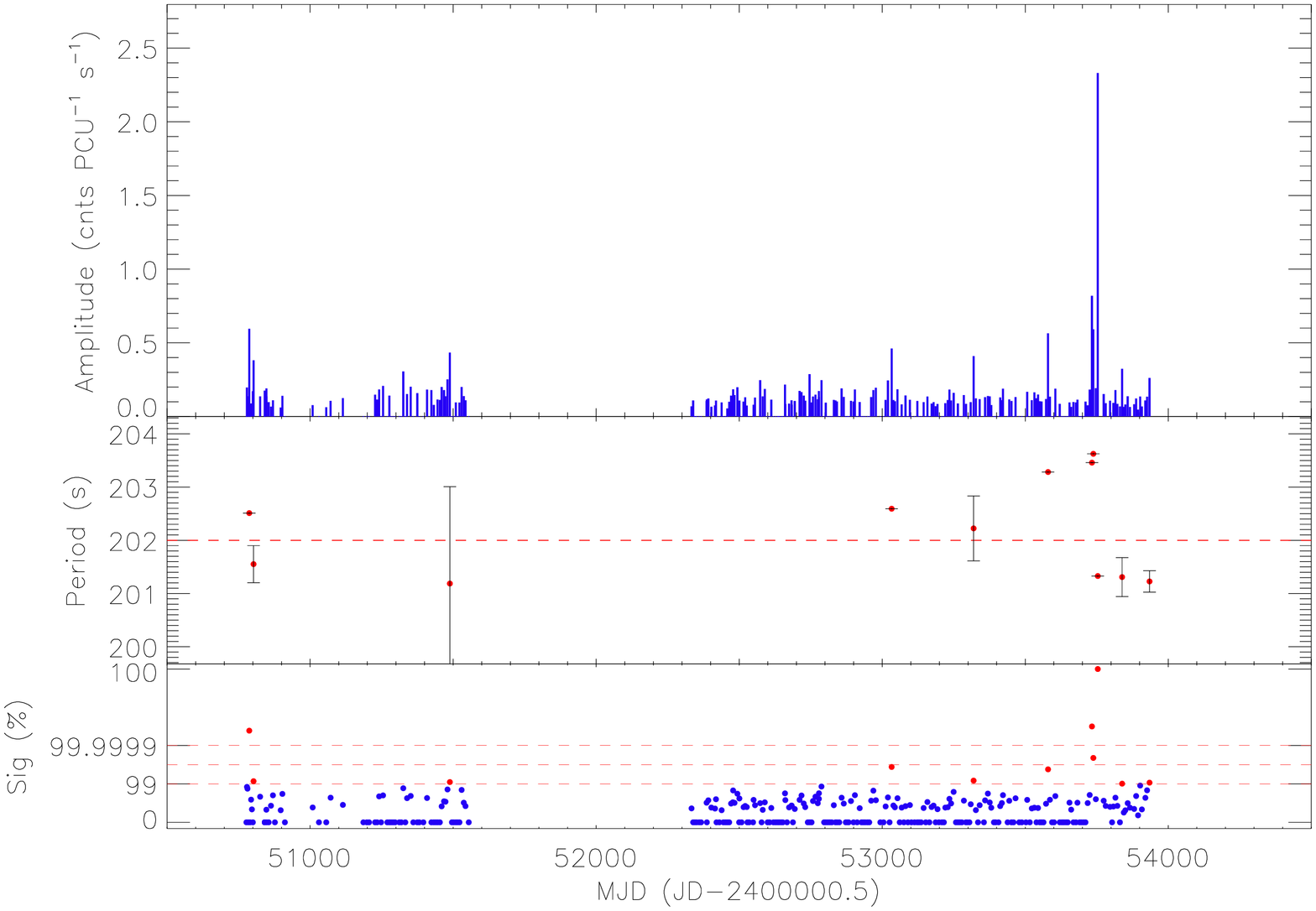}
   \caption{SXP202, \xray\ amplitude light curve.}
   \label{fig_sxp202}
\end{figure}

\sur This source has shown little bright activity throughout the survey except for the outburst in January 2006 (circa MJD 53740,  see \fref{fig_sxp202}). Lomb-Scargle analysis of the data returns no clear period, but we note that a \aprox91\dy\ orbital period would agree with the 6 outburst detections since MJD 53000.

\subsection{SXP264}

\textbf{XMMU J004723.7\minus731226, RX J0047.3\minus7312, AX J0047.3\minus7312 \\
RA 00 47 23.7, dec \minus73 12 25}

\his It was initially reported by \citet{yokogawa2003} from \asca\ observations, although it had previously been detected (yet remained undiscovered) in earlier \xmm\ observations \citep{ueno2004}. It had originally been proposed as a \bex\ binary candidate by \citet{haberl2000} based on its \xray\ variability. \citet{edge2005atel} identified the companion as the emission line star [MA93] 172 and found an optical period of 48.8\pmt0.6\dy, which they propose as the orbital period of the system. Further analysis of the \ogle\ light curve finds an ephemeris of MJD 50592\pmt2 + n\x49.2\pmt0.2\dy\ \citep{williamthesis}. \citet{schmidtkecowley2005} found a 49.1\dy\ period in \ogle\ data.

\begin{figure}
       \includegraphics[angle=90,width=0.95\linewidth]{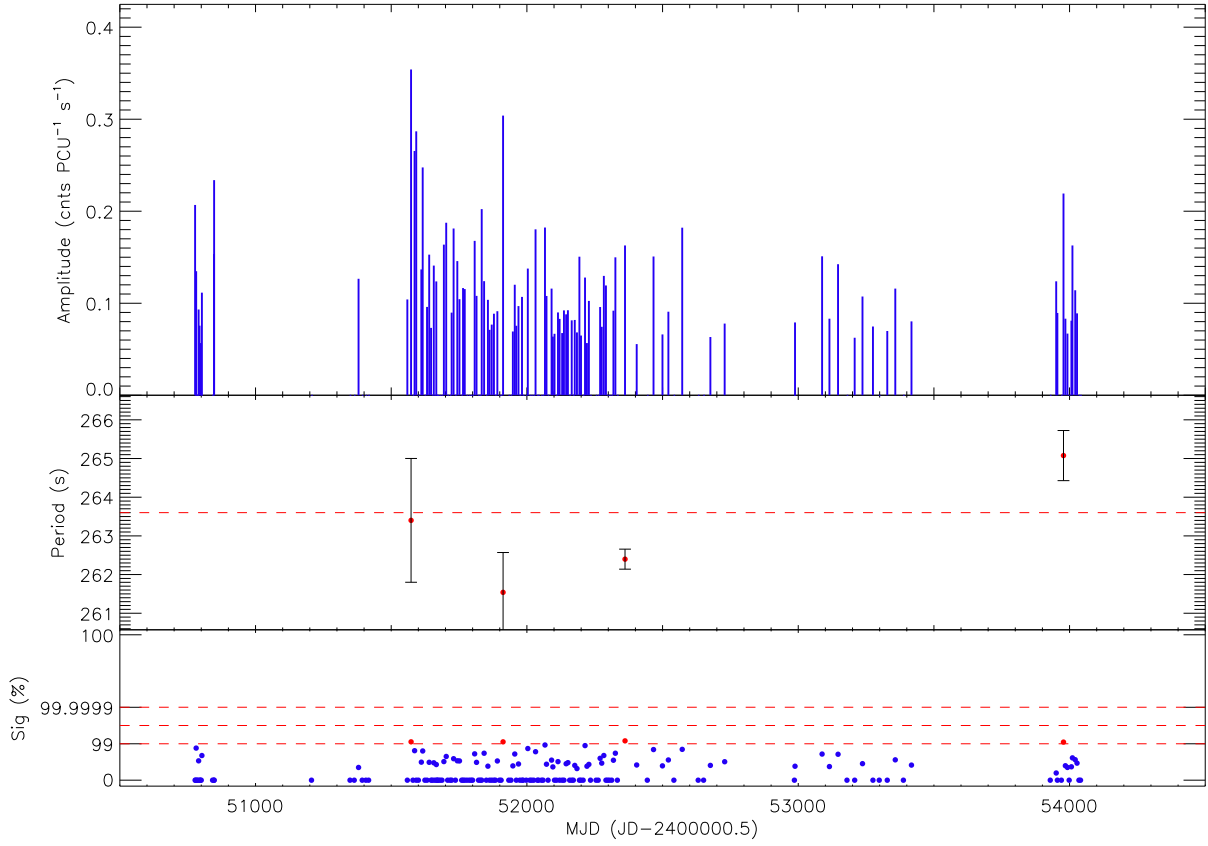}
   \caption{SXP264, \xray\ amplitude light curve.}
   \label{fig_sxp264}
\end{figure}

\sur Due to its location this pulsar was only observed consistently during AO5 and AO6 (see \fref{fig_sxp264}). No major outbursts were detected and the power spectrum shows no significant peak at the optical period, nor at any other. We note that the optical ephemeris does not agree with the \xray\ activity.

\subsection{SXP280}

\textbf{RX J0057.8\minus7202, AX J0058\minus72.0 \\
RA 00 57 48.2, dec \minus72 02 40}

\his  Discovered in March 1998 in an \asca\ observation by \citet{yokogawa98iauc} at a period of 280.4\pmt0.3\s. It is identified with the Be star [MA93] 1036, and \citet{schmidtke2006} find a 127.3\dy\ period in its \ogle\ data with an epoch of maximum brightness at MJD 52194.7. This pulsar was observed on 15 occasions by \einstein, \rosat\ and \asca, but never detected above \lumxeq{6.0}{35} \citep{tsujimoto99}.

\begin{figure}
   \centering
        \includegraphics[angle=90,width=0.95\linewidth]{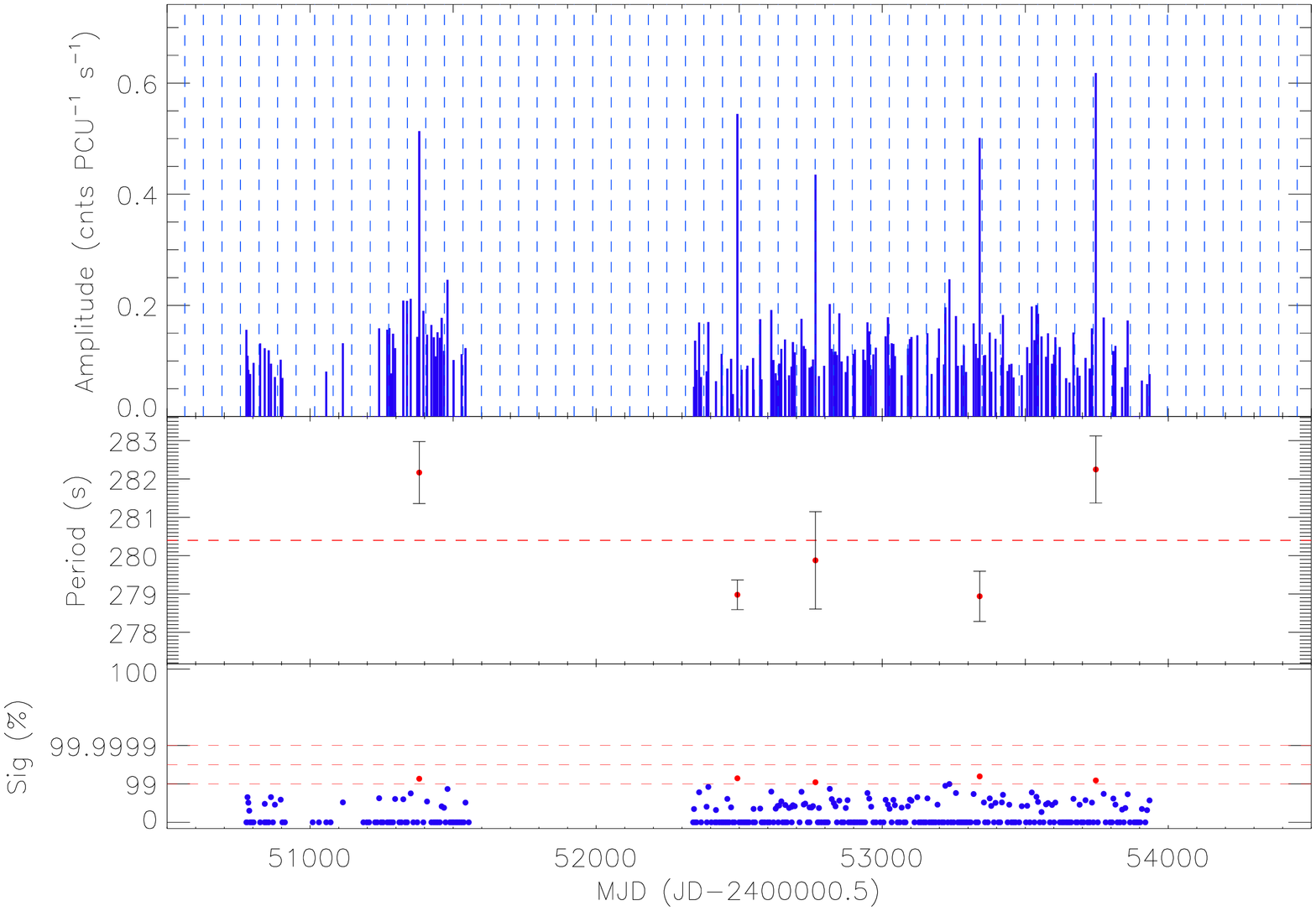}
        \includegraphics[angle=90,width=0.95\linewidth]{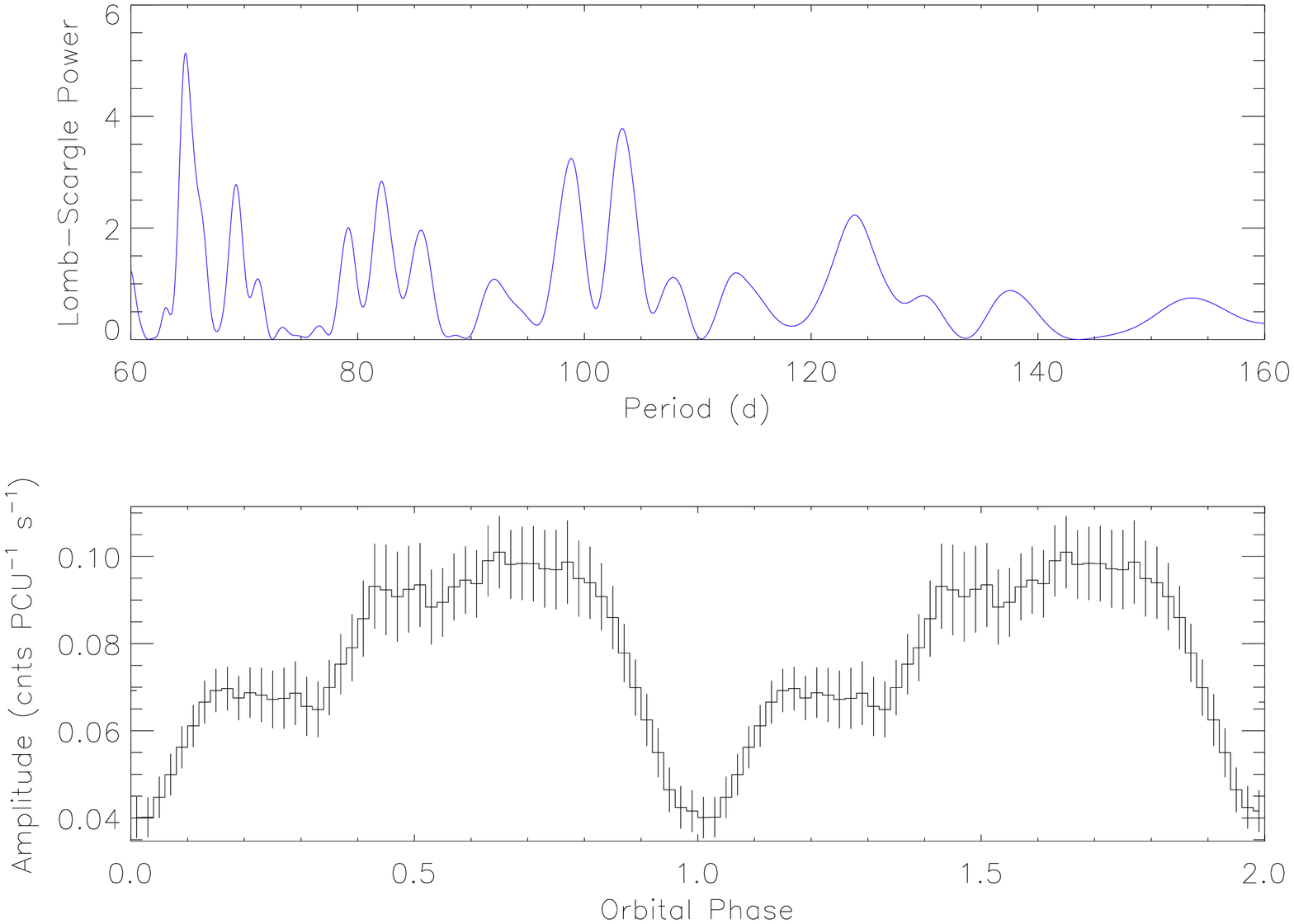}
   \caption{SXP280. a) \textit{Top}: \xray\ amplitude light curve. b) \textit{Middle}: Lomb-Scargle power spectrum; \textit{bottom}: light curve folded at 64.8\dy.}
   \label{fig_sxp280}
\end{figure}

\sur There are only 5 clear detections of this source throughout the survey, and the power spectrum shows no significant periods, although the highest peak is at 64.8\dy, which is \aprox$\frac{1}{2}$ of the optical period. For this reason, the ephemeris lines and folded light curve are shown in \fref{fig_sxp280}. The tentative ephemeris of the \xray\ period is MJD 52312\pmt6 + n\x64.8\pmt0.2\dy, which is within \aprox12\dy\ of the epoch of peak optical flux. A lack of earlier detections of this system in outburst makes it difficult to firmly establish an \xray\ ephemeris.

\subsection{SXP293}

\textbf{XTE J0051\minus727 \\
No position available}

\his Reported by \citet{corbet2004atelc} from observations during this survey of the outburst at MJD 53097.

\begin{figure}
   \centering
        \includegraphics[angle=90,width=0.95\linewidth]{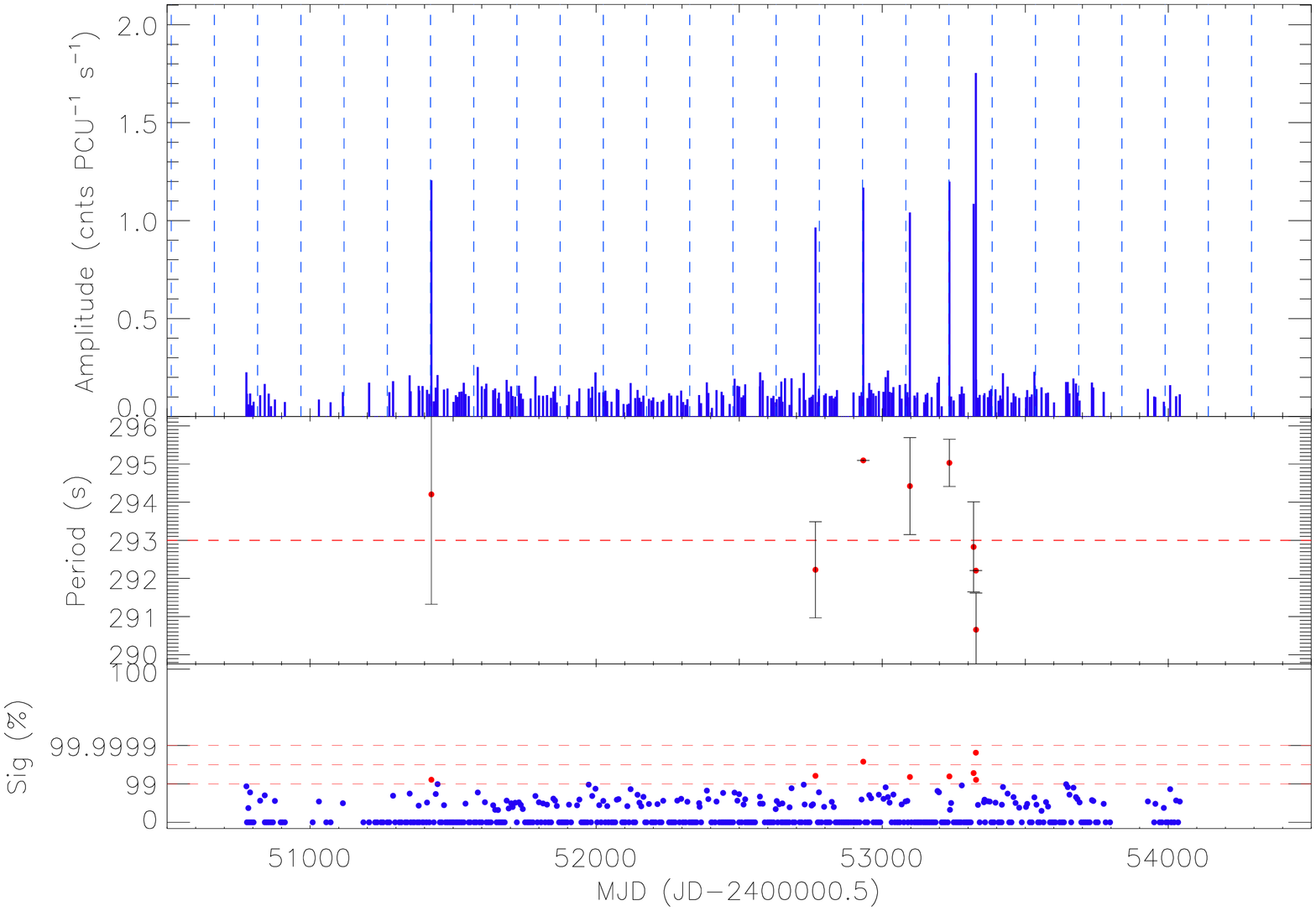}
        \includegraphics[angle=90,width=0.95\linewidth]{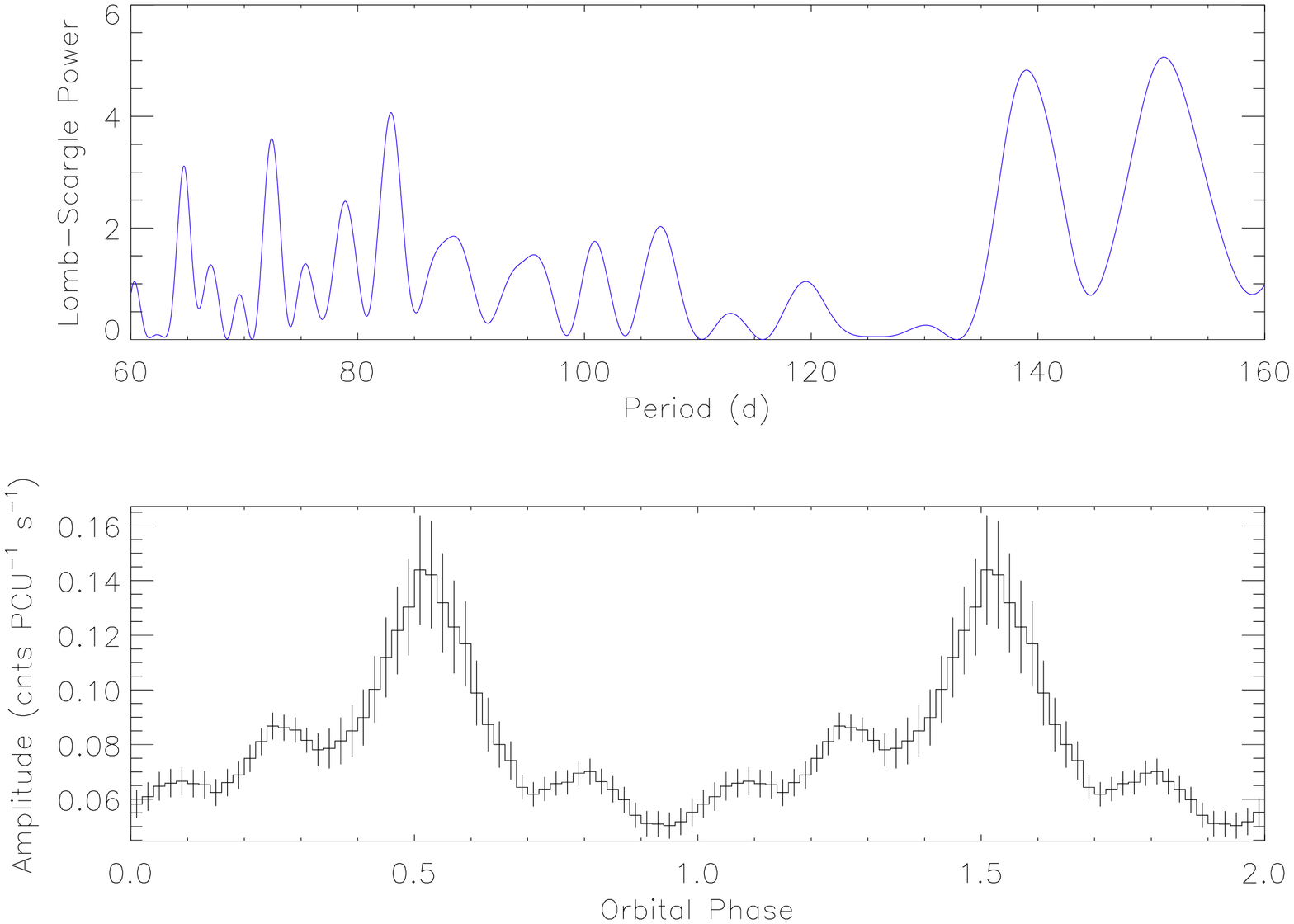}
   \caption{SXP293. a) \textit{Top}: \xray\ amplitude light curve. b) \textit{Middle}: Lomb-Scargle power spectrum; \textit{bottom}: light curve folded at 151\dy.}
   \label{fig_sxp293}
\end{figure}

\sur Only one of the 6 outbursts lasts longer than one week (\fref{fig_sxp293}(a)), which could imply this is a Type~II outburst, and thus not expected to coincide with periastron passage. For this reason, it was removed from the data after initial timing analysis produced no results. The resulting power spectrum, while not showing any significant periods, does have its strongest peak at 151\dy, an ephemeris that agrees well with the remaining 5 detections. We suggest this is the likely orbital period of the system, with an ephemeris of MJD 52327.3\pmt4.5 + n\x151\pmt1\dy.

\subsection{SXP304}

\textbf{RX J0101.0\minus7206, CXOU J010102.7\minus720658 \\
RA 01 01 01.7, dec \minus72 07 02}

\his Discovered in \chandra\ observations at 304.49\pmt0.13\s, the optical counterpart is identified as the emission line star [MA93] 1240 \citep{macomb2003}. The authors measured an unusually high pulse fraction of 90\pmt8\% at a luminosity of \lumxeq{1.1}{34}. \citet{schmidtkecowley2006} suggest there may be a 520\pmt12\dy\ period in \macho\ data of the optical counterpart [MA93] 1240.

\begin{figure}
   \centering
        \includegraphics[angle=90,width=0.95\linewidth]{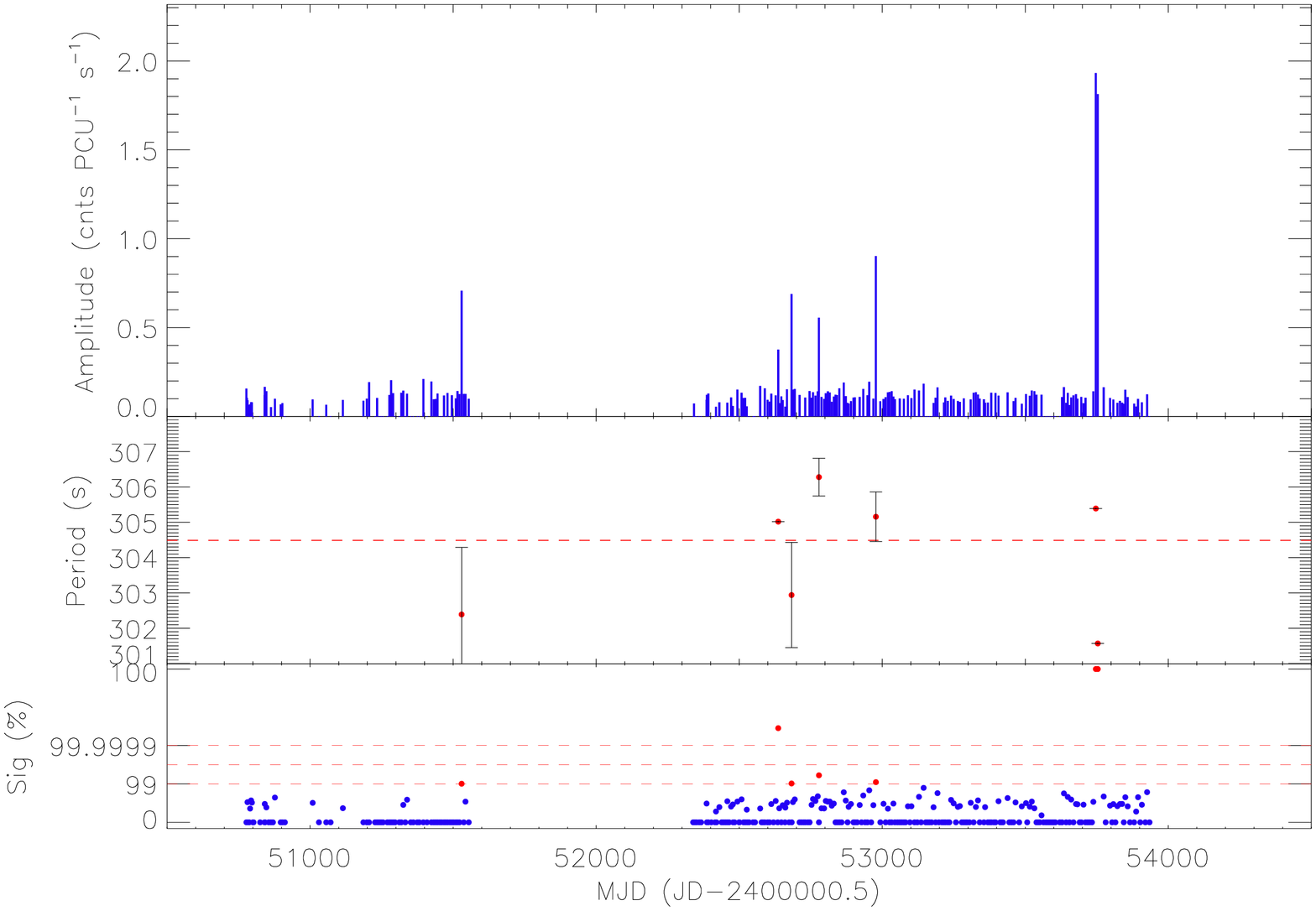}
   \caption{SXP304, \xray\ amplitude light curve.}
   \label{fig_sxp304}
\end{figure}

\sur This source was out of the field of view during AO5 and AO6, so it has only been adequately covered from AO7 onwards. A number of small outbursts were detected during MJD 52600\sd53000 (see \fref{fig_sxp304}), and it was not detected again until recently, in MJD 53747, when it began a 2\sd3 week outburst, the longest and brightest observed so far in this survey. Lomb-Scargle analysis of the data outside of this outburst finds moderate power at \aprox50\dy\ and no significant power at the optical period. It should be noted that this source displays \xray\ activity on timescales much shorter than the reported 520\dy\ optical period.

\subsection{SXP323}

\textbf{RX J0050.7\minus7316, AX J0051\minus733 \\
RA 00 50 44.8, dec \minus73 16 06.0}

\his Pulsations at 323.2\pmt0.4\s\ were detected in November 1997 \citep{yokogawa98iauc} at the coordinates of the \rosat\ source RX J0050.7\minus7316. \citet{cowley97} identified the optical counterpart as a Be star. This system has been found to exhibit optical and IR variability at periods of \aprox0.7 and 1.4\dy\ \citep{coe2002} and 1.695\dy\ \citep{coe2005}. These periods are too short to be the orbital period of the system and are most likely non-radial pulsations in the Be star. \citet{laycock2005} suggest an orbital period of 109\pmt18\dy\ from \xray\ data earlier in this survey.

\begin{figure}
   \centering
        \includegraphics[angle=90,width=0.95\linewidth]{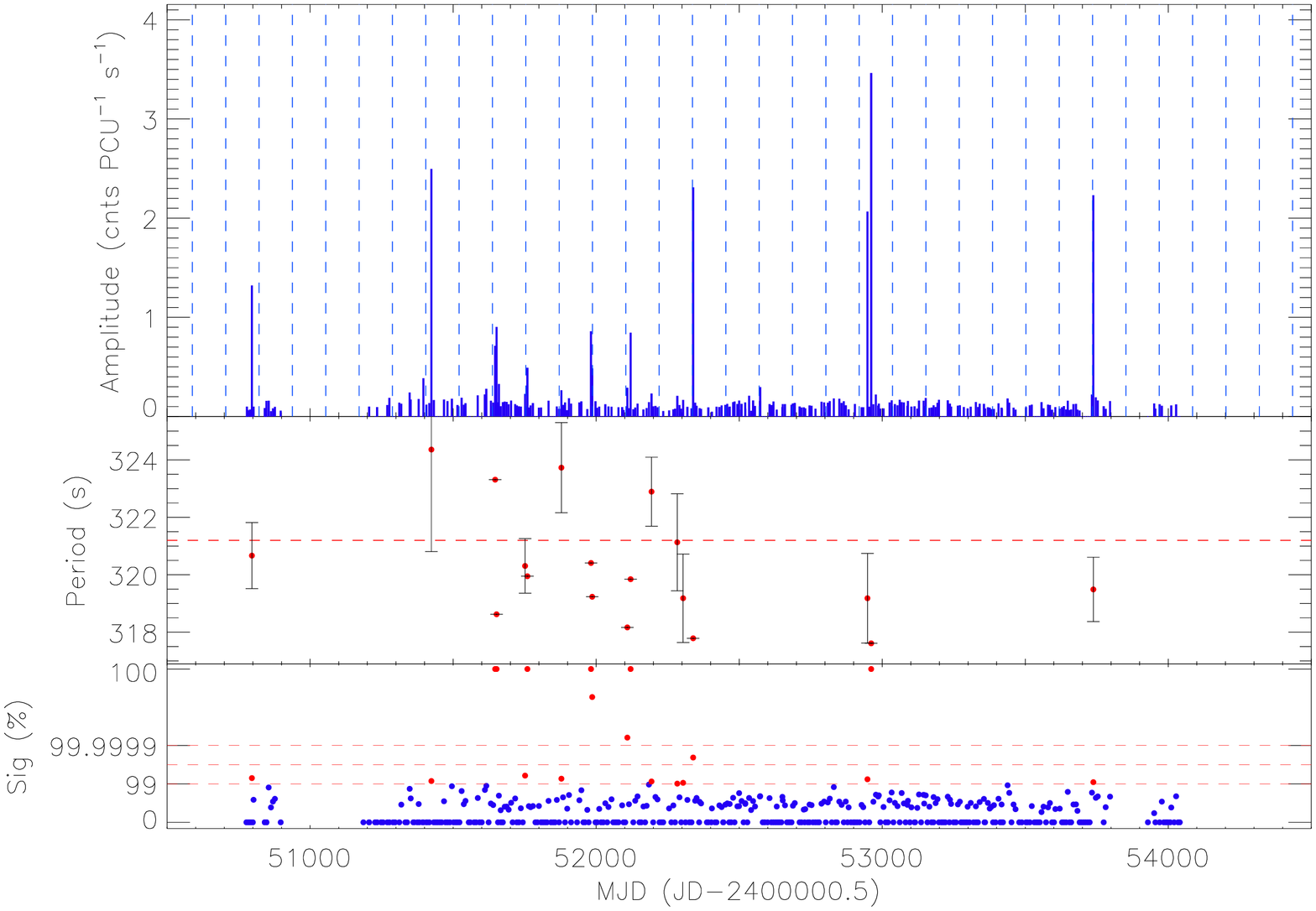}
        \includegraphics[angle=90,width=0.95\linewidth]{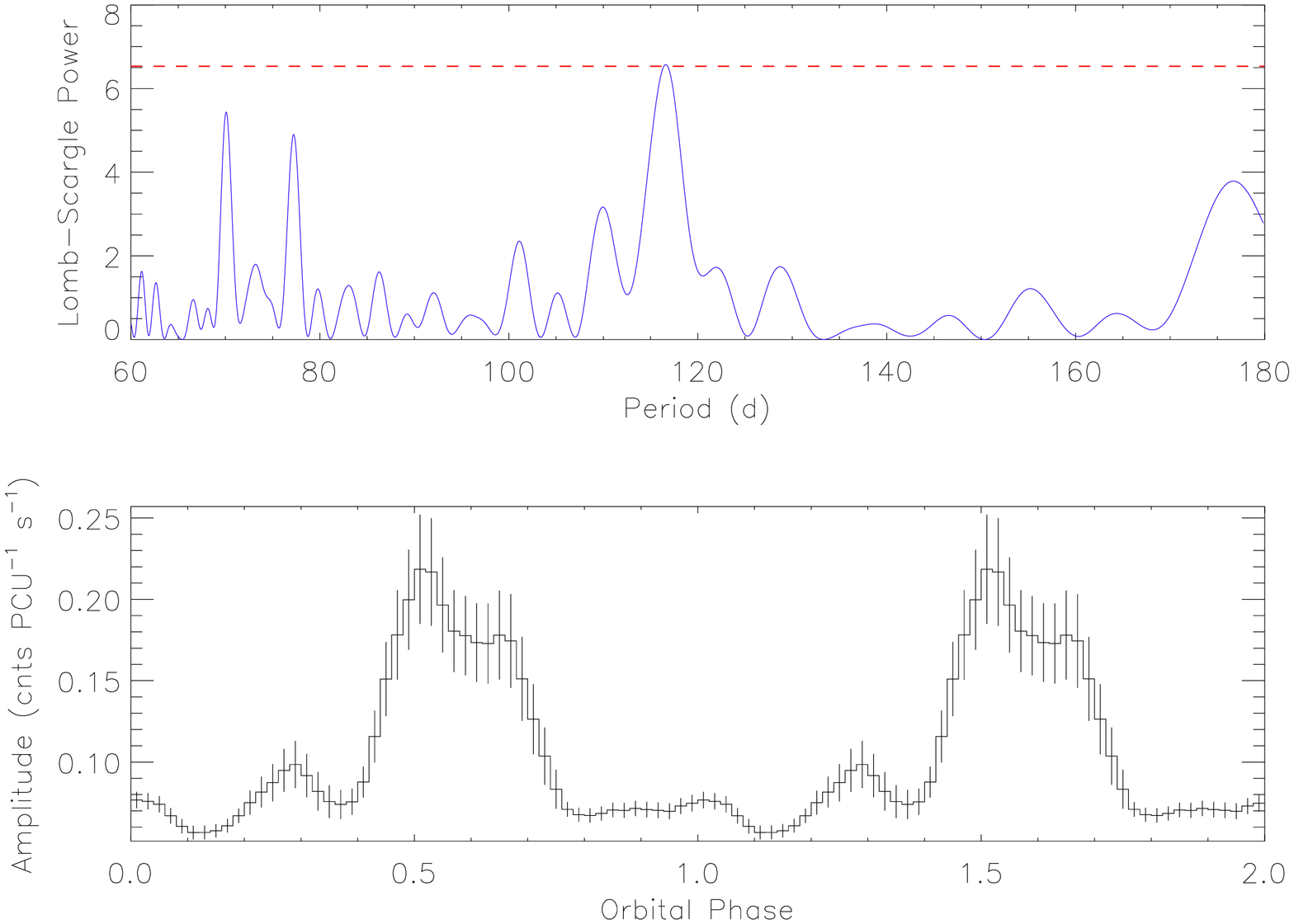}
   \caption{SXP323. a) \textit{Top}: \xray\ amplitude light curve. b) \textit{Middle}: Lomb-Scargle power spectrum; \textit{bottom}: light curve folded at 116.6\dy.}
   \label{fig_sxp323}
\end{figure}

\sur This pulsar showed quite regular, bright activity during the survey (see \fref{fig_sxp323}(a)). We found that the outburst circa MJD 52960 skewed the timing analysis results, which we attribute to it being a Type~II outburst; for this reason it was excluded from the analysis. The ephemeris found for the remaining outbursts is MJD 52336.9\pmt3.5 + n\x116.6\pmt0.6\dy, which is consistent with the orbital period proposed by \citet{laycock2005}.

\subsection{SXP348}

\textbf{RX J0102\minus722, SAX J0103.2 \minus7209, AX J0103.2\minus7209 \\
RA 01 03 13.0, dec \minus72 09 18.0}

\his Pulsations at 345.2\pmt0.1\s\ were detected in \beppo\ observations in July 1998 \citep{israel98iauc}. Subsequently, pulsations at 348.9\pmt0.1\s\ were found in archival \asca\ data from May 1996 implying a \pdoteq{-5.39}{-8} \citep{yokogawa98iaucb}. The \rosat\ source lies in a supernova remnant with a Be counterpart; a weak 93.9\dy\ periodicity has been reported from \ogle\ data \citep{schmidtkecowley2006}. \citet{israel2000} suggest this is a persistent, low luminosity \xray\ system.

\sur Although this source has been detected by a number of different instruments, it was always at low luminosities ($\lesssim$\expt{36}\ergps), so it is not surprising that there are not many detections in the survey {\fref{fig_sxp348}. Furthermore, it has also been detected at a wider range of periods than other pulsars (from 340 to 348\s), which makes it more difficult to pick out in the periodograms from weekly observations. Timing analysis reveals no significant periods.

\begin{figure}
   \centering
        \includegraphics[angle=90,width=0.95\linewidth]{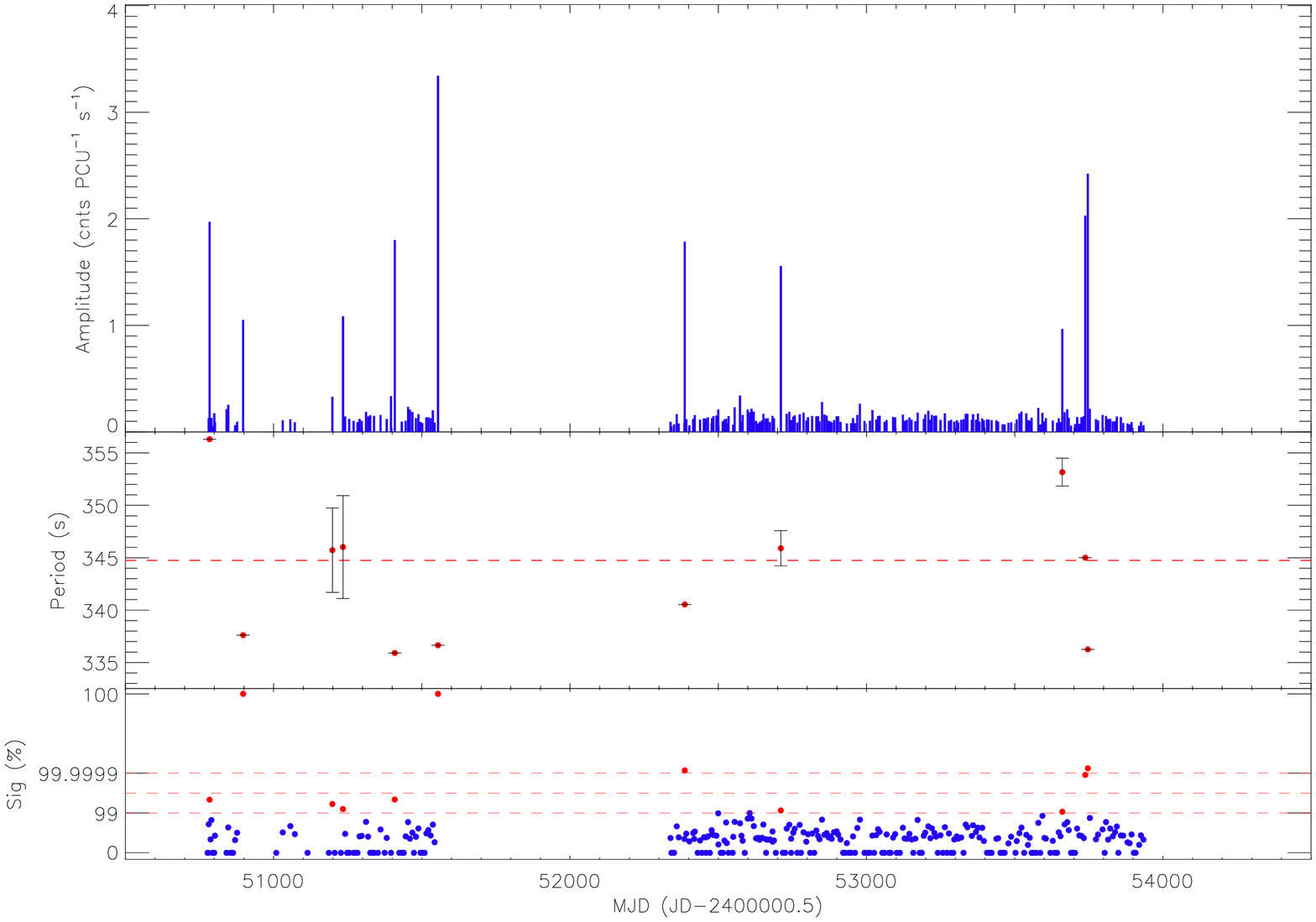}
   \caption{SXP348, \xray\ amplitude light curve.}
   \label{fig_sxp348}
\end{figure}

\subsection{SXP452}

\textbf{RX J0101.3\minus7211 \\
RA 01 01 19.5, dec \minus72 11 22}

\his Was initially proposed as an \xray\ binary by \citet{haberl2000b}. Pulsations were detected in \xmm\ observations during 2001 at 455\pmt2\s\ and in 1993 \rosat\ data at 450\sd452\s\ \citep{sasaki2001}, implying a \pdoteq{2.3}{-8}; the optical counterpart was identified as a Be star \citet{sasaki2001}. \citet{schmidtke2004} propose an orbital period of 74.7\dy\ for this system based on its optical variability.

\sur With only 5 detections throughout the survey (\fref{fig_sxp452}), this source's periodogram shows no periodicities and there is no evidence for the reported 74.7\dy\ optical period.

\begin{figure}
   \centering
        \includegraphics[angle=90,width=0.95\linewidth]{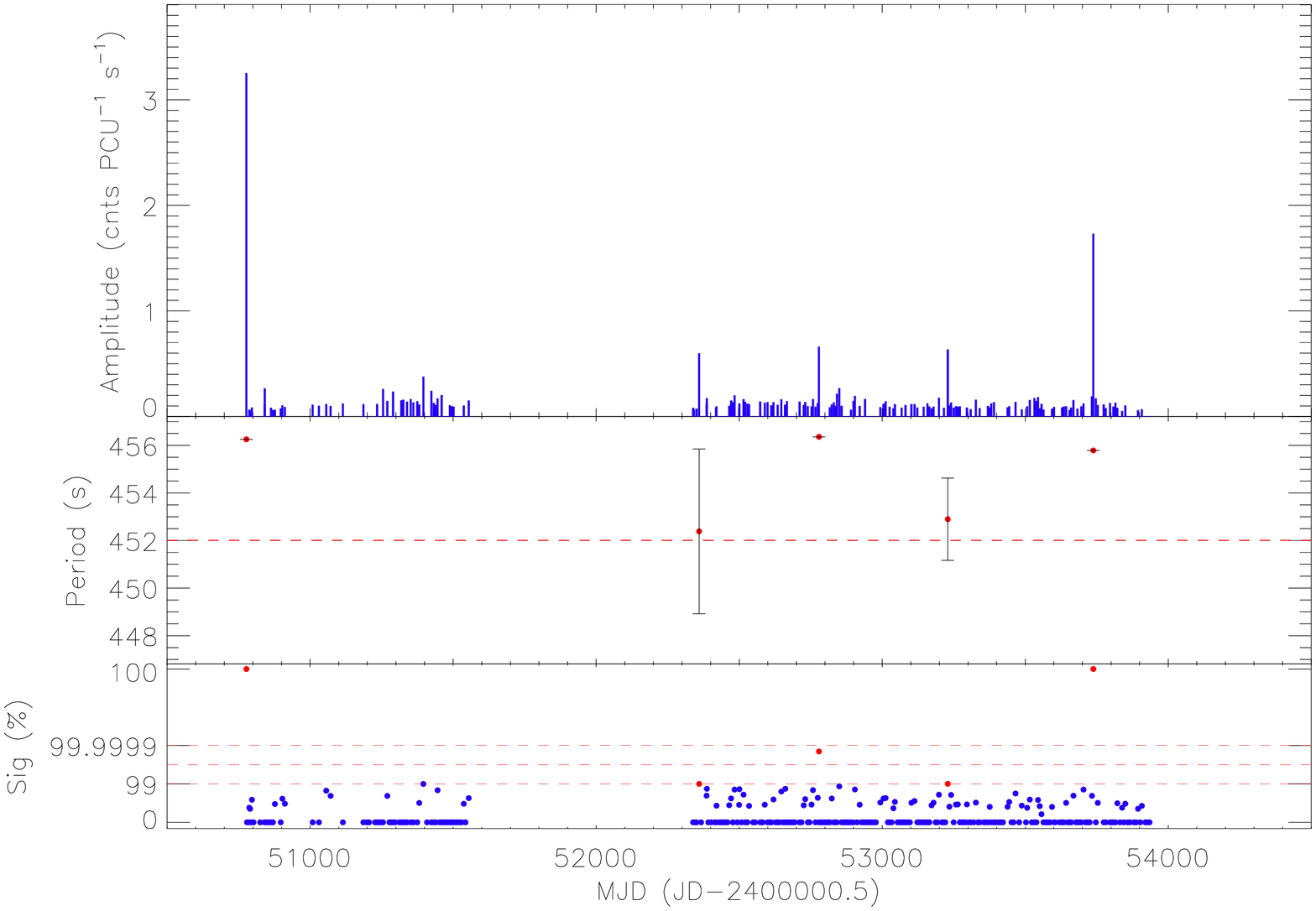}
   \caption{SXP452, \xray\ amplitude light curve.}
   \label{fig_sxp452}
\end{figure}

\subsection{SXP504} \label{sec_sxp504}

\textbf{RX J0054.9\minus7245, AX J0054.87244, CXOU J005455.6\minus724510 \\
RA 00 54 55.6, dec \minus72 45 10}

\his Discovered in archival Chandra data with 503.5\pmt6.7\s\ pulsations \citep{edge2004atelb,edge2004}. Also reported by \citet{haberl2004atel} from an \xmm\ observation at 499.2\pmt0.7\s\ and \lumxeq{4.3}{35}. An orbital period of 268.0\pmt1.4\dy\ was determined from optical (\ogle) data, and corroborated by preliminary \xray\ data from the present survey \citep{edge2005atelb}. The optical ephemeris was later refined to MJD 50556\pmt3 + n\x268.0\pmt0.6\dy\ \citep{edge2005}. \citet{schmidtkecowley2005} found a period of 273\dy\ in \ogle\ data.

\sur The light curve for this system can be seen in \fref{fig_sxp504}(a). Lomb-Scargle analysis of the entire survey (with or without the bright outbursts circa MJD 52440 and 52980) returns a slightly shorter orbital period to the optical one previously reported, but is consistent within errors. Furthermore, \citet{edge2005} find that the epochs of maximum \xray\ flux are coincident with the optical outbursts. The ephemeris we find is MJD 52167.4\pmt8.0 + n\x265.3\pmt2.9\dy. This source displays a lot of activity in between periastron passages, which might be indicative of a low eccentricity orbit and makes timing analysis more difficult.

\begin{figure}
   \centering
        \includegraphics[angle=90,width=0.95\linewidth]{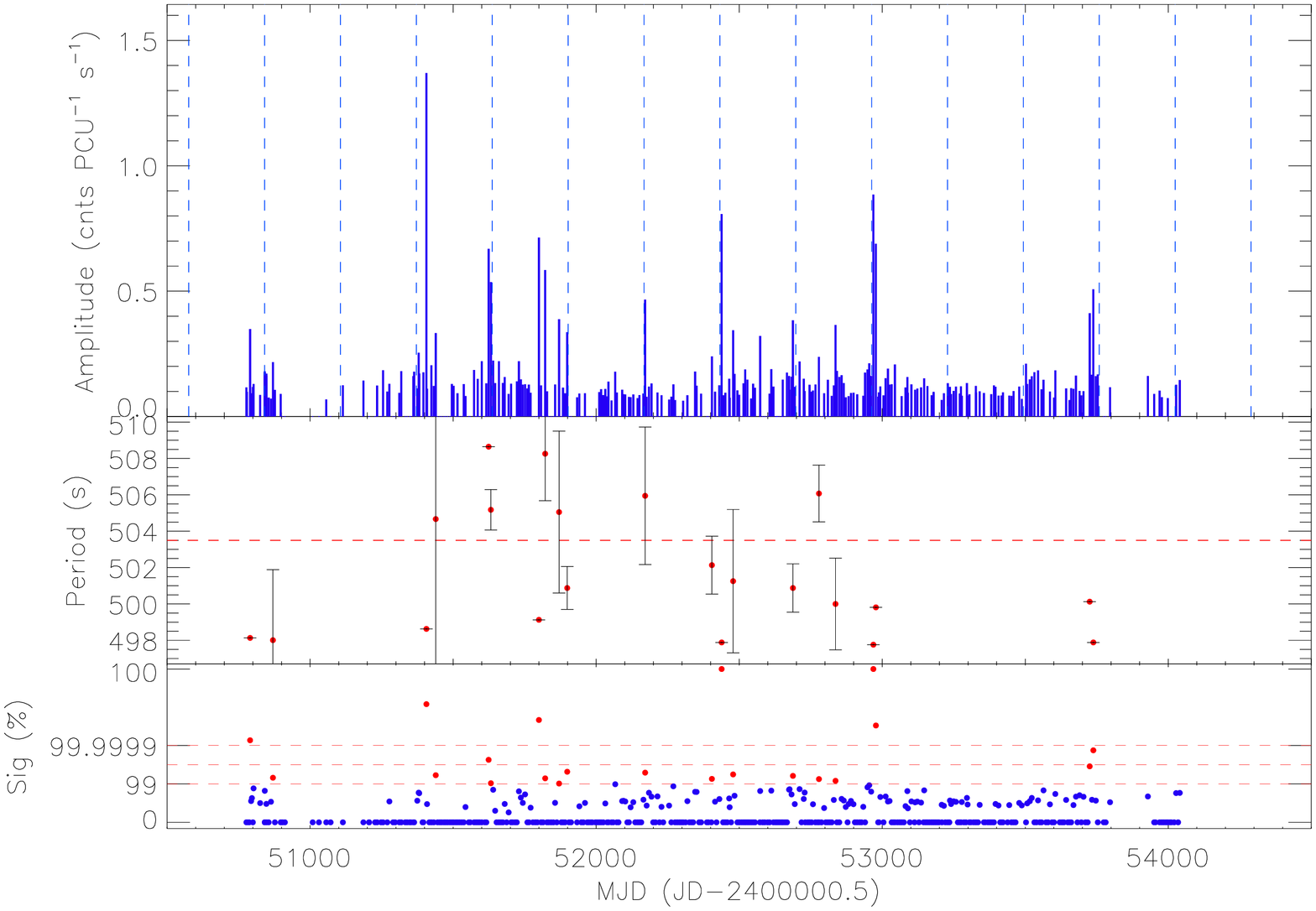}
        \includegraphics[angle=90,width=0.95\linewidth]{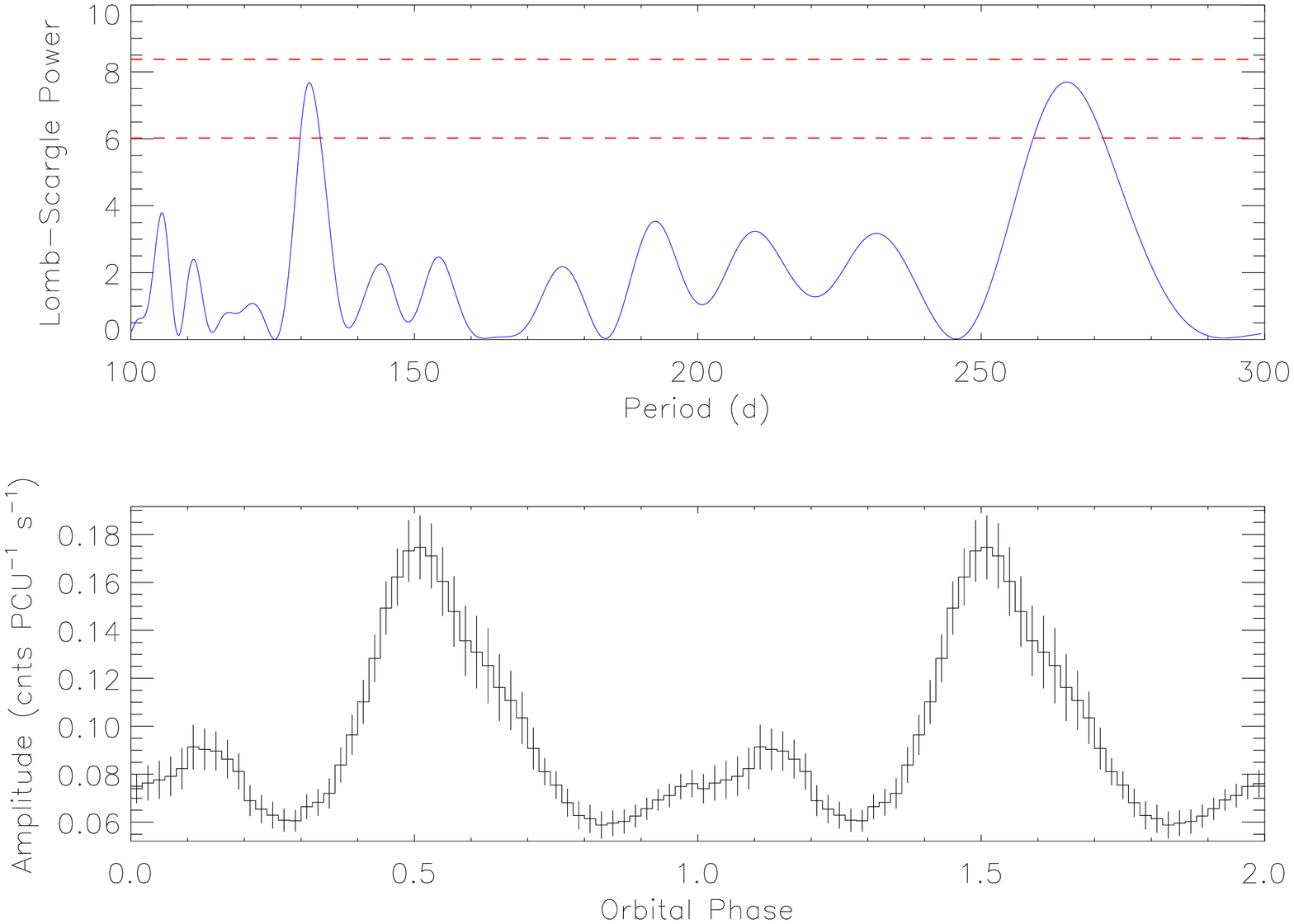}
   \caption{SXP504. a) \textit{Top}: \xray\ amplitude light curve. b) \textit{Middle}: Lomb-Scargle power spectrum; \textit{bottom}: light curve folded at 265.3\dy.}
   \label{fig_sxp504}
\end{figure}

\subsection{SXP565}

\textbf{CXOU J005736.2\minus721934 \\
RA 00 57 36.2, dec \minus72 19 34}

\his Discovered in \chandra\ observations at 564.81\pmt0.41\s\ with a pulse fraction of 48\pmt5\% (\lumxeq{5.6}{34}), the optical counterpart is idenified as the emission line star [MA93] 1020 \citep{macomb2003}. An optical period of 95.3\dy\ has been reported for this system \citep{schmidtke2004}, but this period is not seen in \ogle\ data \citep{williamthesis}.

\sur This source shows a lot of variability, but no clear outbursts (see \fref{fig_sxp565}(a)). The power spectrum returns two significant peaks, with only the higher one providing a convincing orbital profile. The ephemeris for this period is MJD 52219.0\pmt13.7 + n\x151.8\pmt1.0\dy. We note there is no power at the reported optical period nor does the optical ephemeris agree with the brightest \xray\ detections of this source. In view of this, it is possible that the optical counterpart has been misidentified; however, there is no other bright candidate star within 10 arcsec of the very precise \chandra\ position. The folded light curve has a unique shape, showing significant activity around periastron. The asymmetry of the periastron peak is likely due to the presence of an accretion disk around the neutron star that is not completely consumed by the time it reaches apastron, where its lower orbital velocity allows it to ``top up'' its disk from the Be star's wind.

\begin{figure}
   \centering
        \includegraphics[angle=90,width=0.95\linewidth]{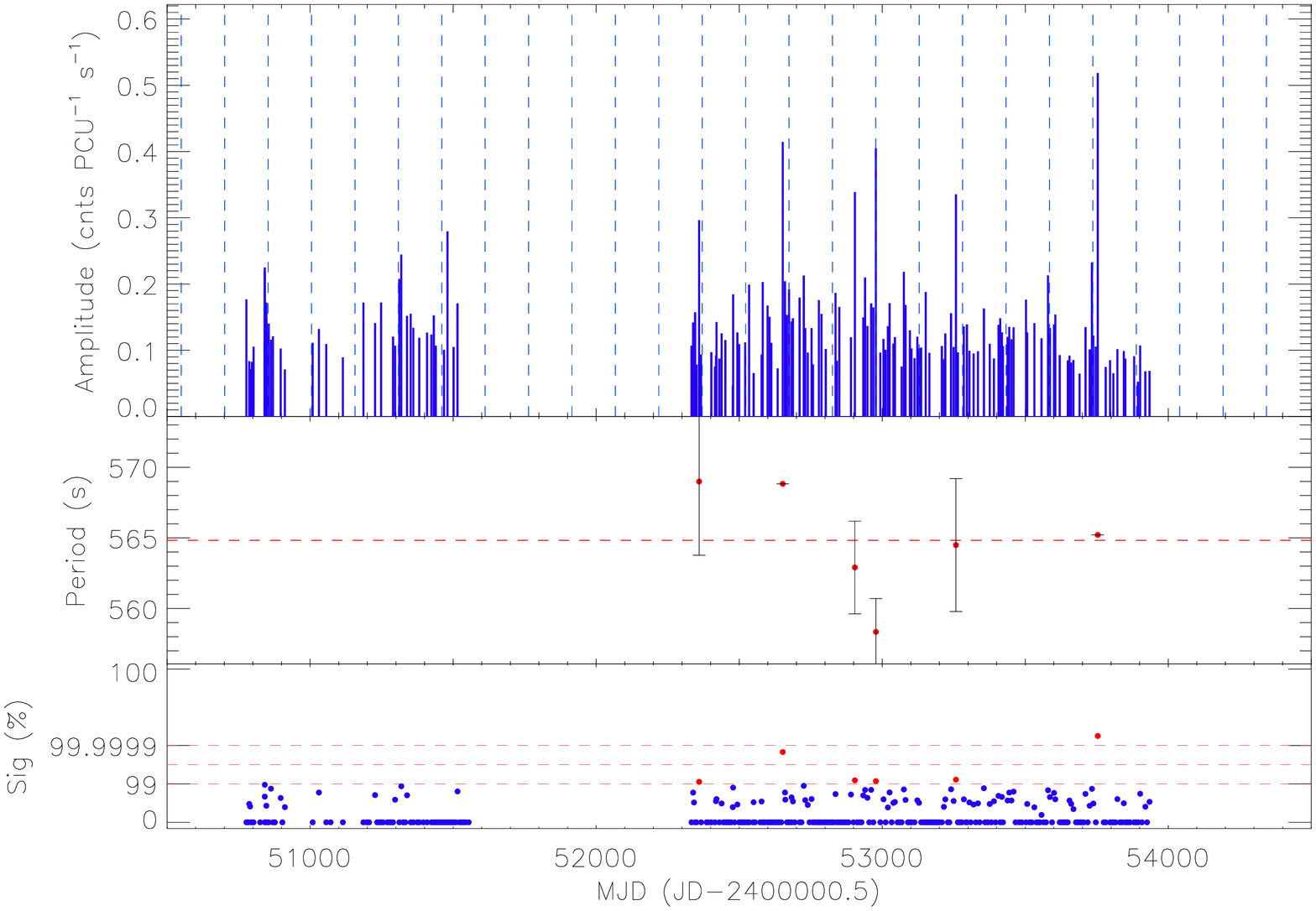}
        \includegraphics[angle=90,width=0.95\linewidth]{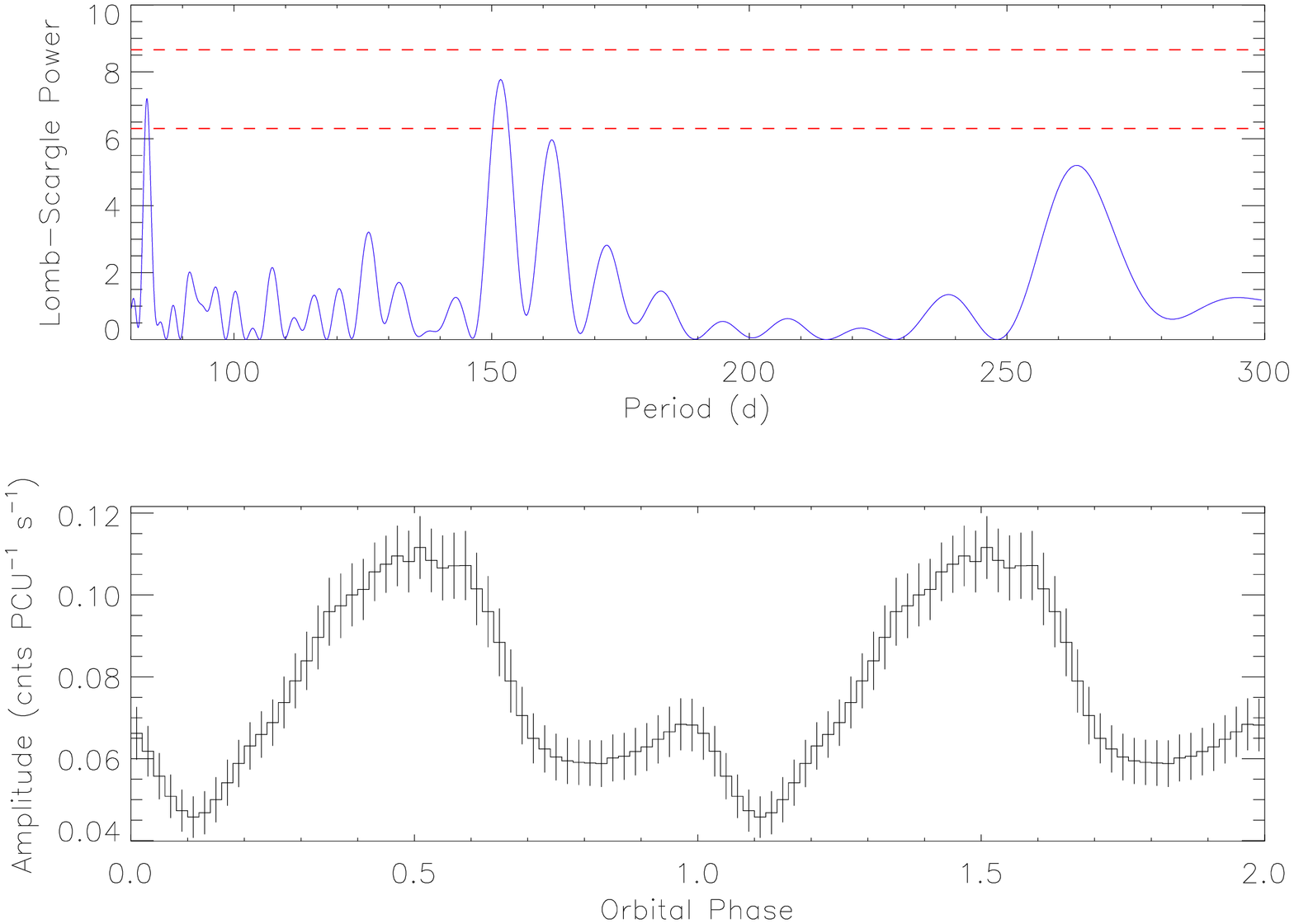}
   \caption{SXP565. a) \textit{Top}: \xray\ amplitude light curve. b) \textit{Middle}: Lomb-Scargle power spectrum; \textit{bottom}: light curve folded at 151.8\dy.}
   \label{fig_sxp565}
\end{figure}

\subsection{SXP700}

\textbf{CXOU J010206.6-714115 \\
RA 01 02 06.69, dec \minus71 41 15.8}

\his Discovered as part of the aforementioned \chandra\ survey of the \smc\ wing reported in \citet{mcgowan2007}. They detected a source with \lumxeq{6.0}{35} (35\pmt5\% pulse fraction) with pulsations at 700.54\pmt34.53\s\ and found its position to coincide with the emission line star [MA93] 1301, a $V = 14.6$ O9 star. The \ogle\ data available for this object show a periodic modulation at 267.38\pmt15.10\dy\ \citep{mcgowan2007}.

\sur Due to its proximity in period to SXP701, limitations of the current analysis script (which cannot work on two pulsars of such similar spin periods) have not allowed the extraction of an \xray\ light curve for this pulsar. However, it has not contaminated the data presented for SXP701 because, even though they are \aprox66\sd arcmin away from each other, they were never within the same field of view.

\subsection{SXP701}

\textbf{RX J0055.2\minus7238, XMMU J005517.9\minus723853 \\
RA 00 55 17.9, dec \minus72 38 53}

\his First observed during an \xmm\ TOO observation at 701.6\pmt1.4\s\ and located within the error circle of a \rosat\ object \citep{haberl2004atel}. \citet{fabrycky2005} finds optical periods of 6.832 and 15.587\hr, which are attributed to stellar pulsations. A weak 412\dy\ period has been seen in \macho\ data \citep{schmidtkecowley2005}.

\sur Similar to SXP565, it displays much \xray\ variability with no bright outbursts (see \fref{fig_sxp701}). Unfortunately, timing analysis provides no clear periodicities.

\begin{figure}
   \centering
        \includegraphics[angle=90,width=0.95\linewidth]{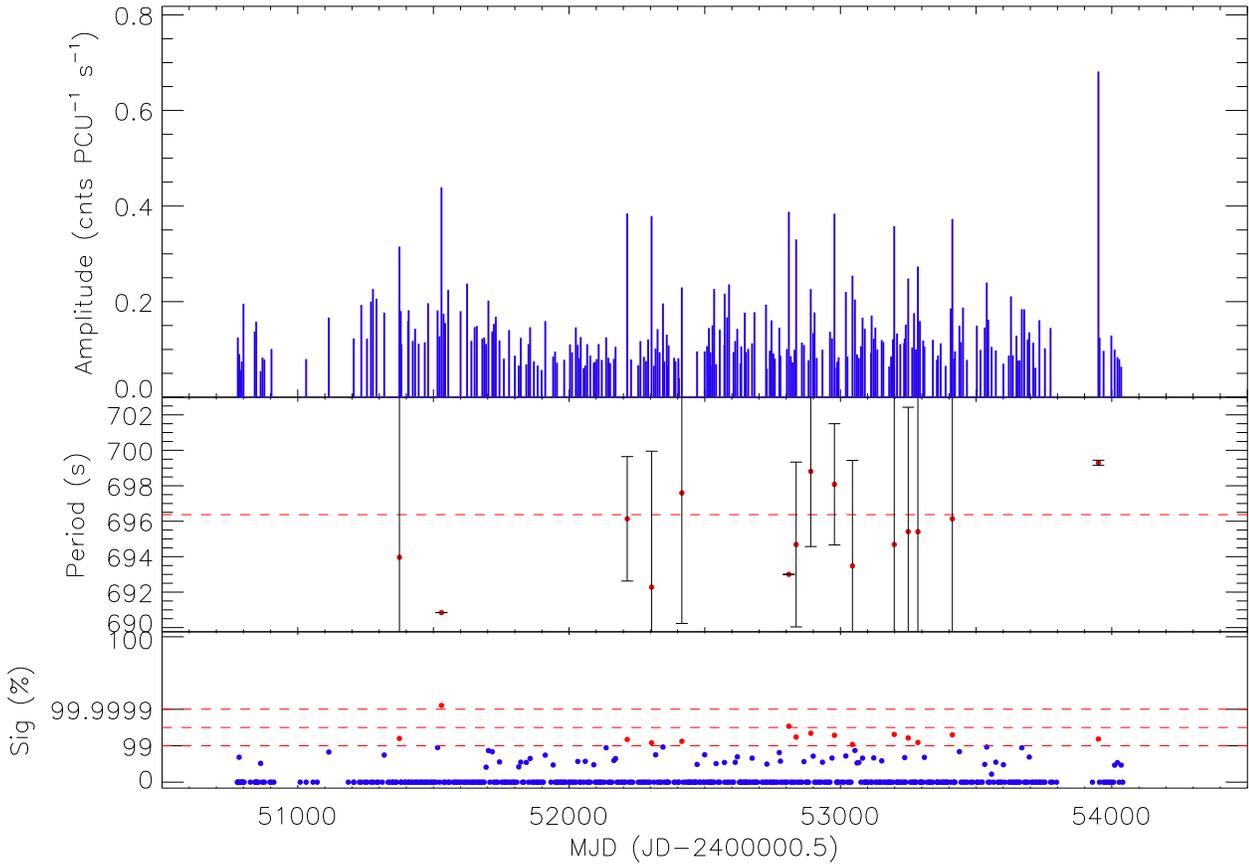}
   \caption{SXP701, \xray\ amplitude light curve.}
   \label{fig_sxp701}
\end{figure}

\subsection{SXP756}

\textbf{RX J0049.6\minus7323, AX J0049.4\minus7323 \\
RA 00 49 42.42, dec \minus73 23 15.9}

\his Pulsations at 755.5\pmt0.6\s\ were detected in a very long (\aprox177\ks) \asca\ observation of the SMC in April 2000, the source was detected at a luminosity of \lumxeq{5}{35} \citep{yokogawa2000d}. The optical counterpart was later established as a V = 15 Be star \citep{edge2003}. \citet{silasthesis} and \citet{laycock2005} find an \xray\ period of 396\pmt5\dy\, while \citet{cowley2003,schmidtke2004} report recurring outbursts at \aprox394\dy\ intervals in \textit{R} and \textit{V} \macho\ data.

\sur Coverage of this pulsar has been sparse due to its southern location in the \smc, only visible to observations of Positions 4 and 5. Despite this, it has been detected various times in outburst, the brightest of which was a 3 week outburst (\fref{fig_sxp756}(a)). We decided to remove the brightest point from this outburst so as not to skew the Lomb-Scargle power spectrum. A strong period is found at 194.9\dy\, which we believe to be the first harmonic of the orbital period. Looking at the light curve it is evident that the bright detections are spaced \aprox400\dy\ apart\footnote{Also note that only one peak appears in the orbital profile in \fref{fig_sxp756}(b). If 194.9\dy\ were the actual orbital period, the plot would show \textit{two} peaks per orbital cycle.}. Moreover, this would be in agreement with the optical period of the system, so we use twice the value of the harmonic as the orbital period and derive the ephemeris MJD 52196.1\pmt3.9 + n\x389.9\pmt7.0\dy. This ephemeris places the first outburst in the data at MJD 51416, which is consistent with the last optical outburst available in the \macho\ data \citep{cowley2003}.

\begin{figure}
   \centering
        \includegraphics[angle=90,width=0.95\linewidth]{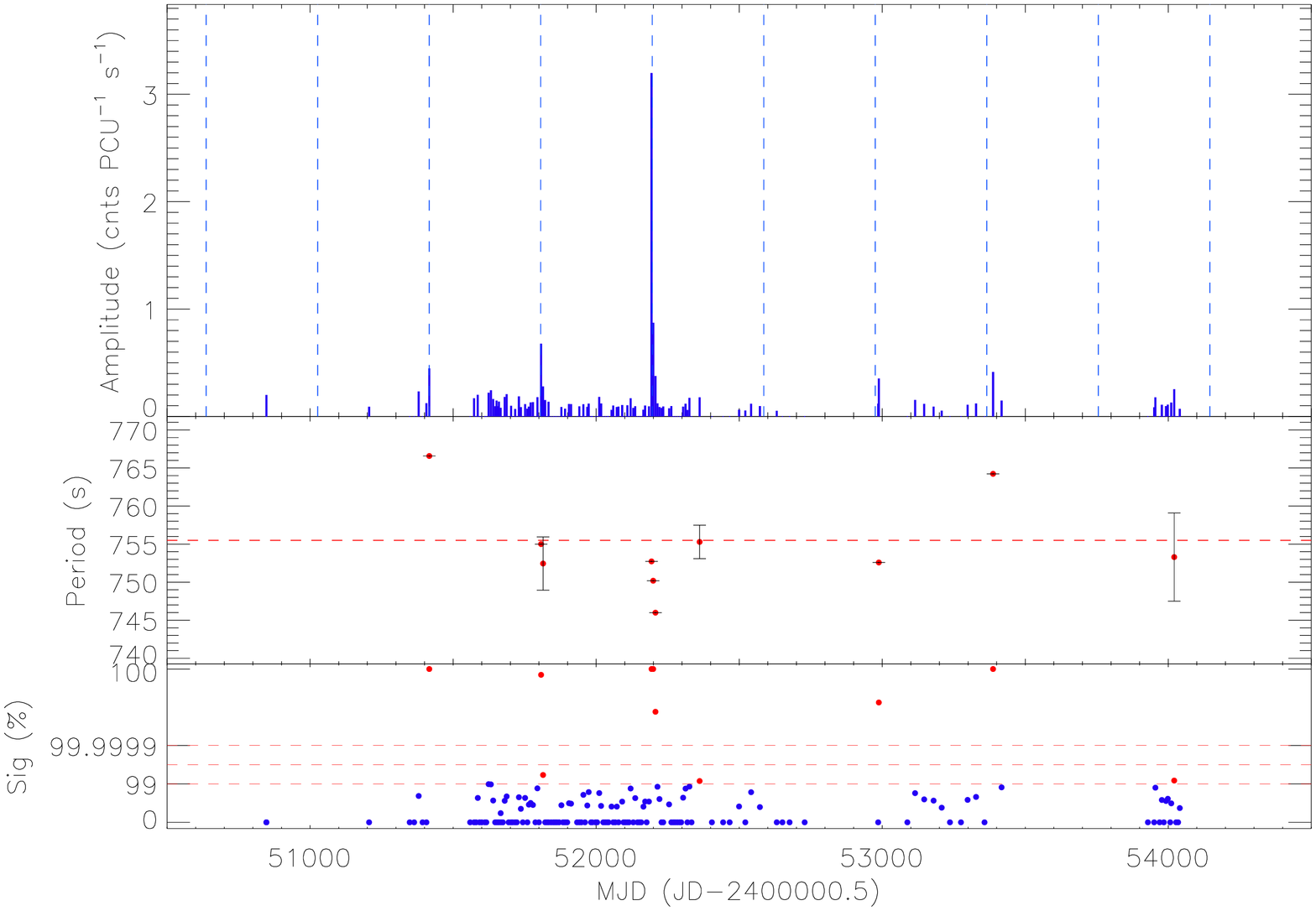}
        \includegraphics[angle=90,width=0.95\linewidth]{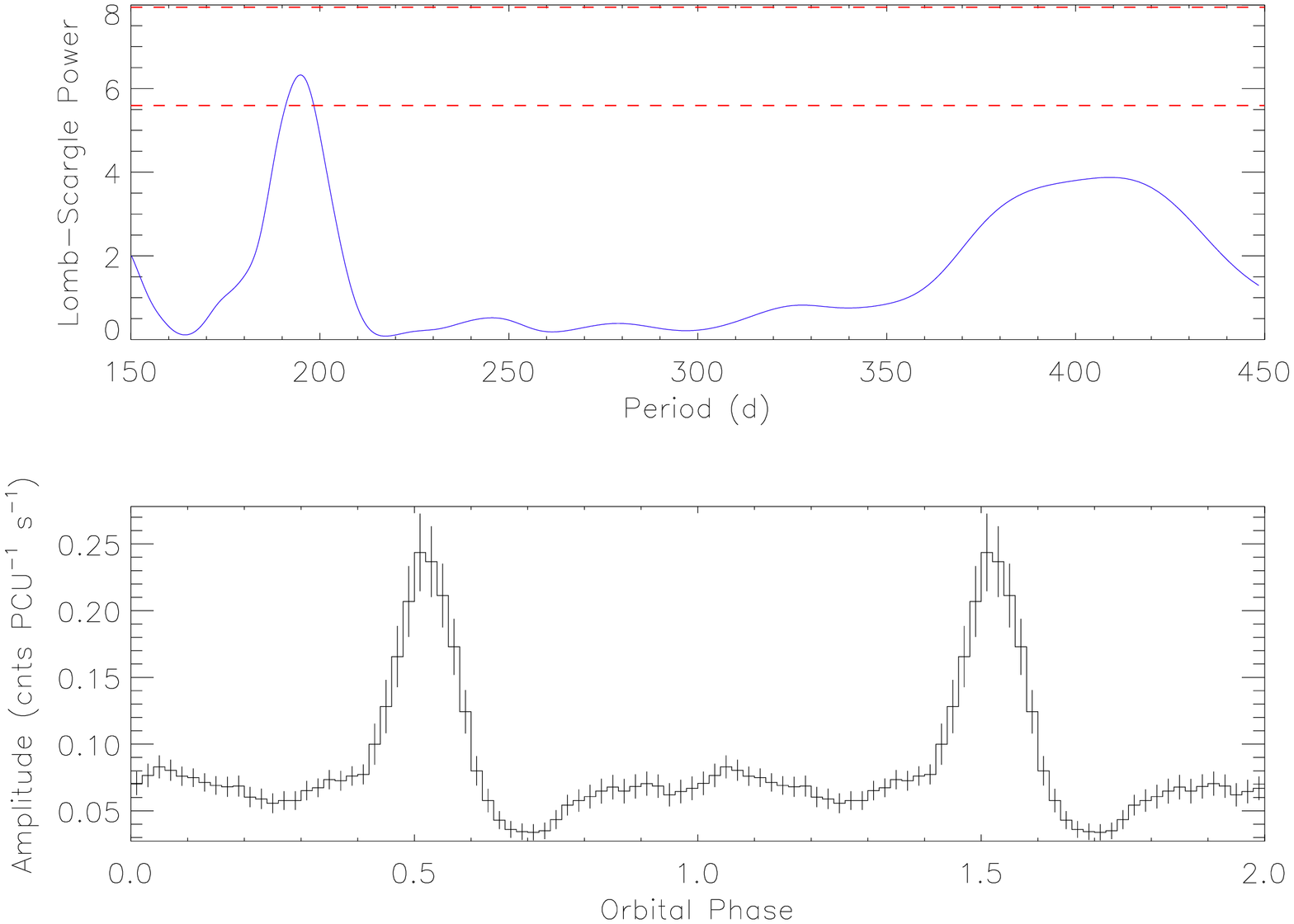}
   \caption{SXP756. a) \textit{Top}: \xray\ amplitude light curve. b) \textit{Middle}: Lomb-Scargle power spectrum; \textit{bottom}: light curve folded at 389.9\dy.}
   \label{fig_sxp756}
\end{figure}

\subsection{SXP1323}

\textbf{RX J0103.6\minus7201 \\
RA 01 03 37.5, dec \minus72 01 33.2}

\his This is the source with the longest known pulse period in the \smc, SXP1323 was reported by \citet{haberlpietsch2005} in a number of archival \xmm\ observations. The authors identify the emission line star [MA93] 1393 (V $\simeq$ 14.6) as the optical counterpart. \citet{schmidtkecowley2006} report three strong periods for this source from \ogle~II data: 0.41\dy, 0.88\dy\ and 26.16\dy; they attribute the first two to non-radial pulsations of the Be star but suggest the latter might be the orbital period.

\sur This source is difficult to detect due to its long period (requiring observations with a baseline longer than \aprox4\ks) and its location near the edge of Position 1/A. Despite these limitations, a number of bright outbursts have been detected but no periodicities can be found in the sparse data (see \fref{fig_sxp1323}).

\begin{figure}
  \centering
       \includegraphics[angle=90,width=0.95\linewidth]{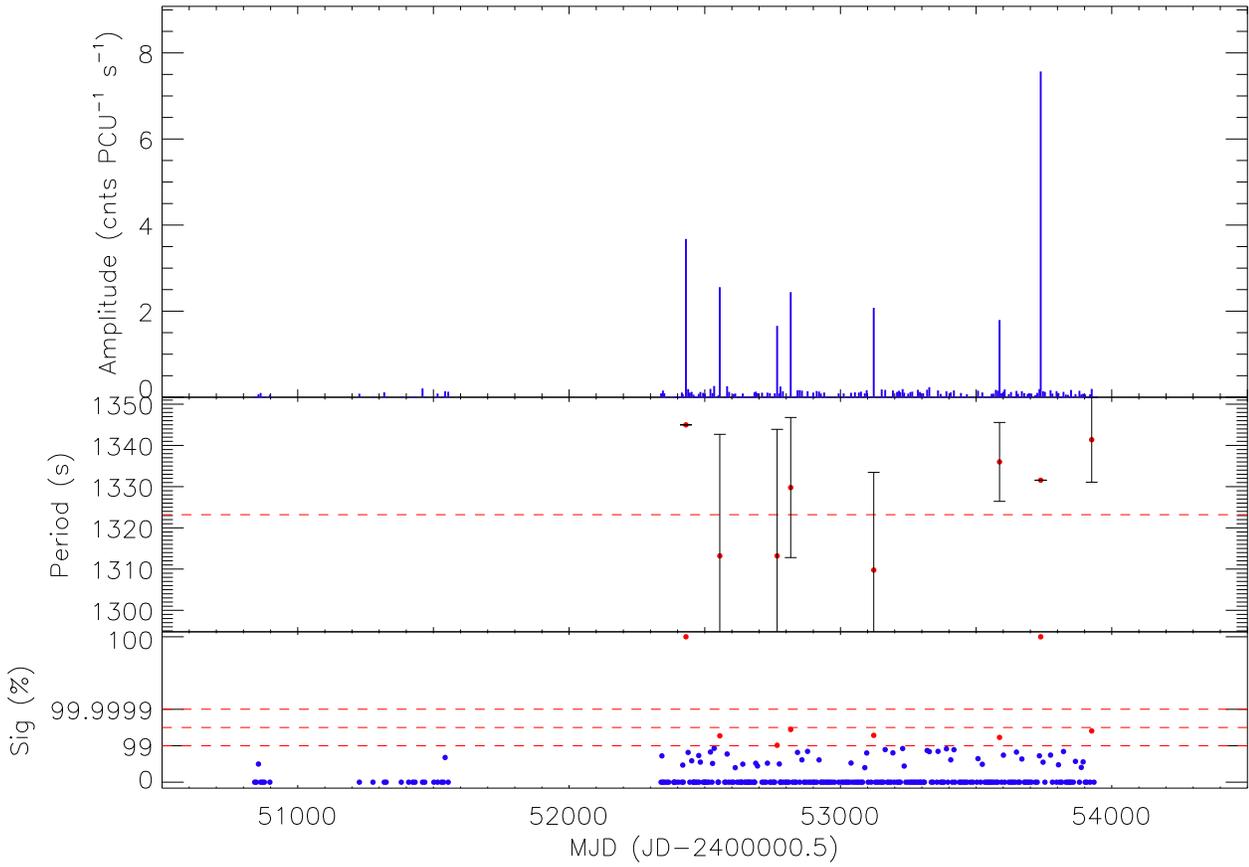}
  \caption{SXP1323, \xray\ amplitude light curve.}
  \label{fig_sxp1323}
\end{figure}

\section{Discussion}

\tref{table_smc_pulsars} presents a summary of the timing analysis results for the \smc\ pulsars, providing orbital periods and ephemerides where available, together with periods observed at other wavelengths. In cases when there is only weak evidence for an orbital period in the \xray\ data, the number is given in parentheses.

Those systems with unequivocal orbital periods have allowed us to study and compare a selection of well-defined orbital profiles. In particular, we consider the profiles of the following systems that have shown regular activity at, or around, periastron pasage: SXP6.85, SXP7.78, SXP46.6, SXP59.0, SXP82.4, SXP144, SXP169 and SXP756. It is notable that their profiles all share a similar shape, even though their orbital periods cover a wide range (from \aprox45\sd390\dy): they are relatively narrow (except for SXP756, all widths are between 0.40 and 0.55 wide in orbital phase), and they are very symmetrical in shape (suggesting accretion takes place fluidly and increases smoothly as the neutron star approaches the deccretion disk, to decrease in an equally smooth manner after periastron). Other systems display different characteristics, some show sharp rises with long declines (e.g.,SXP323, \fref{fig_sxp323}(b)), the opposite (e.g., SXP16.6, \fref{fig_sxp16.6}(b)), or even long, plateau-like profiles (e.g.,SXP280, \fref{fig_sxp280}(b)).

A number of profiles show a small increase in flux half an orbital phase away from the maximum (e.g., SXP7.78, SXP89.0, SXP144, SXP504 and SXP565), suggesting that accretion is taking place on a smaller scale, possibly from the Be star's wind, when the neutron star is travelling slowly enough to accrete from it. Given the range of orbit sizes in the systems that exhibit this behaviour (judging from the orbital periods), it may be the eccentricity of the orbit that determines whether apastron accretion takes place (because the more eccentric the orbit, the slower the neutron star will be travelling at apastron).

Throughout the length of the survey there have been some pulsars whose spin periods have undergone slow, persistent evolution, e.g., SXP7.78 (spin down), SXP46.6 (spin up), SXP59.0 (spin up) or SXP144 (spin down), while others have changed their pulse period over the course of a few weeks, e.g., SXP15.3 (spin up). And still, some other systems have not changed their spin periods significantly despite bright and/or persistent activity, e.g., SXP2.37, SXP4.78 or SXP8.88. This suggests there are a number of pulsars that must be rotating close to, or at, their equilibrium spin periods, as suggested by \cite{corbet84}. A lack of modulation in the spin period throughout a long outburst (as in SXP18.3's outburst circa MJD 54000) further reveals systems that are likely being observed at a low inclination (face on). Unfortunately, despite the number of orbital periods known, it has still not been possible to calculate the complete set of orbital parameters of any \smc\ system (with the exception of SXP0.92 \citep{kaspi93}), so orbital inclination angles and their effects on pulse arrival times remain a matter of speculation.

An interesting case is presented by SXP348; \citet{israel2000} suggest this is a persistent, low luminosity system, and recent Chandra observations \citep{mcgowan2007} tend to support this notion after detecting the pulsar at its shortest recorded pulse period until then, 337.5\s. Given that it was observed in 1996 at 349\s\ in \asca\ data, we can deduce that it has spun down 11.5\s\ in \aprox10 years, implying a \pdoteq{3.6}{-8}. An estimate for the accretion necessary to produce the required torque on this timescale implies a time-averaged \xray\ luminosity of just \lumxge{3.3}{35}, which would explain why it was rarely detected during the present survey. However, \rxte\ detected it a number of times at longer pulse periods than would have been expected; a clear occasion was during the short outburst circa MJD 53740, when the system was detected at a high significance on 2 consecutive weeks with periods of 345.0\s\ and 336.2\s, respectively. If this spin-up were all intrinsic, the expected luminosity would be \lumxge{1.1}{38}, which is not inconceivable for a Type~II outburst. \citet{mcgowan2007} detected it 30\dy\ later (MJD 53777) at \lumxeq{4.2}{37}, but \rxte\ did not detect it 2 days prior to that (MJD 53774). So what is happening in this system? It is not clear whether we are detecting orbitally-induced Doppler-shifted pulse periods or if SXP348 is actually undergoing massive spin up every time it outbursts. Given the pulse periods of the detections in \fref{fig_sxp348}, it is unclear whether the pulsar is, in fact, spinning down (as proposed by \citet{israel2000}); it could very well be spinning \textit{up}.

Another case involving changes in spin is that of SXP59.0 (\fref{fig_sxp59.0}(a)). It went undetected by \rxte\ during MJD 51550\sd52500, during which time its pulse period increased \aprox0.45\s. If this spin-down were a product of reverse accretion torques, we would have expected to see outbursts at \lumxge{2.6}{36}, comparable in strength with those detected during the survey. Although SXP59.0 was not close to the center of the field of view during the spin-down period (as it was for most of the rest of the survey), it should still have been detectable. This suggests that SXP59.0 was spun down through mechanisms other than reverse accretion torques, possibly through the propeller effect.

Yet another example of strange spin behaviour is that of SXP144. Given its pulse period, and based on the Corbet diagram, its orbital period should be \aprox90\dy; so with its clear 59.4\dy\ orbital period, the expectation is that it would spin up during outburst. However, quite the opposite is what \rxte\ has detected. This system went into outburst at least once every orbit over a period of \aprox1160\dy, during which time it spun \textit{down} \aprox1.7\s. It would seem that factors other than just the orbital and spin periods govern the behaviour and evolution of these binary systems, and they need to be taken into account. Orbital eccentricity and/or the magnetic field strength of the pulsar are likely to be important factors in determining the relationship between the spin and orbital periods.

\section{Conclusion}

We have presented the most comprehensive colection of pulsed light curves for 41 probable \bex\ systems in the \smc\ and have determined an \xray\ (likely orbital) ephemeris for 21 of them while presenting possible periods for 6 others. 10 of these ephemerides are from new \xray\ periods, while 6 others are improvements of previously known ephemerides. Of the systems exhibiting periodicities in the \xray, 19 also show optical periodicities, although it is noteworthy that the optical and \xray\ periods only agree on 7 occasions. Of the remaining 12, 3 show an optical period that is half the \xray\ period while 1 exhibits an optical modulation that is twice the \xray\ period. However, there are still 11 systems with a known optical period that show no modulation in \xrays. The systems exhibiting a large discrepancy between the optical and \xray\ periods suggest the possibility that the optical counterpart to the \xray\ system has been missidentified, or that the interaction between the Be star's disk and the neutron star is more complicated than originally thought, although optical periods can be affected by pulsations of the primary star, and \xray\ periods by Type~II outbursts. Nonetheless, for extragalactic \xray\ binaries, long-term monitoring has proven to be the most successful method of study.


\begin{deluxetable}{lcccccc}   
\tabletypesize{\scriptsize}
\tablecolumns{7}
\tablewidth{0pt}
\tablecaption{\xray\ binary systems in the \smc \label{table_smc_pulsars} }
\tablehead{ \colhead{Object} & \colhead{\ps\ (s)} & \colhead{RA} & \colhead{dec} & \colhead{\xray\ \tp\ (MJD)} & \colhead{\xray\ \porb\ (days)} & \colhead{Other \porb\ (days)}}
\startdata
  SXP0.09  &  0.087 &  00  42  35.0   &  -73  40  30.0   &         --         &        --       &    --   \\[2pt]
  SXP0.72  &  0.716 &  01  17  05.2   &  -73  26  36.0   &         --         &        --       &    --   \\[2pt]
  SXP0.92  &  0.92  &  00  45  35.0   &  -73  19  02.0   &         --         &        --       &   51\tablenotemark{a}   \\[2pt]
  SXP2.16  &  2.165 &  01  19  00     &  -73  12  27     &         --         &        --       &    --    \\[2pt]
  SXP2.37  &  2.374 &  00  54  34.0   &  -73  41  03.0   &         --         &        --       &    --   \\[2pt]
  SXP2.76  &  2.763 &  00  59  11.7   &  -71  38  48.0   &         --         &        --       &   82.1\tablenotemark{b}   \\[2pt]
  SXP3.34  &  3.34  &  01  05  02.0   &  -72  11  00.0   &         --         &        --       &    --   \\[2pt]
  SXP4.78  &  4.782 &  00  52  06.6   &  -72  20  44     &         --         &      (34.1)     &   23.9\tablenotemark{c}      \\[2pt]
  SXP6.85  &  6.848 &  01  02  53.1   &  -72  44  33     &  52318.5\pmt7.9    &  112.5\pmt0.5   &   24.82\tablenotemark{d}   \\[2pt]
  SXP7.78  &  7.780 &  00  52  06     &  -72  26  06     &  52250.9\pmt1.4    &  44.92\pmt0.06  &   44.86\tablenotemark{e}, 44.6\tablenotemark{f}, 44.8\tablenotemark{g}   \\[2pt]
  SXP8.02  &  8.020 &  01  00  41.8   &  -72  11  36     &         --         &        --       &    --   \\[2pt]
  SXP8.88  &  8.880 &  00  51  52.0   &  -72  31  51.7   &  52392.2\pmt0.9    &  28.47\pmt0.04  &   33.4\tablenotemark{h}  \\[2pt]
  SXP9.13  &  9.130 &  00  49  13.6   &  -73  11  39     &  52380.5\pmt2.3    &  77.2\pmt0.3    &   40.17\tablenotemark{g}, 91.5\tablenotemark{i}    \\[2pt]
  SXP15.3  &  15.30 &  00  52  14     &  -73  19  19     &         --         &       (28)      &   75.1\tablenotemark{j}    \\[2pt]
  SXP16.6  &  16.52 &       --        &        --        &  52373.5\pmt1.0    &  33.72\pmt0.05  &    --   \\[2pt]
  SXP18.3  &  18.37 &  00  50         &  -72  42         &  52275.6\pmt0.9    &  17.73\pmt0.01  &   34.6\tablenotemark{k}    \\[2pt]
  SXP22.1  &  22.07 &  01  17  40.5   &  -73  30  52.0   &         --         &        --       &    --   \\[2pt]
  SXP31.0  &  31.01 &  01  11  09.0   &  -73  16  46.0   &         --         &        --       &   90.4\tablenotemark{b}  \\[2pt]
  SXP34.1  &  34.08 &  00  55  26.9   &  -72  10  59.9   &         --         &        --       &    --   \\[2pt]
  SXP46.6  &  46.59 &  00  53  58.5   &  -72  26  35.0   &  52293.9\pmt1.4    & 137.36\pmt0.35  &   69.2\tablenotemark{l}, 138.4\tablenotemark{l}   \\[2pt]
  SXP51.0  &  51.00 &       --        &        --        &         --         &        --       &    --   \\[2pt]
  SXP59.0  &  58.86 &  00  54  56.6   &  -72  26  50     &  52306.1\pmt3.7    & 122.10\pmt0.38  &   60.2\tablenotemark{m}  \\[2pt]
  SXP65.8  &  65.78 &  01  07  12.63  &  -72  35  33.8   &         --         &        --       &  110.6\tablenotemark{n}   \\[2pt]
  SXP74.7  &  74.70 &  00  49  04     &  -72  50  54     &  52319.0\pmt3.1    &  61.6\pmt0.2    &   33.4\tablenotemark{o}    \\[2pt]
  SXP82.4  &  82.40 &  00  52  09     &  -72  38  03     &  52089.0\pmt3.6    & 362.3\pmt4.1    &   \aprox380\tablenotemark{p}    \\[2pt]
  SXP89.0  &  89.00 &       --        &        --        &  52337.5\pmt6.1    &  87.6\pmt0.3    &    --      \\[2pt]
  SXP91.1  &  91.10 &  00  50  55     &  -72  13  38     &  52197.9\pmt8.2    & 117.8\pmt0.5    &   88.25\tablenotemark{q}   \\[2pt]
  SXP95.2  &  95.20 &  00  52  00     &  -72  45  00     &         --         &      (71.3)     &    --      \\[2pt]
  SXP101  &  101.40 &  00  57  26.8   &  -73  25  02     &         --         &      (25.2)     &   21.9\tablenotemark{r}   \\[2pt]
  SXP138  &  138.00 &  00  53  23.8   &  -72  27  15.0   &  52400.5\pmt5.2    & 103.6\pmt0.5    &  \aprox125\tablenotemark{s}, 122--123\tablenotemark{t}    \\[2pt]
  SXP140  &  140.99 &  00  56  05.7   &  -72  22  00.0   &         --         &        --       &  197\tablenotemark{t}      \\[2pt]
  SXP144  &  144.00 &       --        &        --        &  52368.9\pmt1.8    &  59.38\pmt0.09  &    --   \\[2pt]
  SXP152  &  152.10 &  00  57  49     &  -72  07  59     &         --         &        --       &    --   \\[2pt]
  SXP169  &  167.35 &  00  52  54.0   &  -71  58  08.0   &  52240.1\pmt2.1    &  68.54\pmt0.15  &   67.6\tablenotemark{u}   \\[2pt]
  SXP172  &  172.40 &  00  51  52     &  -73  10  35     &         --         &       (70)      &   69.9\tablenotemark{v}  \\[2pt]
  SXP202  &  202.00 &  00  59  20.8   &  -72  23  17     &         --         &       (91)      &    --   \\[2pt]
  SXP264  &  263.60 &  00  47  23.7   &  -73  12  25     &         --         &        --       &   49.2\tablenotemark{g}, 49.1\tablenotemark{w}  \\[2pt]
  SXP280  &  280.40 &  00  57  48.2   &  -72  02  40     &    52312\pmt6      &  64.8\pmt0.2    &   127.3\tablenotemark{u}    \\[2pt]
  SXP293  &  293.00 &       --        &        --        &  52327.3\pmt4.5    & 151\pmt1        &    --   \\[2pt]
  SXP304  &  304.49 &  01  01  01.7   &  -72  07  02     &         --         &        --       &   520\tablenotemark{t}   \\[2pt]
  SXP323  &  321.20 &  00  50  44.8   &  -73  16  06.0   &  52336.9\pmt3.5    & 116.6\pmt0.6    &    --   \\[2pt]
  SXP348  &  344.75 &  01  03  13.0   &  -72  09  18.0   &         --         &        --       &   93.9\tablenotemark{v} \\[2pt]
  SXP452  &  452.01 &  01  01  19.5   &  -72  11  22     &         --         &        --       &   74.7\tablenotemark{x}   \\[2pt]
  SXP504  &  503.50 &  00  54  55.6   &  -72  45  10.0   &  52167.4\pmt8.0    & 265.3\pmt2.9    &  268.0\tablenotemark{y}, 273\tablenotemark{w}   \\[2pt]

  SXP565  &  564.83 &  00  57  36.2   &  -72  19  34.0   &  52219.0\pmt13.7   & 151.8\pmt1.0    &   95.3\tablenotemark{q}   \\[2pt]
  SXP700  &  700.5  &  01  02  06.69  &  -71  41  15.8   &         --         &        --       &  267\tablenotemark{z}  \\[2pt]
  SXP701  &  696.37 &  00  55  17.9   &  -72  38  53.0   &         --         &        --       &  412\tablenotemark{{aa}}   \\[2pt]

  SXP756  &  755.50 &  00  49  42.42  &  -73  23  15.9   &  52196.1\pmt3.9    & 389.9\pmt7.0    &  394\tablenotemark{{ab}}   \\[2pt]
  SXP1323 & 1323.20 &  01  03  37.5   &  -72  01  33.2   &         --         &        --       &   26.16\tablenotemark{v}   \\[2pt]
\enddata
\tablenotetext{a}{From radio observations \citep{kaspi93}.}
\tablenotetext{b}{From \ogle~III data \citep{schmidtke2006}.}
\tablenotetext{c}{From \macho\ data of the possible optical counterpart \citep{coe2005}.}
\tablenotetext{d}{From \ogle\ and \macho\ data \citep{schmidtkecowley2007}.}
\tablenotetext{e}{From \macho\ data \citep{cowley2004}.}
\tablenotetext{f}{From \macho\ data \citep{coe2005}.}
\tablenotetext{g}{From \ogle\ data \citep{williamthesis}.}
\tablenotetext{h}{From \ogle\ and \macho\ data \citep{schmidtkecowley2006}.}
\tablenotetext{i}{From \ogle\ data \citep{schmidtke2004}.}
\tablenotetext{j}{From \ogle\ and \macho\ data \citep{williamthesis}.}
\tablenotetext{k}{From \rxte\ data \citep{corbet2004atelb}.}
\tablenotetext{l}{From \ogle\ data \citep{schmidtke2007}.}
\tablenotetext{m}{From \ogle\ and \macho\ data \citep{schmidtkecowley2005}.}
\tablenotetext{n}{From \macho\ data \citep{schmidtkecowley2007b}.}
\tablenotetext{o}{From \ogle\ data \citep{schmidtkecowley2005,williamthesis}.}
\tablenotetext{p}{From \ogle\ data (\pcom{Edge 2006}).}
\tablenotetext{q}{From \macho\ data \citep{schmidtke2004}.}
\tablenotetext{r}{From \ogle\ and \macho\ data \citep{mcgowan2007,schmidtkecowley2007b}.}
\tablenotetext{s}{From \macho\ data \citep{williamthesis}.}
\tablenotetext{t}{From \macho\ data \citep{schmidtkecowley2006}.}
\tablenotetext{u}{From \ogle\ data \citep{schmidtke2006}.}
\tablenotetext{v}{From \ogle\ data \citep{schmidtkecowley2006}.}
\tablenotetext{w}{From \ogle\ data \citep{schmidtkecowley2005}.}
\tablenotetext{x}{From \ogle\ and \macho\ data \citep{schmidtke2004}.}
\tablenotetext{y}{From \ogle\ and \macho\ data \citep{edge2005}.}
\tablenotetext{z}{From \ogle\ data \citep{mcgowan2007}.}
\tablenotetext{{aa}}{From \macho\ data \citep{schmidtkecowley2005}.}
\tablenotetext{{ab}}{From \macho\ data \citep{cowley2003,schmidtke2004}.}
\end{deluxetable}

We thank A. Cowley for pointing out some glaring omissions in the final manuscript and also the anonymous referee for many useful comments that made this paper clearer.

{\it Facilities:} \facility{RXTE (PCA)}.

\appendix

\section{Determining signal significance} \label{sec_appendix_a}

 \citet{scargle82} proposes the normalisation of the periodogram to the variance, $\sigma_n^2$, of the \textit{signal-free} data, where $\sigma_n$ is the standard deviation; as such, Gaussian noise will have a power of 1\footnote{It is clear that most of the data from observations made by \rxte\ during the survey contain contributions from a number of sources in the field of view, and their variance should not be used in the calculations. However, after analysing a large number of observations, it was found that the average power within the calculated power spectra was essentially 1 (likely due to the low S/N), which justifies our use of the light curve variance (including all the pulsar signals) instead of the variance of the noise, which would have been difficult to obtain.}\label{ref_sigma_noise}. Furthermore, the probability function \textit{Prob} associated with the periodogram will be exponentially distributed\footnote{It will be of the form \textit{Prob} = $e^{-Z}$, which is the probability of detecting a peak in the periodogram above a certain power, \textit{Z}.}, and it can be shown that the probability that a periodic signal with power of \textit{Z} is due to noise is

\begin{equation}
  {\rm{FAP}} = 1 - \bigl(1 - e^{-Z}\bigr)^{M}
\end{equation}

\noindent which is the False Alarm Probability, with \textit{M} being the number of independent frequencies, which we (rather conservatively) define as

\begin{equation}
  M = 2 \times n_{f} \times \Delta f \times \tau
  \label{eq_m}
\end{equation}

\noindent where $n_{f}$ is the number of scanned frequencies, $\Delta f$ is the frequency interval used when calculating the periodogram, and $\tau$ is the duration of the observation.

A more useful number may be the \textit{significance} of a detection, or how sure we are that it is real; this is simply 1~\minus~FAP, expressed as a percentage. This is the value that will be used throughout this paper to estimate the importance and believability of a signal, and is given by

\begin{equation}
  Sig (\%) = 100 \times \bigl( 1 - e^{-Z} \bigr)^{M}
\end{equation}

In the Lomb-Scargle periodogram, the peak-to-trough amplitude \textit{A} of the modulation in the signal is related to the power $P_{LS}$ through

\begin{equation}
  A = 4\sqrt{\frac{P_{LS}\,\sigma_n^2}{N}}
  \label{eq_ls_amp}
\end{equation}

\noindent where $N$ is the number of data points \citep{scargle82}.

If the signal detected has any harmonics, its total power will be divided between the individual harmonic peaks in the power spectrum. Using only the fundamental to estimate the amplitude of the signal's pulsations could then severely underestimate it if there were considerable power in any of the harmonics (which is often the case). If the amplitude of all the harmonics is known, it can be shown that the total amplitude of the signal will be given by

\begin{equation}
  A_{total} = \sqrt{\sum_i{A_i^2}}
  \label{eq_ls_totamp}
\end{equation}

From the error in the angular frequency detected at a certain power in the Lomb-Scargle periodogram \citep{horne86}, we derive the error on the period as

\begin{equation}
  dP = \frac{3}{4} \left( \frac{P^2\,\sigma_n}{\sqrt{N}\,\tau\,A} \right)
  \label{eq_dp}
\end{equation}

\noindent where \emph{A} is the Lomb-Scargle amplitude given by \eref{eq_ls_totamp}, and $\sigma_n$ is the standard deviation of the noise, although the standard deviation of the actual data is used, as explained earlier.

Apart from the global significance of a detection, we define an additional quantity, the \textit{local significance}, as the significance of a peak at frequency $f$, within a region of frequency space extending 5\% of $f$ to either side of it.

\section{Luminosity and magnetic field estimation} \label{sec_appendix_b}

If all the matter accreting onto a neutron star is converted into energy, the luminosity that will result is simply the gravitational energy lost by the in-falling mass \citep{fkr}:

\begin{equation}
  L_{{\rm x}} = \frac{GM\dot{M}}{R}
  \label{eq_luminosity}
\end{equation}

\noindent where $\dot M$ is the mass transfer rate. Some manipulation and substitution can provide a more manageable expression:

\begin{equation}
  L_{{\rm x}_{37}} = 8.4 \times 10^9 \left(\frac{M_n \dot{M}}{R_n}\right)
  \label{eq_luminosity1}
\end{equation}

\noindent which will give us the luminosity in terms of 10\supersm{37}\ergps, and where $M_n$ is the mass of the neutron star in units of \msunt, $\dot{M}$ is the mass accretion rate in units of \msunt \yr \supersm{-1}, and $R_n$ is the radius of the neutron star in km.

The angular momentum of a neutron star is given by

\begin{equation}
  \mathcal{L}_n = \frac{2\pi I_n}{P_{\rm s}}
\end{equation}

\noindent where \ps\ is the spin period, and the moment of inertia is given by

\begin{equation}
  I_n = \frac{2}{5}M_n R_n^2
  \label{eq_inertia}
\end{equation}

\noindent with $M_n$ and $R_n$ being the mass and radius of the neutron star in standard units.

The torque experienced by a neutron star spinning up or down is given by

\begin{equation}
  \vert\tau\vert \equiv \left\vert \frac{d\mathcal{L}_n}{dt}  \right\vert = 2\pi I_n\; \frac{\dot{P_{\rm s}}}{P_{\rm s}^2}
  \label{eq_torque1}
\end{equation}

\noindent with \pdot\subs{s} being the rate of change of the spin period \citep{fkr}.

For an accreting pulsar undergoing steady spin up/down, the applied torque will depend on the mass accretion rate, $\dot M$, and the angular momentum of matter in the accretion disc at the magnetospheric radius, $r_m$. This torque is given by

\begin{equation}
  \tau = \dot{M}\sqrt{GM_n r_m}
  \label{eq_torque2}
\end{equation}

The maximum torque possible will occur when $r_m = r_{co}$, where the corotation radius\footnote{The corotation radius is defined as the radius at which matter in the disc is moving at the same speed as the neutron star's surface.} is given by

\begin{equation}
  r_{co} = \left( \frac{GM_n P_{\rm s}^2}{4\pi^2}  \right)^{\frac{1}{3}}
\end{equation}

\noindent Substituting this value in \eref{eq_torque2} will provide the expression for the maximum torque possible:

\begin{equation}
  \tau_{\rm max} = \dot{M} \left[ \frac{G^2M_n^2P_{\rm s}}{2\pi} \right]^{\frac{1}{3}}
  \label{eq_torque_max}
\end{equation}

\noindent Clearly, $\vert\tau\vert \leq \tau_{\rm max}$, so using Eqs. \ref{eq_torque1} and \ref{eq_torque_max}, and substituting the expression for the moment of inertia from \eref{eq_inertia}, we find the accretion rate will be

\begin{equation}
  \dot{M} \geq \frac{2}{5}\,R_n^2\dot{P_{\rm s}} \left[ \frac{16\pi^4M_n}{P_{\rm s}^7G^2} \right]^{\frac{1}{3}} \kg\, {\rm s}^{-1}
\end{equation}

\noindent substituting this value in \eref{eq_luminosity} we finally obtain the lower limit on the luminosity that will be produced through accretion:

\begin{equation}
L_{{\rm x}_{37}} \geq \frac{2R_{\rm n}\dot{P_{\rm s}}}{5\times10^{30}} \left[\frac{16\pi^{4} G M_n^4}{P_{\rm s}^7}\right]^{\frac{1}{3}}
\label{eq_luminosity2}
\end{equation}

\noindent which will be in units of \expt{37}\ergps\ if S.I. units are used (and the neutron star mass is in \msunt). This equation will allow the estimation of the luminosity associated with an outburst if the average spin up/down is measured.

One further value that can be estimated is the magnetic field of the neutron star. Rearranging Eq.~(6.24) of \citet{fkr}, and using the period in place of the frequency, a constraint can be placed on its value:

\begin{equation}
  B_{12} \leq \left[ 3.4\times10^{-4}\, R_n^{-2}\, M_n^{\frac{-10}{7}}\, L_{{\rm x}_{37}}^{\frac{6}{7}}\, \frac{P_{\rm s}^2}{\dot{P}_{\rm s}}\,  \right]^{\frac{-7}{2}}
\end{equation}

\noindent where the magnetic field will be in units of \expt{12}\G\ if $M_n$ is in \msunt\ and $R_n$ in metres. In the present work we assume values for the neutron star's radius and mass of $R_n = 10^4\m$ and $M_n = 1.4\Msun$, respectively.

\label{lastpage}

\end{document}